\documentclass[lettersize,journal]{IEEEtran}
\usepackage{amsmath,amsfonts}
\usepackage{makecell}
\usepackage{amsthm}
\usepackage{amssymb}
\usepackage{algorithmic}
\usepackage{algorithm}
\usepackage{array}
\usepackage[table]{xcolor}
\usepackage{wrapfig}
\usepackage{cite}
\usepackage{supertabular,booktabs}
\usepackage{subfig}
\usepackage{subfloat}
\usepackage{textcomp}
\usepackage{stfloats}
\usepackage{float}
\usepackage{url}
\usepackage{comment}
\usepackage{verbatim}
\usepackage{comment}
\usepackage{graphicx}
\usepackage{cite}
\usepackage[numbers,sort&compress]{natbib}
\usepackage{threeparttable}
\usepackage{tabularx}
\usepackage{booktabs}
\usepackage{xcolor}
\usepackage[a4paper,margin=1.1cm]{geometry}
\usepackage{booktabs}   
\usepackage{tabularx}   
\usepackage{multirow}   
\usepackage{enumitem}   
\usepackage{enumitem}     

\usepackage{tikz}
\usetikzlibrary{plotmarks}
\usepackage{pgfplots}
\usetikzlibrary{spy}
\usepackage{collectbox}
\usepgfplotslibrary{groupplots}
\usetikzlibrary{spy}
\usepackage{pgfplots}
\pgfplotsset{compat=1.18} 

\usepackage{xcolor} 
\newcommand{\Nhan}[1]{\textit{\textcolor{red}{[\underline{Nhan's comment:} #1]}}}

\newcommand{\norm}[1]{\left\lVert#1\right\rVert}

\usepackage{array}
\newcolumntype{L}[1]{>{\raggedright\let\newline\\\arraybackslash\hspace{0pt}}m{#1}}
\newcolumntype{C}[1]{>{\centering\let\newline\\\arraybackslash\hspace{0pt}}m{#1}}
\newcolumntype{R}[1]{>{\raggedleft\let\newline\\\arraybackslash\hspace{0pt}}m{#1}}

\usepackage[hidelinks]{hyperref}
\usepackage{enumitem} 
\setlist[itemize]{label=\textbullet, nolistsep, leftmargin=5pt,
            before*={\mbox{}\vspace{-\baselineskip}}, after*={\mbox{}\vspace{-\baselineskip}}}
\IEEEoverridecommandlockouts
\begin{document}
\theoremstyle{definition}
\newtheorem{exmp}{Example}

\title{Comprehensive Review of Deep Unfolding Techniques for Next-Generation Wireless Communication Systems \vspace{-1.3mm}}

\author{Sukanya Deka, Kuntal Deka, Nhan Thanh Nguyen, Sanjeev Sharma, Vimal Bhatia,  and Nandana Rajatheva
\thanks{Sukanya Deka and Kuntal Deka are with the Department of EEE, Indian Institute of Technology Guwahati, India. Email: \{sukanya.deka, kuntaldeka\}@iitg.ac.in.}
\thanks{Nhan Thanh Nguyen and  Nandana Rajatheva are with Centre for Wireless Communications, University of Oulu, Finland. Email: \{nhan.nguyen, nandana.rajatheva\}@oulu.fi.}
\thanks{Sanjeev Sharma is with the Department of Electronics Engineering, Indian Institute of Technology (BHU) Varanasi, India. Email: sanjeev.ece@iitbhu.ac.in.}
\thanks{Vimal Bhatia is with Department of EE,  Indian Institute of Technology Indore, India. Email: vbhatia@iiti.ac.in.}\vspace{-5mm}
\thanks{Copyright (c) 2026 IEEE. Personal use of this material is permitted.
However, permission to use this material for any other purposes must be
obtained from the IEEE by sending a request to
pubs-permissions@ieee.org.}
}

\maketitle

\begin{abstract}
    The massive surge in device connectivity demands higher data rates, increased capacity with low latency and high throughput. Hence, to provide ultra-reliable, low-latency communication with ubiquitous connectivity for Internet-of-Things (IoT) devices, next-generation wireless communication leverages the incorporation of machine learning tools. 
    However, standard data-driven models often need large datasets and lack interpretability. To overcome this, model-driven deep learning (DL) approaches combine domain knowledge with learning to improve accuracy and efficiency. Deep unfolding is a model-driven method that turns iterative algorithms into deep neural network (DNN) layers. It keeps the structure of traditional algorithms while allowing end-to-end learning. This makes deep unfolding both interpretable and effective for solving complex signal processing problems in wireless systems.
	We first present a brief overview of the general architecture of deep unfolding to provide a solid foundation. We also provide an example to outline the steps involved in unfolding a conventional iterative algorithm. We then explore the application of deep unfolding in key areas, including signal detection, channel estimation, beamforming design, decoding for error-correcting codes, integrated sensing and communication, power allocation, and physical layer security. Each section focuses on a specific task, highlighting its significance in emerging 6G technologies and reviewing recent advancements in deep unfolding-based solutions.
	Finally, we discuss the challenges associated with developing deep unfolding techniques and propose potential improvements to enhance their applicability across diverse wireless communication scenarios.  
	
\end{abstract}

\vspace{-0.1in}
\section{Introduction}
\IEEEPARstart{T}{he} future mobile generations are expected to experience a surge in resource demand due to the exponential growth of connected devices. With the advent of 5G, billions of devices require seamless connectivity, intensifying competition for efficient spectrum allocation \cite{5G,6G,LMIMO}. Each network has unique requirements and constraints, making resource allocation a complex challenge. Given the inherent limitations of available resources, problem formulation must prioritize flexibility, adaptability, and efficient utilization.
Significant advancements have taken place in wireless communication due to various exciting physical layer technologies, including massive multiple-input multiple-output (MIMO), Internet-of-Things (IoT), integrated sensing and communication (ISAC), intelligent reflecting surfaces (IRS), millimeter-wave (mmWave) communication, physical layer security, and next-generation multiple access such as nonorthogonal multiple access (NOMA). Additionally, orthogonal time-frequency space (OTFS) modulation is emerging as a promising alternative to orthogonal frequency-division multiplexing (OFDM), further improving spectral efficiency (SE) and resilience in next-generation networks. 

Deep learning (DL), a machine learning subclass, maps the relationship between input and output without explicitly modeling it in the architecture \cite{Goodfellow}. It involves the use of deep neural network (DNN) with multiple hidden layers (hence ``deep'') to analyze and interpret data. Numerous architectures have been proposed such as multilayer perceptron (MLP), convolutional neural network (CNN), deep reinforcement learning (DRL), generative adversarial networks (GANs), etc. DNNs have been applied to various components of wireless communication, including detection, channel estimation, beamforming designs, and power allocation \cite{Huang_DLPhyLayer, Ye_powerDL, Mashhadi_DL_pilot_chhEst, BF_DNN,Chnnl_DNN}. However, DL architectures exhibit several inherent limitations. These include complex training procedures and a black-box nature of algorithms. The training process for these models requires extensive datasets, considerable time, and intricate fine-tuning. The complexity of implementing DNN architectures often constrains their applicability in real-world systems. Therefore, DL models are highly sensitive to the specific configurations of wireless systems, such as the number of antennas, users, and operating frequencies, which require retraining or reimplementation whenever these parameters change. These challenges are not typically encountered in conventional optimization techniques, underscoring the potential advantages of using classical algorithms \cite{He_model_MIMOdet1, He_model_MIMOdet2}.

 \IEEEpubidadjcol 

Communication problems are inherently influenced by dynamic physical constraints that evolve in real-time. These constraints stem from factors such as the physical environment, hardware limitations, and the specific operational requirements of communication systems. Different communication scenarios impose varying demands on bandwidth, power consumption, and latency, which fluctuate with user density, data rates, and service types. Interpreting and addressing these constraints using DNNs presents significant challenges. Data-driven neural network (NN) architectures were developed primarily to process language and vision data, rendering them suboptimal for wireless communication tasks that involve channels, signals, and outputs governed by complex constraints. Although certain constraints can be handled through the application of specific activation functions, for example, employing the $tanh$ function to enforce a constant modulus constraint in analog beamforming, standard activation functions generally lack the versatility to manage the intricate constraints characteristic of wireless communication systems. To mitigate these challenges, model-driven DL approaches facilitate the development of problem-specific network architectures and customized activation functions, optimized for the unique signal processing requirements of wireless communication \cite{Survey_10Eldar_MLandWC}.



Compared to the conventional fully-connected DL architectures, model-driven DL have fewer trainable parameters, implying less training time and computational complexity. Deep unfolding 
is an approach that falls into the category of model-driven DL \cite{Survey9_DU_Optim}. 
There are several types of model-driven DL approaches, including deep unfolding \cite{Survey6_Monga_AlgoUnroll},\cite{Survey4_DU}, physics informed NN \cite{PINN_1,PINN_2}, learned optimizers \cite{learn_convex, learn_convex_control}, etc.
A deep unfolded network can be trained in an easier and faster fashion with smaller datasets than DNNs. This is because deep unfolding incorporates domain knowledge from iterative algorithms into layers of DNNs, offering improved performance and interpretability \cite{Gregor}.

    \vspace{-0.1in}
\begin{table}[!htbp]
    \centering
    \caption{Acronyms and corresponding full names. }
    \label{table:Acronyms}
    \begin{tabular}{|l|l|}
        \hline
        \textbf{Abbreviation} & \textbf{Description} \\ \hline
        AO & Alternate optimization \\ \hline
        AP & Access point \\ \hline
        AMP & Approximate message passing \\ \hline
        ADC & Analog-to-digital converters \\ \hline
        ADMM & Alternating direction method of multipliers \\ \hline
        BP & Belief propagation \\ \hline
        BS & Base station \\ \hline
        BER & Bit error rate \\ \hline
        CD & Coordinate descent \\ \hline
        CNN & Convolutional neural network \\ \hline
        CSI & Channel state information \\ \hline
        CGD & Conjugate gradient descent \\ \hline
        DL & Deep learning \\ \hline
        DNN & Deep neural network \\ \hline
        DRL & Deep reinforcement learning \\ \hline
        EM & Expectation maximization \\ \hline
        EP & Expectation propagation \\ \hline
        FDD & Frequency division duplexing \\ \hline
        GAN & Generative adversarial network \\ \hline
        GNN & Graph neural network \\ \hline
        HBF & Hybrid beamforming \\ \hline
        IRS & Intelligent reflecting surface \\ \hline
        IoT & Internet-of-things \\ \hline
        ISI & Inter-symbol interference \\ \hline
        ISAC & Integrated sensing and communication \\ \hline
        ISTA & Iterative shrinkage-thresholding algorithm \\ \hline
        LR & Low-resolution \\ \hline
        LS & Least squares \\ \hline
        LLR & Log-likelihood ratio \\ \hline
        LDPC & Low-density parity check \\ \hline
        LMMSE & Linear minimum mean square error \\ \hline
        MF & Matched filter \\ \hline
        ML & Maximum likelihood \\ \hline
        MO & Manifold optimization \\ \hline
        MU & Multi-user \\ \hline
        MAP & Maximum \textit{a posteriori} \\ \hline
        MPD & Message passing detector \\ \hline
        MSE & Mean square error \\ \hline
        MMSE & Minimum mean square error \\ \hline
        MIMO & Multiple input multiple output \\ \hline
        MISO & Multiple input single output \\ \hline
        MSER & Minimum symbol error rate \\ \hline
        mMTC & Massive machine type communication \\ \hline
        mmWave & millimeter-wave \\ \hline
        NN & Neural network \\ \hline
        NOMA & Non-orthogonal multiple access \\ \hline
        NMSE & Normalized mean square error \\ \hline
        OAMP & Orthogonal approximate message passing \\ \hline
        OFDM & Orthogonal frequency division multiplexing \\ \hline
        OTFS & Orthogonal time frequency space \\ \hline
        PGA & Projected gradient ascent \\ \hline
        PGD & Projected gradient descent \\ \hline
        QAM & Quadrature amplitude modulation \\ \hline
        QoS & Quality of service \\ \hline
        QPSK & Quadrature phase shift keying \\ \hline
        RF & Radio frequency \\ \hline
        RNN & Recurrent neural network \\ \hline
        RSMA & Rate splitting multiple access \\ \hline
        SD & Sphere decoder \\ \hline
        SE & Spectral efficiency \\ \hline
        SER & Symbol error rate \\ \hline
        SCA & Successive convex approximation \\ \hline
        SGD & Stochastic gradient descent \\ \hline
        SIC & Successive interference cancellation \\ \hline
        SNR & Signal-to-noise ratio \\ \hline
        SSCA & Stochastic successive convex approximation \\ \hline
        SINR & Signal-to-interference and noise ratio \\ \hline
        THz & Terahertz \\ \hline
        TPGD & Trainable projected gradient descent \\ \hline
        WMMSE & Weighted minimum mean square error \\ \hline
        ZF & Zero forcing \\ \hline        
    \end{tabular}
\end{table}
\vspace{-0.05in}
\subsection{Related Works}
The growing application of deep unfolding in communication systems highlights the need for a comprehensive survey in this area. There are limited survey articles that comprehensively discuss deep unfolding applications in the context of the physical layer of communication. Here, we present a few related survey articles on deep unfolding that explore some particular task such as detection, channel estimation, or precoder design. Wang \textit{et al.} in \cite{Survey1_DL_Phy_layer} presented a review of DL and deep unfolding approaches that address physical layer signal processing challenges, including MIMO detection, channel estimation, and channel decoding. The authors of \cite{Survey2_Phy} provided an overview of deep unfolding applications in transceiver design, MIMO detection, and channel state information (CSI) estimation and feedback. Similarly, in \cite{Survey3_DL_in_PhyLayer}, the authors highlighted the role of DL in communication, with a focus on DNNs, GANs, recurrent neural network (RNN), and end-to-end learning. They also explored deep unfolding for Bayesian optimal estimation, using MIMO detection and channel estimation as case studies. The authors in \cite{Survey4_DU} focused primarily on MIMO detection while exploring its use in multiuser MIMO precoding and belief propagation-based channel decoding. A survey on model-based, DL, and hybrid approaches in the context of wireless communication networks is provided in \cite{Survey5_Wireless_Both}. In \cite{Survey6_Monga_AlgoUnroll}, Monga \textit{et al.} conducted a survey of deep unfolding algorithms specifically implemented for applications in signal and image processing. The authors in ~\cite{Survey7_Redefine} reviewed the implementation of deep unfolding in signal processing tasks. A brief review on deep unfolding in MIMO detection is presented in \cite{Survey8_Model_guidelines}. The authors in \cite{Survey9_DU_Optim} have presented a general framework for optimization that leverages the benefit of deep unfolding with signal detection as one of the case studies. In \cite{Survey_10Eldar_MLandWC}, a chapter is dedicated to model-based machine learning for communication. The deep unfolding networks have been discussed for signal detection, channel estimation, and channel coding. Machine learning-based optimization framework is discussed in \cite{Survey11_ML_Optim} with an emphasis on deep unfolding in specific design paradigms such as signal detection. Quoc \textit{et al.} have presented a survey on application of deep unfolding in signal detection, channel estimation and beamforming in \cite{IntelligentRadio}. The authors in \cite{Survey_DU_24} explored the application of learning-based optimization techniques to enhance the capacity of communication networks. The survey reviews key domains such as encoding-decoding, equalization, power control, and beamforming, emphasizing the impact of optimization through learning on network performance. A comparative study between purely data-driven and deep unfolding is conducted with complete emphasis on HBF optimization in \cite{Survey12_AI_MIMO} The significant rise in application of deep unfolding has led to development of theoretical framework to validate and support practical experiments. There are several works \cite{ISTA_theoretical, LISTA_theoretical, ALISTA_theoretical} which primarily focus on the convergence rate of the unfolded iterative shrinkage-thresholding algorithm (ISTA) and its variants such as learned ISTA (LISTA) \cite{LISTA_theoretical} and analytic LISTA (ALISTA) \cite{ALISTA_theoretical}. The authors of \cite{L2O_survey} discuss the motivation and design methodology of deep unfolding, using LISTA as a representative case study. They have studied the design of learned parameter, techniques to design the architecture of the network and parameterize the resulting unfolded network. Certain open mathematical problems pertaining to deep unfolding are also explored in \cite{L2O_survey}. 
\subsection{Contributions}
Recently, there has been a surge of application of deep unfolding in various areas of physical layer of wireless communication. The existing survey contributions such as \cite{Survey4_DU, Survey1_DL_Phy_layer, Survey2_Phy}, all published before 2020, focus on some particular areas of the physical layer of wireless communication, such as detection, channel estimation, precoder design, or channel coding. In~\cite{Survey12_AI_MIMO}, the authors have provided a survey on application of deep unfolding only in the area of hybrid beamforming design. However, the realm of deep unfolding has recently expanded into emerging areas such as sensing and communication, which have not been addressed in any of the previous works. Most of the previous literature discuss the general approach of deep unfolding without dedicated analysis concerning communication perspective. Only a few works such as \cite{Survey6_Monga_AlgoUnroll, Survey11_ML_Optim} have given a tutorial description but with a biased emphasis towards signal processing tasks. Therefore, we provide a detailed tutorial describing the deep unfolding in communication perspective and discuss convergence proofs that impact algorithmic stability. In addition, in this work, the application of deep unfolding is explored in the context of critical 6G technologies that are not addressed in existing survey articles such as~\cite{Survey4_DU, Survey1_DL_Phy_layer, Survey2_Phy}. Therefore, there is a lack of coverage of recent studies addressing critical 6G technologies, such as mmWave spectrum, new air interface that includes massive MIMO, IRS, next-generation multiple access, and are often limited to specific application areas. Consequently, a comprehensive survey is needed to encompass the broad applications of deep unfolding techniques across various aspects of the physical layer in wireless communication which has not been addressed in existing survey articles~\cite{Survey4_DU, Survey1_DL_Phy_layer, Survey2_Phy, Survey12_AI_MIMO}. This paper provides an in-depth review of deep unfolding methods applied to the physical layer of wireless communication. 
The contributions are summarized below:
\begin{figure}[t]
	\vspace{-0.1in}
	\centering
	\includegraphics[width=1\linewidth]{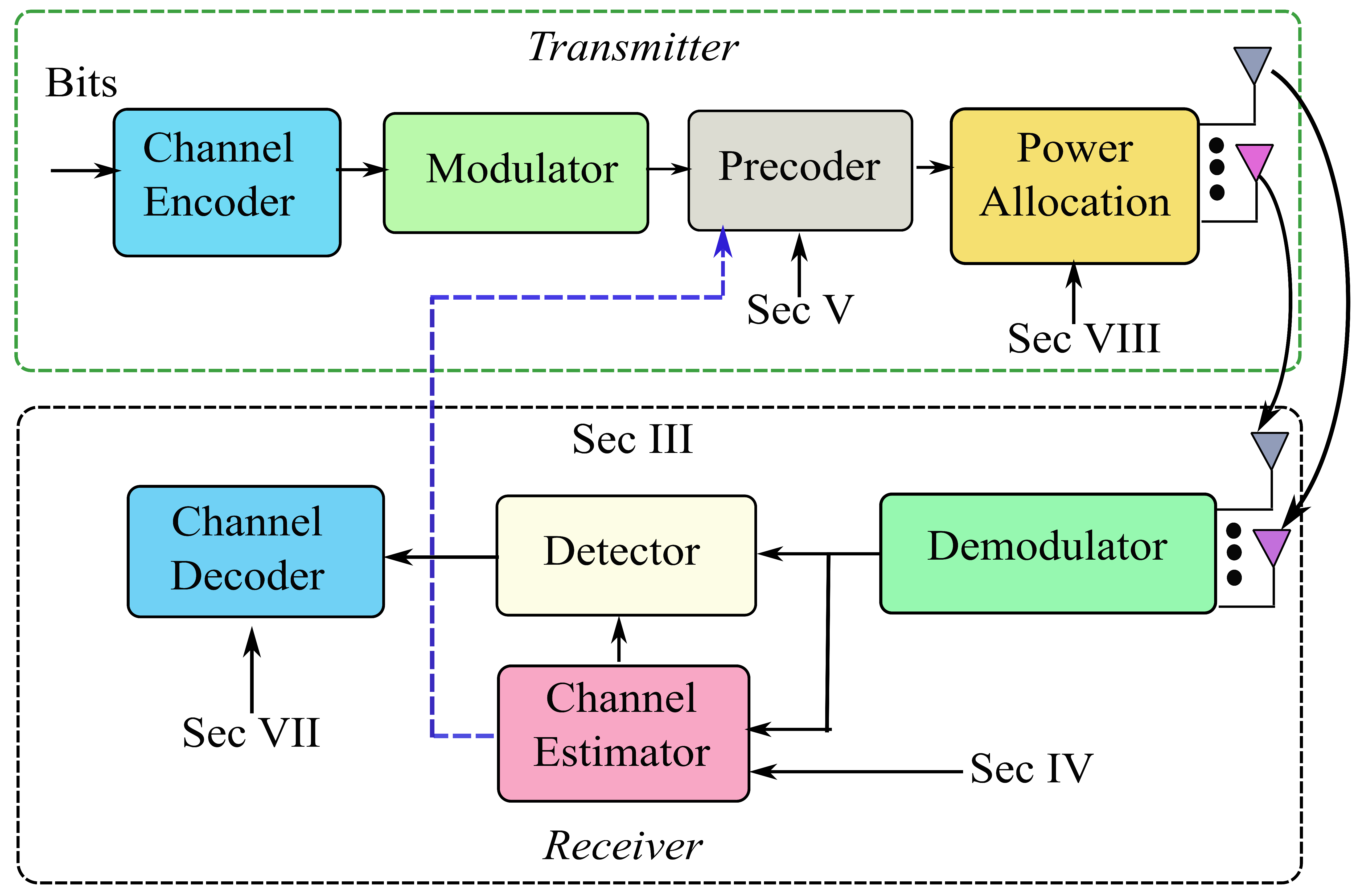}
	\caption{Typical block diagram of a modern wireless communication system.  The section number covering a particular component is shown as well. }
    
	\label{fig:block_diagram}
	\vspace{-0.1in}
\end{figure}

\begin{enumerate}
	\item 
This paper presents a systematic, step-by-step approach for developing deep unfolding algorithms, with a detailed tutorial on the alternating direction method of multipliers (ADMM)-based MIMO detection. The tutorial offers insights into the ADMM unfolding process, identification of the trainable parameters, and demonstrates the design of the resulting unfolded network. Furthermore, we discuss the theoretical foundation of the convergence properties  and guidelines for selecting hyperparameters that impact algorithmic stability of deep unfolding. We have highlighted that the convergence properties can be studied under three categories: convergence order, generalization error and normalized mean squared error (NMSE).
	\item Fig.~\ref{fig:block_diagram} depicts the principal components of the physical layer in contemporary wireless communication systems. Considering this block diagram, we undertake a systematic survey of each component in details. In particular, we investigate the role of deep unfolding for signal detection, channel estimation, precoder design, integrated sensing and communication (ISAC), decoding, power allocation, and physical-layer security. For each domain, we highlight the integration of deep unfolding with key enabling technologies for sixth-generation (6G) networks, such as intelligent reflecting surfaces (IRS), millimeter-wave (mmWave) communications, and the Internet of Things (IoT). The survey aims to provide a comprehensive perspective on the design of adaptable and forward-looking frameworks for next-generation wireless systems.
    \item In addition, each section concludes with a cross method comparison table that evaluates prominent unfolded networks across key metrics: computational complexity (floating point operations (FLOPs) and big $\mathcal{O}$ complexity), memory footprint (number of trainable parameters) and convergence speed. These tables highlight the practical trade-offs and deployment boundaries associated with each architecture. Furthermore, we discuss the key takeaways to provide readers with deeper insights into the practical applicability and design considerations of deep unfolding in physical layer communication. 
	\item This paper concludes the survey by outlining a set of challenges in developing deep unfolding models and propose future directions that could guide advancements in deep unfolding techniques to enhance their performance. We point out that challenges in designing a deep unfolding model mainly emerge from scenarios where these models does not offer significant gain in performance-complexity tradeoff. However, we also provide potential solutions that deep unfolding models can benefit from to overcome those challenges. 
    \item We also provide open source representative works with its github links for different design problems such as detection, channel estimation, ISAC, among others.  We believe that this will enable readers to reproduce and develop new unfolding methods.
\end{enumerate}

\subsection{Paper Outline}
\begin{figure}[!htbp]
	\vspace{-0.1in}
	\centering
	\includegraphics[width=1\linewidth]{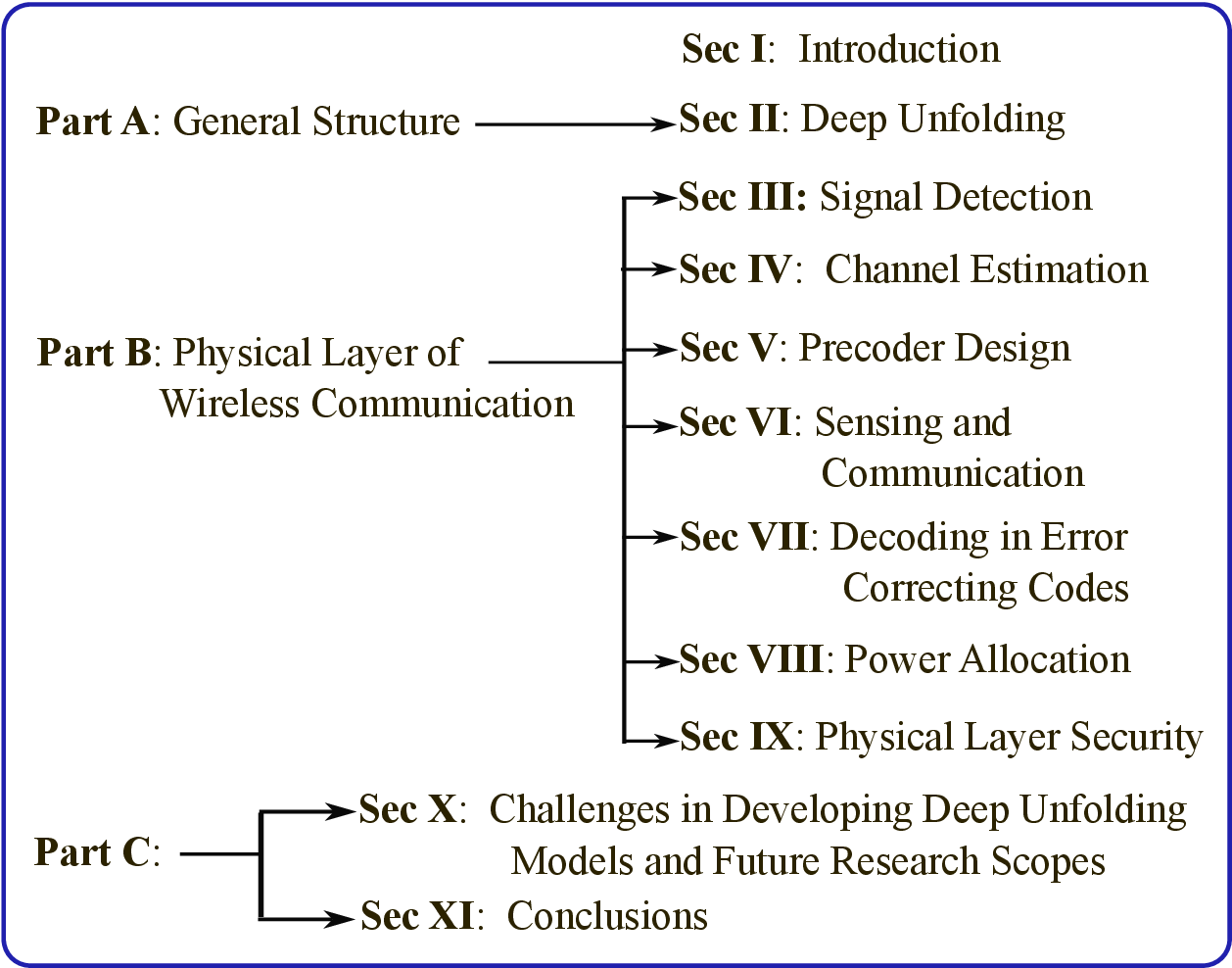}
	\caption{Outline of the paper.}
	\label{fig:outline}
	\vspace{-0.1in}
\end{figure}

The skeleton of this paper is depicted in Fig. ~\ref{fig:outline}.
This survey paper is organized as follows:

\begin{itemize}
	\item \textbf{Part A}: 
    Section~\ref{sec::deep_unfolding} presents the deep unfolding framework, which combines classical iterative algorithms with DNNs. The understanding of deep unfolding is exemplified via ADMM based MIMO detection. A step-by-step approach is presented for developing the unfolding of ADMM specifically for MIMO detection. Also a brief understanding on how trainable parameters impact the convergence order of the deep unfolded network is discussed.  
	\item \textbf{Part B}: This part explores deep unfolding applications in physical layer tasks: signal detection (Section~\ref{sec::signal_detection}), channel estimation (Section~\ref{sec::channel_estimation}), precoder design (Section~\ref{sec::precoder_design}), ISAC (Section~\ref{sec::sensing_communication}), decoding for error correcting code (Section~\ref{sec::decoding_ECC}), power allocation (Section~\ref{sec::power_allocation}), and security (Section~\ref{sec::security}). 
	\item \textbf{Part C }: {Section~\ref{sec::challenge} discusses the limitations of existing deep unfolding approaches and the challenges associated with their adoption in wireless communications. We further outline potential directions and solutions to address these limitations, highlighting promising avenues for future research.}
	Section~\ref{sec::conclusion} concludes the paper. 
\end{itemize}

\subsection{Notation and Acronyms}
Lower case, bold-face lower case, and bold-face upper case letters denote scalars, vectors, and matrices respectively. $(\cdot)^T$ and $(\cdot)^H$ denote transpose and Hermitian transpose, respectively.  $\Vert \cdot \Vert$ denotes the Euclidean norm of a vector. 
$\mathbb{R}^n$ and $\mathbb{C}^n$ denote $n$-dimensional real and complex vector spaces, respectively. $\mathcal{N}(\mu,\Sigma)$ denotes the multivariate Gaussian distribution with mean vector $\mu$ and covariance matrix $\Sigma$, while $\mathcal{CN}(\mu,\Sigma)$ denotes the complex multi-variate Gaussian distribution. Table~\ref{table:Acronyms} lists the main acronyms used in this paper.
\section{Deep Unfolding: General Architecture and Fundamentals} \label{sec::deep_unfolding}
\begin{figure}
\centering
\includegraphics[width=1.03\linewidth]{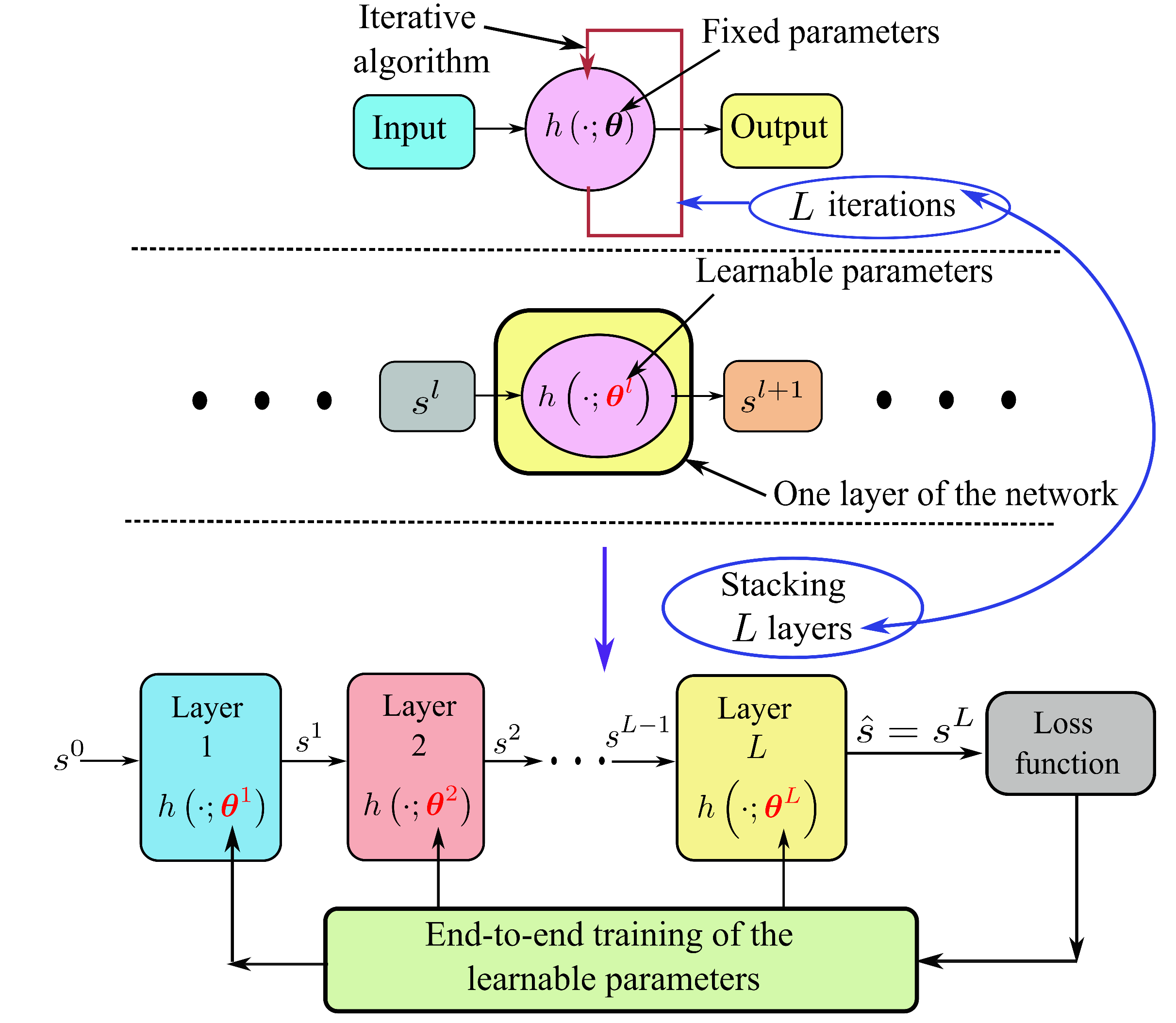}
\caption{General framework of training for  deep unfolding. The learnable parameters are shown in \textcolor{red}{red color}. }
\label{fig:deep_unfolding_only}
\end{figure}

Deep unfolding, or algorithm unrolling \cite{Survey6_Monga_AlgoUnroll}, bridges model-based iterative algorithms and data-driven DNNs \cite{He_model_MIMOdet1}, \cite{Survey2_Phy}, \cite{JADC_Qiang}, \cite{Liao_MassiveMIMODEt}, \cite{Guo_CSIfeedbackMassiveMIMO}, \cite{Li_ExpectOTFS}, \cite{Zhao_underwater_OFDM}, \cite{He_DU_Turbo}. It maps each iteration of  an iterative algorithm to a DNN layer, enabling end-to-end learning while retaining interpretability. In deep unfolding, the learnable parameters include model-specific variables and regularization coefficients, optimized layer by layer. Fig.~\ref{fig:deep_unfolding_only} illustrates the process of unfolding an iterative algorithm.
In general, for any iterative algorithm, the update equation for each iteration is given as:
\begin{equation}
{\bf{s}}^{l+1}=h({\bf{s}}^{l};{\boldsymbol{\theta}}^{l})
\end{equation}
where $\left\{{\bf{s}}^{l}\right\}_{l=1}^L \in {\mathbb{R}}^{n}$ denotes the sequence of variable vector updated across $L$ layers of the algorithm. At each layer, the update is performed by executing $h(\cdot;\cdot)$. Repeated application of this update rule over $L$ layers, yields the unfolded network architecture, $h(\cdot;{\boldsymbol{\theta}}^{l})_{l=1}^{L}$.
The trainable parameters, $\left\{\boldsymbol{\theta}^{l}\right\}_{l=1}^L$
(shown in red fonts in Fig.~\ref{fig:deep_unfolding_only}) includes the model parameters and the step-size for the regularization coefficients of the algorithm \cite{Survey11_ML_Optim}. 
To learn the trainable parameters $\left\{\boldsymbol{\theta}\right\}_{l=1}^L$ in an end-to-end approach, the loss $\ell({\bf{s}}^{L},\bf{s})$ is computed, 
where ${\bf{s}}^{L}$ is the output of the network and $\bf{s}$ is the ground truth target. The key idea behind end-to-end learning is that all parameters across $L$ layers are simultaneously adjusted through backpropagation, where the loss is minimized using gradient-based optimization.
Feeding the data forward through the unfolded network with trainable parameters in the testing stage is similar to executing the parameter-optimized iterative algorithm for a finite number of iterations. 
Thus, the number of layers in deep unfolding models is fixed, resulting in a predefined computational complexity. Within these fixed layers, the model is trained to optimize the trainable parameters, enabling the problem to be addressed using a structured algorithm with parameters tailored for optimal performance. 

\begin{exmp}
 To develop an understanding on deep unfolding, we take ADMM as an example. ADMM decomposes a complex optimization problem into simpler subproblems, which are solved iteratively. The algorithm combines the principles of dual ascent and the method of multipliers. Each iteration involves solving the subproblems for the primal variables in an alternating manner, followed by updating the dual variables (Lagrange multipliers) to enforce consistency with the original problem's constraints \cite{ADMM}. This iterative process ensures that the subproblem solutions converge to the global solution of the main optimization problem. ADMM in its unfolded version has been widely used in MIMO signal detection \cite{Survey7_Redefine}, \cite{Inexact_ADMM_det_2021} and decoding in error correction codes \cite{Wei_DUADMM_Linear, IrregularLDPC_2023}. Here, we show the unfolding of ADMM for MIMO detection as an illustrative example.  
\begin{figure*}[!htbp]
\centering
\includegraphics[width=1\linewidth, height=0.25\linewidth]{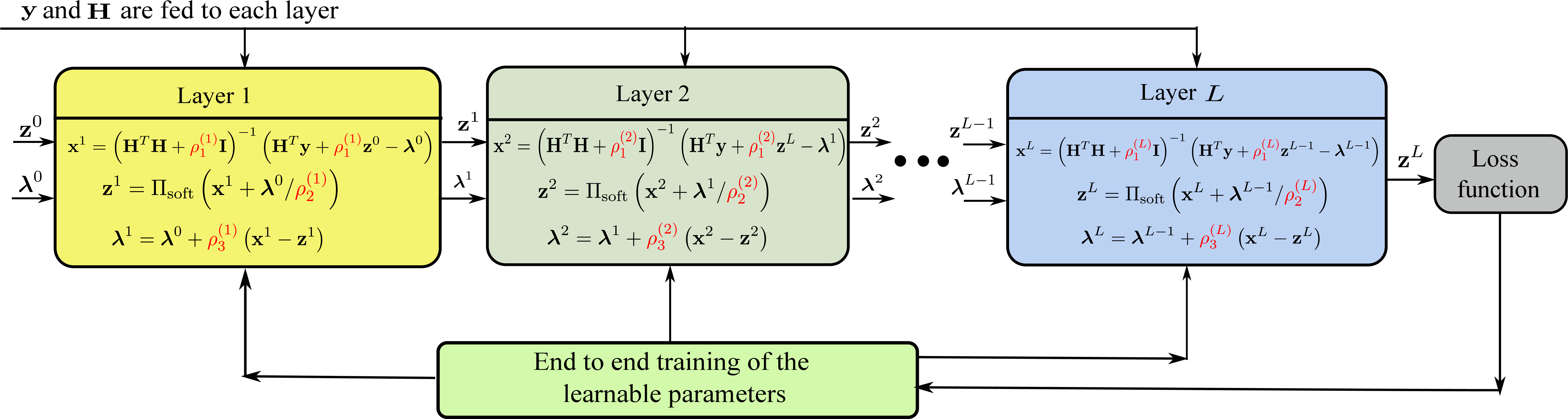}
\caption{Deep unfolding training framework for ADMM-based MIMO detection.  The learnable parameters are shown in \textcolor{red}{red color}. }
\label{fig:unfold_admm_det}
\end{figure*}

Consider a MIMO system with $M$ transmit antennas and $N$ receive antennas. The received signal vector at the receiver can be given by
\begin{equation}
\label{rxd}
 {\bf{\bar y}} = {\bf{\bar H\bar x}} + {\bf{\bar n}} \nonumber
\end{equation}
where 
${\bf{\bar{H}}} \in \mathbb{C}^{N \times M}$ represents the channel matrix, ${\bf{\bar{x}}} \in {\cal{\bar{X}}}^{M}$ is the transmitted symbol vector, and ${\bf{\bar{n}}} \in \mathbb{C}^N$ denotes additive white Gaussian noise (AWGN) vector at the receiver. Here, $\cal{\bar{X}}$ represents the complex constellation set. In this example, we consider the QPSK modulation scheme, and thus, $\left( {{\cal{\bar{X}}} = \left\{ {1 + i, - 1 + i, - 1 - i,1 - i} \right\}} \right)$. The maximum likelihood (ML) detection rule is given as
\begin{equation}
\label{ML_eq_c}
 \min _{ \bf{\bar x} \in {\cal{\bar X}}^{M}} \norm{ {\bf{\bar y}} - {\bf{\bar H \bar x}}}_{2} ^{2}
\end{equation}
Denote the following real-valued vectors and matrix:
\begin{equation*}
\mathbf{y} \triangleq
\begin{bmatrix}
\mathfrak{R}(\bar{\mathbf{y}}) \\ 
\mathfrak{I}(\bar{\mathbf{y}})
\end{bmatrix}, \quad
\mathbf{x} \triangleq 
\begin{bmatrix}
\mathfrak{R}(\bar{\mathbf{x}}) \\ 
\mathfrak{I}(\bar{\mathbf{x}})
\end{bmatrix}, \quad
\mathbf{n} \triangleq
\begin{bmatrix}
\mathfrak{R}(\bar{\mathbf{n}}) \\ 
\mathfrak{I}(\bar{\mathbf{n}})
\end{bmatrix},
\end{equation*}
\begin{equation*}
\mathbf{H} \triangleq
\begin{bmatrix}
\mathfrak{R}(\bar{\mathbf{H}}) & -\mathfrak{I}(\bar{\mathbf{H}}) \\ 
\mathfrak{I}(\bar{\mathbf{H}}) & \mathfrak{R}(\bar{\mathbf{H}})
\end{bmatrix},
\end{equation*}
where $\mathfrak{R} (\cdot)$ and $\mathfrak{I} (\cdot)$ denote the real and imaginary part of a complex vector or matrix, respectively. 
With this transformation, each element of \( \mathbf{x} \) is drawn from the discrete space  \{-1, 1\} . We then relax this space to the continuous domain \( {\cal{X}} = [-1, 1] \). Consequently, (\ref{ML_eq_c}) can be relaxed to the following problem in the real domain: $\min _{ {\bf{x}} \in {{\cal{X}}}^{2M}} \norm{ {\bf{y}} - {\bf{Hx}}}_{2} ^{2}.$ 
%
We further introduce an auxiliary variable $\mathbf{z}$ to reformulate this problem as:

\begin{equation}\label{opt_prob}
\begin{IEEEeqnarraybox}[][c]{rCl}
\underset{{\bf{x}} \in {{\cal{X}}}^{2M}}{\min}
& & \quad \quad \norm{{\bf{y}} - {\bf{Hx}}}_2^2 \\
\text{subject to}
& & \quad \quad \mathbf{x} = \mathbf{z}, \quad \mathbf{z} \in {{\cal{X}}}^{2M}.
\end{IEEEeqnarraybox}
\end{equation}

The augmented Lagrangian is given as:
\begin{equation}
\label{lagrangian}
\mathcal{L}_\rho(\mathbf{x}, \mathbf{z}, \boldsymbol{\lambda}) = \norm{ \mathbf{y} - \mathbf{Hx}}_{2} ^{2} +  \boldsymbol{\lambda}^{T} (\mathbf{x} - \mathbf{z} ) \\
+ \frac{\rho}{2} \left\| \mathbf{x} - \mathbf{z}  \right\|_2^2
\end{equation}
where $\boldsymbol{\lambda} \in \mathbb{R}^{2M}$ is the dual variable (or Lagrange multiplier) and $\rho > 0$ is the penalty parameter. The iterative steps of the ADMM involve alternating minimization of the augmented Lagrangian with respect to the primal variables $\bf x$ and $\bf z$, followed by an update of the dual variable, $\boldsymbol{\lambda}$ \cite{Efficient_QAM_ADMM, OTFS_withADMM}. The closed-form solutions for $\bf x$ and $\bf z$ can be obtained as:
%
\begin{align}
{\mathbf{x}}^{l+1} &=  \arg \min_{\mathbf{x}} \mathcal{L}_{\rho} (\mathbf{x}, \mathbf{z}^{l}, \boldsymbol{\lambda}^{l}) \nonumber\\ 
&= \arg\min_{\bf{x}}\norm{ \mathbf{y} - \mathbf{Hx}}_{2} ^{2} + {(\boldsymbol{\lambda}^{l})}^{T}({\mathbf{x}}-{\mathbf{z}}^{l})+\frac{\rho}{2}\norm{ {\mathbf{x}} - {\mathbf{z}}^{l}}_{2} ^{2} \nonumber \\
&= \left({\mathbf{H}}^{T} \mathbf{H} + \rho \mathbf{I}\right)^{-1} \left({\mathbf{H}}^{T} \mathbf{y} + \rho {\mathbf{z}}^{l} - {\boldsymbol{\lambda}}^{l}\right) ,\label{eq:x_update}\\
{\mathbf{z}}^{l+1} &= \arg \min_{\mathbf{z}} \mathcal{L}_{\rho} (\mathbf{x}^{l+1}, \mathbf{z}, {\boldsymbol{\lambda}}^{l}) \nonumber \\ 
&= \arg\min_{\bf{z}} - {(\boldsymbol{\lambda}}^{l})^{T}{\mathbf{z}}+\frac{\rho}{2}\norm{ {\mathbf{x}}^{l+1} - {\mathbf{z}}}_{2} ^{2} \nonumber \\
&= \arg\min_{\bf{z}} \frac{\rho}{2}\norm{ {\mathbf{z}} - ({\mathbf{x}}^{l+1}+\frac{{\boldsymbol{\lambda}}^l}{\rho})} _{2}^{2} \nonumber \\
&= \Pi_{[-1,1]}\left(\mathbf{x}^{l+1} + \frac{\boldsymbol{\lambda}^l}{\rho}\right) , \quad \quad {\text{and}}\label{eq:z_update} \\
{\boldsymbol{\lambda}}^{l+1} &= {\boldsymbol{\lambda}}^{l} + \rho \left({\mathbf{x}}^{l+1} - {\mathbf{z}}^{l+1}\right) \quad \label{eq:dual_update} \forall l
\end{align}
where $\Pi_{[-1,1]}({\boldsymbol{a}})$ denotes an element-wise projection of $\boldsymbol{a}$ onto the set $[-1,1]$. More specifically, if ${\boldsymbol{b}} = \Pi_{[-1,1]}(\boldsymbol{a})$, we have
\[
b_i =
\begin{cases}
+1, & \text{if } a_i > 1 \\
-1, & \text{if } a_i < 1 \\
a_i, & \text{otherwise}.
\end{cases}
\]

The detailed procedure for the unfolded ADMM approach is provided in Algorithm~\ref{algo:unfolded_ADMM}. In deep unfolding, to ensure gradient flow during backpropagation in the training stage, all operations must be differentiable. 
To enable differentiable approximation, a soft projection is used that allows gradient to flow while keeping outputs near the desired constellation points. Consequently, (\ref{eq:z_update}) becomes:
\begin{align}
{\mathbf{z}}^{l+1} &= \Pi_{\text{soft}}\left(\mathbf{x}^{l+1} + \frac{\boldsymbol{\lambda}^l}{\rho}\right). 
\label{eq:z_update_soft} 
\end{align}
Furthermore, the layer-wise unfolding of ADMM for MIMO detection is illustrated in Fig.~\ref{fig:unfold_admm_det}. The received symbol vector $\mathbf{y}$ and the channel matrix $\mathbf{H}$ are treated as fixed inputs and are provided to every layer of the network. Each layer of the unfolded network corresponds to one iteration of the ADMM procedure, specifically the update steps in equations (\ref{eq:x_update}), (\ref{eq:dual_update}), and (\ref{eq:z_update_soft}). Corresponding to each of these equations, a specific $\rho$ is considered to enable learning.  
These parameters form the trainable set ${\boldsymbol{\theta}} = \left\{ \textcolor{red}{\rho_{1}^{(l)}},\textcolor{red}{\rho_{2}^{(l)}}, \textcolor{red}{\rho_{3}^{(l)}} \right\}_{l=1}^{L}$, where each layer $l$ has its own set of penalty coefficients. In the final layer $L$, the network produces the estimate ${\mathbf{z}}^{L}=\hat{\bf{z}}$.
For $n=1, \dots ,N$, data samples, the output $\hat{\bf{z}}_{n}({\bf{y}}_{n},{\bf{H}};\rho_{1}^{(l)},\rho_{2}^{(l)},\rho_{3}^{(l)})$ is retrieved from the network as the predicted transmitted symbol for ${\bf{y}}$. The network training loss function formed between the predicted output $\hat{\bf z}$, and ground truth transmitted symbol ${\bf z}_{n}$ is given as:
\begin{align}
    \ell(\rho_{1}^{(l)},\rho_{2}^{(l)},\rho_{3}^{(l)})=\frac{1}{N}\sum_{n=1}^{N}\norm{\hat{\bf{z}}_{n}({\bf{y}}_{n},{\bf{H}};\rho_{1}^{(l)},\rho_{2}^{(l)},\rho_{3}^{(l)})-{\bf z}_{n}}_2^{2}.
\end{align} The network is trained by minimizing the loss with gradient-based learning techniques such as stochastic gradient descent (SGD) to learn $\rho_{1}^{(l)}$, $\rho_{2}^{(l)}$ and $\rho_{3}^{(l)}$.
\begin{algorithm}[ht]
\caption{Unfolded ADMM for MIMO detection.}
\label{algo:unfolded_ADMM}
\begin{algorithmic}[1]
	\STATE \textbf{Input:} Initial values $\mathbf{z}^{(0)}$, ${\boldsymbol{\lambda}}^{(0)}$
    \STATE \textbf{Given:}  $\mathbf{y}$ and $\mathbf{H}$
    \STATE \textbf{Learnable parameters} $\boldsymbol{\theta}=\left\{\textcolor{red}{\rho_{1}^{(l)}},\textcolor{red}{\rho_{2}^{(l)}},\textcolor{red}{\rho_{3}^{(l)}}\right\}_{l=1}^{L}$ 
	\FOR{$l = 1$ to $L$} 
	\STATE $\mathbf{x}^{l+1} = \left(\mathbf{H}^{T} \mathbf{H} + \textcolor{red}{\rho_{1}^{(l)}} \mathbf{I}\right)^{-1} \left(\mathbf{H}^{T} \mathbf{y} + \textcolor{red}{\rho_{1}^{(l)}} \mathbf{z}^{l-1} - {\boldsymbol{\lambda}}^{l}\right)$
	\STATE $\mathbf{z}^{l+1} = \Pi_{\text{soft}}\left(\mathbf{x}^{l+1} + \frac{{\boldsymbol{\lambda}}^{l-1}}{\textcolor{red}{{\rho_{2}^{(l)}}}}\right)$
	\STATE ${\boldsymbol{\lambda}}^{l+1} = {\boldsymbol{\lambda}}^{l} + \textcolor{red}{\rho_{3}^{(l)}} \left({\mathbf{x}}^{l+1} - {\mathbf{z}}^{l+1}\right)$
	\ENDFOR
	\STATE \textbf{Output:} Final estimate $ \hat{\bf{z}}={\bf{z}}^{L}$
\end{algorithmic}
\end{algorithm}

\end{exmp}
\begin{table}[htb!]
\centering
{\footnotesize
\caption{Simulation Parameters.}
\label{table:sim_param}
\begin{tabular}{|p{1.7in}|p{1in}|} 
    \hline
    \textbf{Parameter Name} & \textbf{Value}  \\
    \hline \hline
    Number of transmit antennas & 16\\
    \hline
    Number of receive antennas & 16\\
    \hline
    Constellation & 4 QAM\\
    \hline
    Propagation model & EVA \cite{3gpp36.104}\\
    \hline   
    Training data size
    & 10000 \\ 
    \hline
    Testing data size
        & 10000\\
    \hline
    Epochs
        & 20\\
    \hline
    Batch size
        & 300\\ \hline
    Number of Layers 
        & 5 \\
    \hline
    $\Pi_{\text{soft}}$ & $tanh$ \\
    \hline
\end{tabular}
}
\end{table}
\pgfplotsset{every semilogy axis/.append style={
				line width=1 pt, tick style={line width=0.7pt}}, width=8.5cm,height=6cm, tick label style={font=\small},
			label style={font=\small},
			legend style={font=\small},
			legend pos= south west}
			\begin{figure}[!htpb]
				\centering
				\begin{tikzpicture}
					\begin{semilogyaxis}[xmin=0, xmax=20, ymin=1e-4, ymax=1,
						xlabel={\small $E_b/N_0$ (dB)},
						ylabel={\small BER},
						grid=both,
						grid style={dotted},
						legend style={at={(0.01,0.25)},anchor=west},
						legend cell align=left,
						legend entries={ZF, MMSE , Classical ADMM, Unfolded ADMM, SD }
						]
						\addplot [green,thick,mark=triangle,mark size=3pt] table [x={x}, y={y1}] {ber_all.txt};
						\addplot [blue,thick,mark=diamond,mark size=3pt] table [x={x}, y={y2}] {ber_all.txt};
						\addplot [red,thick,mark=square,mark size=3pt] table [x={x}, y={y3}] {ber_all.txt};
						\addplot [black,thick,mark=oplus,thick,mark size=3pt, mark options={scale=1,solid}] table [x={x}, y={y4}] {ber_all.txt};
					\addplot [brown,thick,mark=triangle,mark size=3pt] table [x={x}, y={y5}] {ber_all.txt};
					\end{semilogyaxis}
				\end{tikzpicture}		
				\vspace*{-0.1cm}
				\caption{\footnotesize{BER performance comparison for 16$\times$16 MIMO system  in EVA model \cite{3gpp36.104}. For soft projection, $tanh$ is used. }} 
				\label{ber}
			\end{figure}
Fig.~\ref{ber} presents the bit error rate (BER) performance comparison of the unfolded ADMM detector with zero forcing (ZF), minimum mean squared error (MMSE), ADMM and sphere decoder (SD), under QPSK modulation. For the unfolded ADMM, the penalty parameters and step sizes are considered as learnable parameters. We consider a $16 \times 16$ MIMO system in extended vehicular (EVA) propagation model. Table~\ref{table:sim_param} summarizes the key simulation parameters. The power delay profile of EVA propagation model~\cite{3gpp36.104} is given by [$0,-1.5,-1.4,-3.6,-0.6,-9.1,-7.0,-12.0,-16.9$] dB against the excess tap delays [$0,30,150,310,370,710,1090,1730,2510$] ns. The results demonstrate that the unfolded ADMM detector outperforms ZF, MMSE, and classical ADMM detectors. The unfolded ADMM detector performs better due to the use of deep unfolding, where learnable parameters: penalty parameters and step sizes, are adaptively trained. The SD achieves near ML performance and is used often as a benchmark scheme for validating the performance of various detectors. This example has been included to provide the readers with an illustrative understanding of the deep unfolding process. The comparative analysis highlights the advantage of the unfolded method over classical approaches. Additional performance improvements can be achieved by optimizing or introducing trainable parameters in the unfolded network.  

The introduction of trainable parameters make it challenging to derive theoretical insights for unfolded algorithms as compared to classical optimization algorithms. For any optimization method, it is crucial to perform convergence analysis. In the literature, the convergence studies of various methods can be classified into three broad categories:
\begin{enumerate}
    \item \textbf{Convergence order:} 
    In this category, the convergence properties of unfolded networks are explored to provide a comparison with classical methods. Under idealized choices of trainable parameters, different unfolded architectures are studied that yield an accurate solution with a faster convergence order. It is well known that the classical ISTA exhibits sublinear convergence \cite{FISTA_amir}. The work in \cite{ISTA_theoretical} discusses how learned ISTA (LISTA) or unfolded ISTA produces a sequence of trainable parameters per layer exhibiting a linear convergence order. As discussed in \cite{LISTA_theoretical}, in the unfolded ISTA, due to the transformation of trainable parameters into learned weights, a linear convergence order is achieved. 
    The authors in \cite{L2O_survey} point out that in deep unfolding, the most essential step in achieving better empirical performance and acceleration is to design the learnable parameters and analyze how the learnable parameters influence each other. For instance, in \cite{LISTA_theoretical}, the parameters related to dictionary based matrices in sparse signal recovery, are not shared across the LISTA network's layers. 
    The learnable parameters are instead trained in each layer, similar to untying them across the layers. The increased complexity in parameterization enables the network to learn complex mapping and exhibit better acceleration in converging towards the solution. 
    In \cite{ALISTA_theoretical}, the authors have proposed analytic LISTA (ALISTA), in which the weight matrices are precomputed through an optimization procedure, thereby eliminating the need for learning them from data. Only a small number of scalar parameters: the layer-wise step size, threshold, and momentum factor are learned. Building upon this framework, the subsequent HyperLISTA model dynamically adapts these parameters in response to the reconstruction error, retraining them for each instance to match the current problem state. This instance adaptive strategy yields a superlinear convergence order, a property that cannot be achieved when a uniform set of optimal parameters is learned across the entire dataset.
    Although theoretical studies have been attempted on the dependency of parameters on convergence order and stability of the algorithm for ISTA, there is a significant scope to explore theoretical studies on how other classical algorithms such as alternating direction of method of multipliers (ADMM), approximate message passing (AMP), etc., perform compared to their respective unfolded versions. For instance, channel estimation for integrated sensing and communication (ISAC) in intelligent transportation system is proposed in~\cite{Converge_ISAC}. An unfolded network for channel estimation based on ISTA is designed with trainable parameters, that includes the parameters related to dictionary based matrices in each ISTA iteration and the learnable threshold parameter used in the soft-thresholding operator. In this work, the authors theoretically analyze the proposed unfolded network and establish it achieves a linear convergence order.
    \item \textbf{Generalization error:}  A fundamental question for any learned model is its reconstruction ability. Especially, the degree to which it can accurately recover a parameter from incomplete or corrupted data. Generalization error or bounds give a probabilistic estimate on the difference between the true error (evaluated over the underlying data distribution) and the empirical error computed from the available training samples for a given hypothesis function. Thus, such bounds predict how well a learned model performs on unseen data and assess whether the training process has led to appropriate generalization~\cite{gen_error}. The underlying machine learning algorithm of a learned model can be viewed as stochastic transformation which maps training data to hypotheses. The algorithm stability largely depends on the generalization capability of learning algorithm, and thus is of theoretical interest~\cite{StabilityGen_oliver}.
    As described in~\cite{gen_error}, based on the training sequence $\mathcal{S}$ and given hypothesis space $\mathcal{H}$, a learning algorithm yields a function $h_{S}\in \mathcal{H}$. The function $h_{S}$ is defined as $h_{S}:\mathcal{X} \rightarrow \mathcal{Y}$ for a training sequence from a distribution $\mathcal{D}$ on $\mathcal{X} \times \mathcal{Y}$. Let us consider a training sequence $\mathcal{S}=\{({\mathbf{x}}_{i},{\mathbf{y}}_{i})\}_{i=1,\dots,m}$ with $m$ i.i.d. samples drawn from the distribution $\mathcal{D}$. Therefore, the empirical loss $\hat{\mathcal{L}}(h)$ of a hypothesis $h$ is the reconstruction error on the training sequence $\mathcal{S}$, as given below:
    \begin{equation}
        \hat{\mathcal{L}}(h)=\frac{1}{m}\sum_{i=1}^{m} \ell(h,{\mathbf{x}}_{i},{\mathbf{y}}_{i}).
    \end{equation}
    The loss function $\ell$ measures the reconstruction error. Mean squared error (MSE) is the most commonly used loss function.
    The true loss $\mathcal{L}(h)$ defined as the risk of a hypothesis $h$ is computed by evaluating expectation, $\mathbb{E}[\cdot]$ over a distribution $\mathcal{D}$ as shown below:
    \begin{equation}
         \mathcal{L}(h)=\mathbb{E}_{\mathbf{x},\mathbf{y}\sim \mathcal{D}}[\ell(h, \mathbf{x},\mathbf{y})].
    \end{equation}
    The generalization error of the hypothesis $h_{\mathcal{S}}$ is based on the training samples $\mathcal{S}$. It is defined as the difference between the empirical loss and the true loss as shown below:
    \begin{equation}
        \text{GE}(h_{\mathcal{S}})=|\hat{\mathcal{L}}(h_{\mathcal{S}})-\mathcal{L}(h_{\mathcal{S}})|.
    \end{equation}
    In~\cite{StabilityGen_oliver}, the authors have identified several definitions to quantify the stability of a learning algorithm such as: hypothesis stability, point-wise hypothesis stability, error stability, and uniform stability. \\ The stability of a learning algorithm is quantified by the bound on the generalization error which must be tight. The trainable parameters in deep unfolding are trained via DL tools so it is imperative that algorithmic stability will be impacted by the choice of these parameters. Thus, the notion of algorithmic stability can offer a theoretical bridge between optimization and learning in deep unfolding.  A proper parameterization is fundamental to the empirical performance and acceleration achieved during training of the network. Hence, to maintain convergence and algorithmic stability with robust generalization, the learnable parameters must be carefully designed. Consequently, exploring stability aware parameterization and training strategies presents a promising direction for advancing the theoretical foundation of deep unfolding method.\\
   Moreover, in~\cite{gen_error}, the authors have discussed generalization error that relies on Rademacher complexity analysis. The result presented in~\cite{gen_error}, suggested that deep unfolding models may exhibit better generalization capabilities than purely data driven neural networks, in line with empirical results. For example, the generalization error bound for LISTA depends on the number of layers only logarithmically, whereas generalization error bound for purely data driven neural network scale exponentially in the number of layers. There have been works~\cite{gen_rademecher,DECONET_learning} that discuss generalization error and Rademacher complexity, involving different learnable parameters and a variety of network architectures. 
    \item \textbf{NMSE:} The convergence rates of different approaches: classical, purely data driven and deep unfolding are examined through the NMSE with respect to the number of layers or iterations required to achieve the optimal solution. Some existing works that use NMSE as a performance metric  are as follows, for channel estimation~\cite{Hu_DDPG, TrainableChnnlest_2023, RIS_chnnlEst_MU_2021}, detection~\cite{Wei_ConjugateDescent}, precoder design~\cite{Wu_RSMA_THz}, ISAC~\cite{ISAC_Net_2024}, joint activity detection and channel estimation~\cite{Algounroll_JADC,random_access_2024, Mao_est_and_det}. Generally, NMSE evaluates the performance of any approach by evaluating the MSE between ground truth, $\mathbf{x}$ and estimated value, $\hat{\mathbf{x}}$ followed by normalization as given below:
\begin{equation}
       \text{NMSE} = \frac{\mathbb{E}[|| {\hat{\mathbf{x}}-\mathbf{x}}||^{2}]}{\mathbb{E}[||{\mathbf{x}}||^{2}]}.
\end{equation}
    NMSE is used as a unified performance metric to compare classical iterative algorithms, purely data-driven models, and deep unfolding methods. It provides a direct measure of convergence speed in terms of the number of layers or iterations required to reach a target accuracy. The NMSE-vs-layers plots reported in~\cite{Algounroll_JADC, Hu_DDPG, TrainableChnnlest_2023, random_access_2024, RIS_chnnlEst_MU_2021, Wu_RSMA_THz, Mao_est_and_det, Wei_ConjugateDescent, ISAC_Net_2024} visually illustrate the convergence behavior of unfolded networks relative to benchmark algorithms.
These results consistently show that NMSE decreases as the number of layers or iterations increases. Classical iterative methods exhibit slow NMSE reduction because they typically require many iterations to converge. In contrast, deep unfolding achieves the same NMSE performance with significantly fewer layers, outperforming both purely data-driven models and classical approaches~\cite{Hu_DDPG, Algounroll_JADC}.\\
While NMSE provides a direct measure of reconstruction accuracy with respect to the number of layers, it does not capture characteristics such as the convergence order of the algorithm or its susceptibility to overfitting. These aspects are instead characterized by convergence order analysis and generalization error, which together describe the theoretical convergence behavior and algorithmic stability of deep unfolding methods. However, NMSE can be complemented with these analyses to obtain an understanding of the efficiency and the reliability of unfolded architectures.
  \end{enumerate}
 
The existing theoretical studies have examined the convergence order~\cite{ISTA_theoretical, LISTA_theoretical, ALISTA_theoretical} and generalization error~\cite{gen_error, StabilityGen_oliver, gen_rademecher} of unfolded networks. However, the authors in~\cite{ISTA_ADMM_optim} have provided optimization guarantees showing that unfolded networks can achieve near-zero training loss as the number of learning epochs increases. Their analysis focuses on how the training loss decreases with more epochs while keeping the number of layers fixed. In their numerical results, they show that purely data-driven networks require significantly more parameters to achieve the same near-zero training loss compared to LISTA and ADMM-based compressive sensing networks. This highlights the parameter efficiency and training advantages of unfolded architectures.

In the following sections, we provide a comprehensive survey of deep unfolding algorithms for a variety of critical wireless communication tasks. 


\vspace{-1em}
\section{Signal Detection } \label{sec::signal_detection}
This section discusses the applications of deep unfolding for signal detection. We segregate these signal detection techniques into various key technologies, as described in the following. 


\subsection{Massive MIMO} \label{massive_MIMO}
Massive MIMO uses numerous antennas at both the transmitter and receiver, improving spectral and energy efficiency, but making symbol detection more complex due to inter-symbol interference (ISI) \cite{LMIMO}. Designing low-complexity, high-performance detectors is still challenging. MIMO detection can be linear or non-linear. Linear detectors, like matched filter (MF), ZF, and MMSE, involve matrix inversion and can be enhanced with lattice reduction \cite{MassiveMIMO_det}. Non-linear detectors, such as successive interference cancellation (SIC) and parallel interference cancellation (PIC), remove interference in stages \cite{LMIMO}, \cite{MIMO_det_survey}. SIC schemes (e.g., MF-SIC, ZF-SIC, MMSE-SIC) use linear detectors to detect symbols, estimate interference, and remove it before detecting the next symbol. SD, a non-linear method, approximates ML detection by searching a bounded hypersphere around the received lattice point, offering a more efficient search than exhaustive methods \cite{sph}, \cite{SD1}, \cite{SD2}.


The projected gradient descent (PGD) method is a popular optimization-driven approach for MIMO detection. In each iteration, PGD performs gradient descent to minimize detection error, then projects the solution onto a feasible, valid constellation set \cite{TPGD}.
One of the first papers to use deep unfolding in detection is the DetNet architecture \cite{Samuel_learningtodetect}. To achieve ML optimization, the PGD algorithm is unfolded where the gradient's step size is learned across layers. The authors in \cite{Samuel_learningtodetect} claimed that the computational complexity is reduced compared to the SD. DetNet nearly matches SD's performance (within 0.5 order of BER at 18 dB). The computational complexity is linear ($\mathcal{O}(L.K^2)$), where $L$ and $K$ are the number of layers and constellation size, respectively. On the other hand, the SD exhibits ($\mathcal{O}(2^K)$) for worst-case complexity. The FLOPs count for QPSK, 8PSK and 16 QAM constellation with varying transmitted symbol size, show that DetNet has lower FLOP count than those for semidefinite relaxation (SDR) based MIMO detection but has higher FLOP count than the AMP detector. However, the proposed network is not flexible to changes in the size of the constellation or the number of users. 
To address MIMO detection under imperfect CSI, He \textit{et al.} introduced model-driven DL networks, orthogonal approximate message passing (OAMP) network, OAMP-Net \cite{He_model_MIMOdet1}, and its enhanced variant OAMP-Net2 \cite{He_model_MIMOdet2}. OAMP-Net2 outperforms OAMP-Net by adapting to varying channel conditions. However, the robustness of the OAMP-Net2 under hardware impairment has not been verified in \cite{He_model_MIMOdet2}. The OAMP algorithm is unfolded, and the linear MMSE (LMMSE) channel estimator is used to design the receiver in \cite{He_model_MIMOdet2}. Learnable parameters, such as the step-size for updating the prior mean and variance in the MMSE estimator, are trained using SGD. These parameters in OAMP-Net2 help compensate for channel estimation errors, improving performance compared to DetNet \cite{Samuel_learningtodetect}. 

Furthermore, for overloaded massive MIMO systems, low-complexity detectors are essential. The trainable PGD (TPGD) algorithm \cite{TPG_2021} unfolds the PGD method with trainable parameters (step size) optimized via backpropagation and SGD. To further improve PGD with an increase in convergence speed, Yun \textit{et al.} in \cite{MomentNet_2024} added momentum components to the iterative hidden layers. Momentum helps guide parameter updates in the right direction and reduces oscillations during PGD optimization \cite{Moment_PGD}. Both networks in \cite{TPG_2021, MomentNet_2024} achieve lower complexity and outperform DetNet \cite{Samuel_learningtodetect} and OAMP-Net2 \cite{He_model_MIMOdet1}.
A detector based on the Hubbard–Stratonovich transformation \cite{HS_2024} avoids matrix inversion, offering faster training and lower execution cost compared to TPGD \cite{TPG_2021} and OAMP-Net2 \cite{He_model_MIMOdet2}.
MMNet \cite{MMNet} unfolds ISTA, alternating linear detection and nonlinear denoising, designed for ill-conditioned and non i.i.d. channels. MMNet balances complexity and flexibility, outperforming DetNet \cite{Samuel_learningtodetect}, OAMP-Net \cite{He_model_MIMOdet1}. The MMNet outperforms the classical LMMSE detector with a performance gain of approximately $4-6$ dB at a symbol error rate (SER) of $10^{-3}$ for different channel conditions.

The CONCRETE MAP method in \cite{CMD_2021} solves the maximum \textit{a posteriori} (MAP) detection problem via CONtinuous relaxation of disCRETE random variables, approximating MAP solutions using gradient descent unfolding. It generates soft outputs for integration with soft-input decoders and outperforms DetNet \cite{Samuel_learningtodetect}, OAMP-Net2 \cite{He_model_MIMOdet2}, and MMNet \cite{MMNet} in BER under correlated channels. In \cite{Wei_ConjugateDescent}, conjugate gradient descent (CGD) iterations are unfolded to design the network, LcgNet. The authors also proposed a quantized version, QLcgNet, which minimizes memory usage with negligible performance degradation. LcgNet surpasses DetNet \cite{Samuel_learningtodetect} in SER, complexity, and memory cost, making it suitable for massive MIMO system. Olutayo \textit{et al.} \cite{CG_2024} improved LcgNet by adding a preconditioner during training, enhancing convergence speed by optimizing the receiver filter’s spectrum. Preconditioning improves the condition number of a matrix \cite{Preconditioner}, optimizing the receiver filter's spectrum. This enhancement significantly increases the convergence speed compared to the original LcgNet in \cite{Wei_ConjugateDescent}. A dynamic conjugate gradient network further adapts to time-varying channels using self-supervised learning with forward error correction codes.

Physical layer signal processing handles continuous-time analog waveforms, which must be quantized for hardware implementation. Low-resolution (LR) one-bit analog-to-digital converters (ADC) are widely used at radio frequency (RF) ends for high sampling rates, low cost, and low power \cite{ADC_imp}.
Khobahi Khobahi \textit{et al.} \cite{LoRD_2021} developed an LR detection network (LoRDNet), which unfolds the PGD method for symbol detection using the ML estimator, learning channel behavior implicitly from training data without requiring CSI. LoRDNet outperforms classical ML \cite{nML} and DL networks like DeepSIC \cite{DeepSIC}, even with small datasets ($\sim500$ samples).
A generalized PGD algorithm was proposed in \cite{PGD_2024}, where multiple gradient steps are performed before projection to balance convergence and performance. The self-correcting auto detector (SAD) unfolds the PGD algorithm, by combining a self-correcting gradient module and a learnable denoising autoencoder projection.
SAD \cite{PGD_2024} achieves better detection-complexity trade-off than DetNet \cite{Samuel_learningtodetect}, using only $4$ projection steps versus DetNet’s $15$ projection steps, significantly reducing latency and computational cost.

The MIMO fading channel matrix can be modeled using graphs, such as Bayesian belief networks, Markov random fields, and factor graphs, which represent dependencies between variables for efficient detection \cite{LMIMO}. Iterative message-passing detectors (MPD), like belief propagation (BP), approximate message passing (AMP), and channel hardening-exploiting message passing (CHEMP), reduce computational complexity. Expectation Propagation (EP) extends BP for efficient detection in large MIMO systems by approximating posterior distributions iteratively in polynomial time \cite{EP_1, EP_2}.
Tan \textit{et al.} \cite{Tan_massiveMIMO_det} unfolded MPD iterations by learning prior probabilities and correction factors (damping, scaling, LLR rescaling) via mini-batch SGD, improving performance in correlated channels and higher-order modulation over classical MPDs.
A low-complexity AMP-graph NN (GNN) model (AMP-GNN) that avoids matrix inversion and outperforms OAMP-Net2 \cite{He_model_MIMOdet2} is proposed in \cite{MP_GNN_2024}. Due to high complexity, MPDs are not suited for single-carrier frequency division multiple access (FDMA) systems, where a low peak-to-average power ratio is a key design objective for efficient transmission. Hence, an unfolded network in \cite{MP_detector_2024}, simplifies generalized AMP iterations as discussed in \cite{Generalized_AMP}, by approximating per-subcarrier statistics through statistical averaging. The unfolded network \cite{MP_detector_2024} maintains high performance even with channel mismatches and achieves $0.83$ dB at BER$=10^{-3}$ gain over classical generalized AMP \cite{Generalized_AMP}. For EP-based detection, a modified EP MIMO detector (MEPD) \cite{AlgoParam_2021} improved parameter selection via a modified moment-matching scheme, outperforming OAMP-Net2 \cite{He_model_MIMOdet2} in high modulation cases. In \cite{EP_2021}, an approximate EP inspired by \cite{Approx_EP} is unfolded, achieving better detection performance and robust convergence.
To address high computational complexity of SD, Nguyen \textit{et al.} \cite{SP_DNN} proposed a fast DL-aided SD algorithm inspired by \cite{FSNet}. The DNN generates reliable initial candidates for SD, prioritizing candidates closer to the DNN's output for testing. This approach accelerates sphere shrinkage, reducing computational cost. The sparse network architecture eliminates matrix inversion, relying only on element-wise matrix multiplications. The proposed method performs better in highly correlated channels than DetNet \cite{Samuel_learningtodetect}, OAMP-Net2 \cite{ He_model_MIMOdet2}, MMNet \cite{MMNet}, and LcgNet \cite{Wei_ConjugateDescent}.
A MIMO detector based on annealed Langevin dynamics is introduced in \cite{AL_2023}. It replaces the traditional two-step iterative framework (continuous-space optimization followed by projection) with a single-step approximation of the MAP estimator by sampling from Langevin dynamics. This approach shows superior performance in correlated channels compared to DetNet \cite{Samuel_learningtodetect}, OAMP-Net2 \cite{He_model_MIMOdet2}, and MMNet \cite{MMNet}.
To mitigate the computational costs of matrix inversion in classical detectors like ZF and MMSE, Berra \textit{et al.} \cite{DU_2023} proposed unfolding accelerated Chebyshev successive over-relaxation algorithm. This network requires fewer trainable parameters and features a fast, stable training scheme. The unfolded detector achieves better performance compared to DetNet \cite{Samuel_learningtodetect}, OAMP-Net2 \cite{He_model_MIMOdet2}, TPGD-detector \cite{TPG_2021}, MEPD \cite{AlgoParam_2021}, EP-Net \cite{Li_ExpectOTFS}, and other methods like \cite{Liao_MassiveMIMODEt}. 
\begin{table*}[t!]
\centering
\renewcommand{\arraystretch}{1.4}
\caption{
\centering
Comparison of Classical vs Unfolded methods for Signal Detection. 
\newline
$T$: Layers, $I$: Iterations, $M$: Symbol size, $N_t$: No. of transmit antenna, $N_r$: No. of receive antenna.}
\label{tab_det}
\begin{tabular}{|p{1.5cm}|p{2.5cm}|p{2.5cm}|p{2.5cm}|p{3.5cm}|p{2cm}|p{1.5cm}|}
\hline
\multirow{2}{*}{\textbf{Unfolded}} & 
\multicolumn{2}{c|}{\textbf{Computational Complexity}} & 
\textbf{Number of Parameters (Unfolded)} & 
\multicolumn{3}{c|}{\textbf{Convergence speed (Number of layers)}} \\
\cline{2-3} \cline{5-7}
\textbf{Model} & \textbf{Classical} & \textbf{Unfolded} &  & \textbf{Simulation Setup} & \textbf{Classical} & \textbf{Unfolded} \\
\hline
DetNet \cite{Samuel_learningtodetect} & Sphere Decoder: $\mathcal{O}(M^{N_{t}})$
& $\mathcal{O}(TN_{t}^{2})$ & $(6N_{r}N_{t}+2N_{r}+N_{r})T$
& QPSK, 8-PSK, 16-QAM, $0.55$-Toeplitz channel, $N_{t}=\{15,20,30\}$, $N_{r}=\{20,25,30\}$. 
& $\sim 1000$ iterations & $30$ Layers \\
\hline
LcgNet \cite{Wei_ConjugateDescent} & CGD: $\mathcal{O}(I(8N_{t}^{2}+14N_{t}+8))$
& $\mathcal{O}(T(4N_{t}^{2}+6N_{t}))$
& $(6N_{t}+4N_{t}^{2}+4T)$
& MIMO $N_{r}=32$, $N_{t}=64$, Rayleigh fading channel.
& $\sim 40$ iterations & $7$ Layers \\
\hline
OAMP-Net2 \cite{He_model_MIMOdet2} & OAMP: $\mathcal{O}(N_{t}^{3})$
& $\mathcal{O}(TN_{t}^{3})$
& $4T$
& MIMO Rayleigh fading channel, QPSK and 16-QAM, $N_t=N_r=4$ and $N_t=N_r=8$.
& $1000$ iterations & $4$ Layers \\
\hline
MMNet \cite{MMNet} & MMSE: $\mathcal{O}(N_{r}^{3})$
& $\mathcal{O}(N_{r}^{3})$
& $2N_{t}(N_{r}+1)$
& Massive MIMO, QAM modulation, 3GPP 3D channel, $N_{t}=\{16,32\}$, $N_{r}=64$.
& $50$ iterations & $10$ Layers \\
\hline
ADMM-Net \cite{Inexact_ADMM_det_2021} & ADMM: $\mathcal{O}(M^{3})$
& $\mathcal{O}(M^{2})$
& $5T$
& MIMO Rayleigh fading channel, BPSK, 4-QAM and 16-QAM, $N_t=N_r=50$.
& $30$ iterations & $<20$ Layers \\
\hline
\end{tabular}
\vspace{-1em}
\end{table*}

To avoid the computational burden incurred due to matrix inversion, the authors in \cite{DU_sparse_2022} have proposed a sparse refinement architecture for symbol detection in uplink massive MIMO. The authors incorporated deep unfolding for components in matrix inversion, to reduce the detection complexity. The sparse weight matrix associated with the Jacobian iteration is trained across the layers of the network.
Liao \textit{et al.} \cite{Liao_MassiveMIMODEt} incorporated interference cancellation into DNN layers. Inspired by \cite{Mandloi_ISD}, their sequential detector trains residual error coefficients with MSE loss, reducing complexity by avoiding matrix inversion and relying on matrix addition and multiplication. Compared to DetNet \cite{Samuel_learningtodetect}, which requires $90$ layers for convergence, their model achieves convergence with just $8$ layers, offering significant computational simplicity.
The authors in \cite{Inexact_ADMM_det_2021} developed an unfolded ADMM-based network for MIMO detection. They employed an inexact ADMM update as in \cite{ADMM, linearizedADMM} to reduce complexity while achieving near-optimal performance. Compared to DetNet \cite{Samuel_learningtodetect}, the proposed network \cite{Inexact_ADMM_det_2021} requires fewer trainable parameters, resulting in improved performance in terms of symbol error. Since, in \cite{Inexact_ADMM_det_2021}, perfect CSI was assumed for MIMO detection, the authors in \cite{Robust_2023}, proposed an ADMM-based unfolded algorithm with channel estimation errors. A set of penalty and relaxation parameters are learned layer wise. 
Moreover, the complexity of the proposed network is reduced as compared to OAMP-Net2 \cite{He_model_MIMOdet2}. The authors in~\cite{Arikawa} have proposed an unfolded coherent detector for optical communication. They have developed a MIMO filter architecture to compensate for hardware impairments such as transmitter and receiver in-phase/quadrature skew. The learnable parameters include the filter coefficients of the MIMO filter and are trained using the method in~\cite{TPG_2021}. The results showed that the unfolded network performed well in the presence of chromatic dispersion, polarization rotation, and frequency offset accumulated over the fiber channel.





\subsection{Modern Multicarrier Modulation}

A cellular network must simultaneously support a growing number of active subscribers within a limited time-frequency spectrum \cite{6G_guide}. By increasing bandwidth the symbol period is reduced which worsens the ISI in multi-path fading channels. OFDM mitigates ISI by splitting wideband signals into orthogonal narrowband subcarriers \cite{OFDM_basics}.
MIMO-OFDM detectors faces challenges due to high-dimensional signal space. An unfolded OAMP-based low-complexity receiver for cyclic prefix-free MIMO-OFDM is explored in \cite{Chnnel_det_OFDM_2019}. The network was tested in over the air experiments and it performed reliably with low synchronization errors under imperfect CSI. Similarly, detection of coarsely quantized or LR signals in one-bit OFDM-based MIMO systems is a challenge. One-bit ADCs severely quantize received signals, discarding amplitude information, and hence, ML detection becomes intractable due to nonlinear quantization effects. Shao \textit{et al.} \cite{EM_2024} proposed an unfolded expectation maximization (EM) algorithm that iteratively estimates unquantized symbols, outperforming ZF detection at high SNR. Furthermore, Ullah \textit{et al.} \cite{SO_2024} unfolded a likelihood ascent search algorithm for joint detection and soft-output estimation. The likelihood ascent search algorithm is a neighborhood search-based detector for the MIMO system. The algorithm begins with an initial symbol estimate from linear detectors (ZF or MMSE) and iteratively refines the solution by flipping bits in the detected symbols to minimize the likelihood cost function. Near-optimal detection is achieved with the unfolded network, $1.2$ dB gap with 16-QAM as compared to SD at BER of $10^{-5}$ at polynomial complexity. Zhang \textit{et al.} \cite{Adaptive_2022} unfolded expectation propagation (EP) with trainable layer-specific damping factors, providing robust detection under varying channels and noise, validated in over-the-air tests.
In underwater acoustic communication, where radio or light waves fail, Zhao \textit{et al.} in \cite{Zhao_underwater_OFDM} proposed UDNet, an unfolded MMSE equalizer that generalizes well with low complexity in doubly-selective channels. In terms of performance, UDNet achieved better generalization and maintained lower computational complexity than DetNet \cite{Samuel_learningtodetect}.
Li \textit{et al.} \cite{Li_ExpectOTFS} proposed EP-Net for OTFS detection, training step-size, damping, and posterior distribution parameters to improve detection performance.

In \cite{OTFS_Est_2024}, an unfolded OAMP-based detector for OTFS is designed to handle channel estimation errors. This detector is trained using imperfect CSI and optimizes three trainable parameters: step-sizes for updating the linear estimate, nonlinear estimate, and noise variance. Numerical analysis demonstrates that at SNR of $30$ dB, the proposed detector achieves a BER of $2.4 \times 10^{-4}$ while EP-Net \cite{Li_ExpectOTFS} achieves BER of $1.2 \times 10^{-3}$.

\subsection{Next Generation Multiple Access}

Multiple access techniques allow multiple users to share the same communication channel and resources, such as time, frequency, or code, to transmit data efficiently. Usually multiple access schemes achieve this by dividing signaling dimensions orthogonally across time, frequency, or code \cite{MA_6G}. This orthogonality enables low-complexity multi-user detection at the receiver. However, orthogonal multiple access cannot achieve the desired sum-rate capacity for a multiuser wireless system. To address this, NOMA has been introduced, allowing more users to be served than the number of orthogonal resource units by leveraging non-orthogonal resource sharing \cite{6G_guide}. NOMA is categorized into power-domain NOMA and code-domain NOMA \cite{PD_CD_NOMA}. Power domain NOMA multiplexes users by assigning different power levels on the same resource, such as time, frequency, or code, and relies on SIC at the receiver. Whereas, code domain NOMA multiplexes users by assigning different non-orthogonal spreading codes on the same resource block.


A deep unfolded detector, named the learned preconditioned CGD network with SIC, is proposed for downlink MIMO-NOMA systems in \cite{DTL_2023}. The algorithm incorporates a data transfer learning strategy to achieve a robust model with low data set requirements. This approach efficiently adapts the model to variations in environmental factors, such as channel models, modulation schemes, or power allocation, ensuring consistent performance in diverse scenarios. Tanner-graph-based transmission schemes for NOMA are discussed in \cite{Tanner_decoding_2023}. To enhance detection efficiency, a message-passing detector is unfolded, allowing the detector to improve as more information about transmitted data is learned during training epochs.  

In sparse code division multiple access (CDMA), each user transmits using sparse signature codes to reduce interference and enable efficient multiuser access \cite{SCDMA,LDS}. Takabe \textit{et al.} proposed the complex sparse TPGD (C-STPG) detector in \cite{Sparse_CDMA}, achieving superior detection over LMMSE with low complexity and minimal training due to few trainable parameters. They also introduced gradual sparsification, a deep unfolding method that jointly learns sparse signature matrices by minimizing MSE, tuning step size and a softness parameter.

\subsection{Discussions and Takeaways}

Deep unfolding has proven to be a compelling solution for signal detection in complex and high-dimensional problems, such as those in massive MIMO, low-resolution ADC, and time-varying systems. By transforming iterative algorithms such as gradient descent variants, ISTA, ADMM, and AMP into structured networks, unfolding approaches bridge interpretability and performance. In Table~\ref{tab_det}, we summarize the computational complexity, number of trainable parameters and convergence speed of the most popular deep unfolding models against those of classical methods for signal detection. 

While pioneering unfolding architectures such as DetNet \cite{Samuel_learningtodetect} and LcgNet \cite{Wei_ConjugateDescent} have demonstrated the advantages of data-driven signal detection with explainable DNN structures, they exhibit limited generalization in challenging scenarios, such as correlated channels. More recent unfolding-based approaches, including MMNet \cite{MMNet} and ADMM-Net \cite{Inexact_ADMM_det_2021}, address these limitations by offering robustness to both channel correlation and imperfect CSI. However, methods like MMNet \cite{MMNet} and LoRDNet \cite{LoRD_2021} introduce additional latency and training complexity due to their reliance on online training and periodic retraining, making them less suitable for fast-varying channel environments.


The research community has actively tested deep unfolded detectors in real-world scenarios. Over-the-air test platforms  \cite{Adaptive_2022, Chnnel_det_OFDM_2019, Zhao_underwater_OFDM} have been employed to validate the performance of unfolded detector networks under practical conditions. These real-world experiments confirm the feasibility of model-driven deep unfolding networks that are trained offline and subsequently deployed on software-defined radios or software-based receivers for real-time inference. The results demonstrate that deep unfolding techniques in signal detection can achieve robust BER performance in practical indoor environments, outperforming both traditional and purely data-driven methods.     

Although the aforementioned unfolded detectors achieved improved performance with reduced complexity under both perfect and imperfect CSI, their robustness over practical hardware and synchronization impairments have not been investigated. The performance analysis in this direction for different unfolded detectors are worth investigating in future works.
\renewcommand{\arraystretch}{1.0}

\section{Channel Estimation} \label{sec::channel_estimation}
Channel estimation is essential for signal detection because it allows the receiver to compensate for distortions introduced by the transmission medium such as fading, delay, and phase shifts. 
In the following, we discuss deep unfolding aided channel estimation methods. 
\subsection{Massive MIMO}

MMSE-based channel estimation is commonly used in MIMO systems to estimate the channel response in noisy environments. 
In massive MIMO systems with frequency division duplexing (FDD), reducing CSI feedback overhead is challenging. A method in \cite{DLwithCSI_2019} combines DL and superposition coding to address this. Downlink CSI is encoded, superimposed on uplink user data, and sent to the BS. A multi-task NN unfolds the MMSE process to reduce interference and reconstruct the CSI efficiently.

ISTA is commonly used for sparse channel estimation, enabling efficient CSI feedback. Each ISTA iteration includes a linear operation followed by a nonlinear soft-thresholding step, similar to the rectified linear unit (ReLU) activation function \cite{Gregor}. 
In \cite{Inverse_chnnlest_2021}, Chen \textit{et al.} proposed a trainable DL architecture for sparse recovery problems using an adaptive depth approach. Instead of a fixed number of layers, the method dynamically adjusts the depth based on the problem's complexity. Highly sparse channels use fewer layers, while more complex channels require additional layers.
This adaptive approach improves accuracy and lowers computational costs compared to classical methods and fixed-depth networks. Results in \cite{Inverse_chnnlest_2021} demonstrate that the method reduces reconstruction error and speeds up channel estimation, making it more efficient.
A scenario-adaptive CSI compression framework is proposed in \cite{CSIFeedback_DU_2024}, requiring minimal channel measurements for training. The authors introduced a CSI feedback architecture designed for efficient encoder updates in dynamic wireless environments, which is essential for downlink precoding in massive MIMO systems.
The encoding network focuses on improving the robustness and efficiency of CSI compression. It employs a deep unfolding-based feedback network to iteratively refine the feedback process and enhance feature extraction. Similarly, the decoding network uses a deep unfolding structure inspired by ISTA. This iterative approach improves CSI recovery accuracy, making the process more efficient and adaptable to changing conditions.

In massive MIMO, where channels exhibit sparsity, majorization-minimization (MM) is effective for sparse channel estimation \cite{MM_Chnnel_est}. In \cite{chnnlest_2024}, the authors propose a two-timescale joint uplink/downlink dictionary learning and channel estimation framework. The short-term subproblem estimates sparse channel vectors by unfolding the MM algorithm and treating these vectors as trainable parameters. The long-term subproblem updates the dictionary using a constrained stochastic successive convex approximation (SSCA) method without unfolding. Simulation results show reduced pilot overhead and superior performance in both static and dynamic scenarios, especially with mobile users.

The matching pursuit algorithm is a greedy method used for sparse channel estimation, iteratively identifying channel components by correlating the residual with dictionary elements. Yassine \textit{et al.} \cite{Yassine_mpNet} proposed mpNet, a deep unfolded matching pursuit based network for unsupervised multiuser massive MIMO channel estimation, initialized with a dictionary of imperfect steering vectors. The network dynamically adjusts its number of layers based on the received data, providing efficient estimation in realistic settings. Building upon the unfolded network: mpNet~\cite{Yassine_mpNet}, Chatelier \textit{et al.}~\cite{Baptiste} have considered hardware impairments in the shifting of the subcarrier frequency to achieve SISO-OFDM channel estimation.

The basis pursuit algorithm is a widely used method for sparse signal recovery, particularly in compressed sensing. It identifies the sparsest solution to an underdetermined system by minimizing the $\ell_1$-norm of the signal. This approach is crucial in scenarios where data is acquired under constraints, and the signal is assumed to be sparse in certain domains.
In \cite{BasispursuitChnnlEst_2022}, the authors observed that existing sparse channel estimation schemes for downlink CSI often deliver suboptimal reconstruction due to the use of random measurement matrices. To address this, they proposed deep unfolding-based autoencoders, that optimize the measurement matrix by unfolding the basis pursuit algorithm. This customized design enhances reconstruction performance in sparse channel estimation tasks. 


Sparse Bayesian learning is another effective approach for channel recovery \cite{BayesianCS}. It estimates the non-zero elements of the channel matrix while utilizing prior knowledge of sparsity.
Hu \textit{et al.} \cite{Hu_DDPG} introduced a deep unfolding framework called deep deterministic policy gradient (DDPG) for MU-MIMO systems. DDPG, a variant of deep Q-networks, is designed to handle problems in continuous action spaces \cite{DDPG}. 
Unlike conventional deep unfolding NNs with fixed depth, the authors leveraged DDPG to dynamically adjust the network depth based on the complexity of the problem. This adaptive design enhances the efficiency and performance of sparse channel recovery in varying scenarios.
The channel estimation problem is formulated using an off-grid sparse Bayesian learning model. The sparse Bayesian learning algorithm is unfolded with trainable parameters grid points and off-grid variables.
Simulation results show that the proposed method significantly outperforms fixed-depth models, achieving a 30\% reduction in the number of layers while improving  NMSE performance.

\subsection{mmWave Massive MIMO}
mmWave massive MIMO operates in the high-frequency range of 30 GHz to 300 GHz, offering abundant spectral resources to alleviate the bandwidth crunch at sub-6 GHz frequencies \cite{6G_guide}. However, mmWave signals suffer from severe propagation and penetration losses. To compensate for these losses, large-scale antenna arrays with tens to hundreds of elements are required to achieve high power gains. The short wavelength (1–10 mm) of mmWave signals allows compact integration of these arrays into small transceivers. These features introduce unique challenges in designing physical-layer transmission algorithms \cite{mmWaveBF_challenge}. The BS, equipped with numerous antennas, must acquire downlink CSI. However, the requirement of large amount of feedback creates significant overhead, posing a challenge in channel estimation in mmWave massive MIMO systems.

Compressive sensing techniques are widely used to reduce feedback overhead by exploiting channel sparsity \cite{Compressive_Sensing_survey}. The limited scattering properties of mmWave propagation and the power-focusing capability of lens antenna arrays transform the spatial domain channel into a sparse beamspace representation. Consequently, the channel estimation problem becomes a sparse signal recovery task, often addressed through compressed sensing methods. However, performing beamspace channel estimation in massive MIMO systems with limited RF chains remains a challenging task.

Compressive sensing aided deep unfolding is explored in \cite{LAMP}, where the authors apply Onsager correction for AMP as explained in \cite{MP_donoho}. The architecture of the unfolded model, learned AMP (LAMP) \cite{LAMP} is similar to unfolded ISTA in \cite{Gregor}. A learned denoising-based AMP network (LDAMP) is proposed in \cite{LDAMP} to learn beamspace channel estimates. The authors demonstrate that the iterative sparse signal recovery algorithm with denoising achieves robustness in channel estimation particularly in scenarios with large antenna arrays and limited RF chains.
Yi \textit{et al.} in \cite{AMP_beamspace} introduced an unfolded network called learned AMP with deep residual learning (LampResNet). In this model, the AMP algorithm is unfolded to estimate the channel, and deep residual learning is employed to mitigate channel noise. NMSE analysis shows that the proposed network in \cite{AMP_beamspace} outperforms LAMP \cite{LAMP} and achieves performance comparable to LDAMP \cite{LDAMP} in terms of estimation accuracy.
However, exploiting channel sparsity does not always yield optimal estimation performance. To address this, the authors of \cite{TrainableChnnlest_2023} investigated the sparsity of the mmWave channel in the beamspace. They formulate the estimation problem as a sparse signal recovery task and improve convergence speed by unfolding the iterative PGD algorithm. In this approach, the step-size of the descent step is treated as a trainable parameter.
In \cite{ChannelEst_mmWave_MassiveMIMO_2024}, an unfolded iterative trimmed ridge regression algorithm is proposed for beamspace channel estimation. The method supports parallel computation for near zero latency and is applicable to both narrowband and wideband massive MIMO systems, ensuring versatile channel estimation performance.

\vspace{-1em}
\subsection{Modern Multicarrier Modulation}

Yiyun \textit{et al.} \cite{CSIrecoveryMIMO_ISTA_2020} proposed a CSI recovery network for FDD downlink massive MIMO-OFDM by unfolding ISTA, with compression and reconstruction modules for CSI feedback. Hu \textit{et al.} \cite{LORA_2024} improved it with learnable optimization and regularization algorithm (LORA), by learning the regularization term via end-to-end training to adapt to different channel estimation errors.
In \cite{CSI_ISTA_2024}, Yangyang \textit{et al.} exploited the correlation between real and imaginary parts of CSI differential terms, enabling a shared model to reduce complexity and improve encoder efficiency. Each of the networks proposed in \cite{CSIrecoveryMIMO_ISTA_2020, LORA_2024,CSI_ISTA_2024} enhance CSI feedback performance for FDD massive MIMO-OFDM system.
Guo \textit{et al.} \cite{Guo_CSIfeedbackMassiveMIMO} proposed FISTA-Net, unfolding fast ISTA with momentum, trainable step-sizes, and thresholds for faster convergence and accurate low-rank OFDM mmWave CSI feedback.
For quantized CSI feedback, the authors in \cite{DU_CSI_FB_2023} introduced TiLISTA, a deep unfolding bit-level feedback network using tied learned ISTA, reducing learnable parameters and enhancing robustness to quantization errors. Building upon ISTA, they used tied learned ISTA as discussed in \cite{liu2018alista}. Since the trainable parameters are shared across all layers, the number of learnable parameters are reduced, offering a faster and more accurate solution to the signal reduction problem.
FISTA-Net \cite{Guo_CSIfeedbackMassiveMIMO} focuses on faster convergence while TiLISTA \cite{DU_CSI_FB_2023} targets robustness to quantization errors.

In \cite{DL_Chnnel_est_2023}, an unfolded sparse Bayesian learning method is proposed for mmWave massive MIMO-OFDM with hybrid beamforming, effectively capturing channel sparsity and reducing NMSE from $0.1203$ to $0.033$. Jianqiao \textit{et al.} \cite{Recovery_2023} introduced an unfolded inverse-free variational Bayesian learning framework to avoid costly matrix inversions, improving sparse channel recovery by learning the prior parameters of the dictionary.
In \cite{ChnnelEst_2023}, a generalized expectation consistent algorithm is unfolded for sparse channel estimation in lens based beamspace massive MIMO-OFDM, mitigating severe mmWave attenuation. In \cite{Recovery_2023} and \cite{ChnnelEst_2023} accurate channel recovery with few RF chains is achieved.
Additionally, in \cite{ChannelAMP_2021} the author proposed an unfolded multiple measurement vector AMP network that exploits angle-domain sparsity to reduce uplink pilot overhead, jointly training the phase-shift network and channel estimator for improved estimation performance.

Gao \textit{et al.} \cite{Gao_ComNet} proposed a network, ComNet, that unfolds an OFDM receiver into two subnetworks: channel estimation and signal detection. The LMMSE and ZF are unfolded for channel estimation and signal detection, respectively. The weights of the LMMSE and ZF matrices are trained layer-wise. The simulation results show that the BER performance is better compared to the classical LMMSE estimation. Furthermore, Wang \textit{et al.} in \cite{OTFS_chnest} proposed an unfolded least squares (LS) estimation to enhance channel estimation for OTFS MIMO system under fractional Doppler condition. The network, CENet, is structured as parallel sub-networks, each responsible for estimating one row (delay tap) of the delay-time channel matrix. At BER= $3 \times 10^{-3}$, CENet achieved $\sim5$ dB SNR gain over LS estimation algorithm.   

\vspace{-1em}
\subsection{IRS}
To mitigate the blockage effect and extend the connectivity range for mmWave communication, IRS  is a key technology proposed to enhance both energy and SE. IRS consists of many reflecting elements that steer the propagation of electromagnetic waves by adjusting their reflection coefficients. However, to unlock the full potential of IRS, accurate channel estimation is required for both direct and cascaded links. This is a challenging task because IRS-based channel estimation involves estimating multiple channels simultaneously: the direct channels between the BS and each user, the channel between the BS and IRS elements, and the channel between IRS elements and users.

In \cite{He_learning_estimate_RIS}, Jiguang \textit{et al.} proposed a method for estimating the rank-deficient cascaded channel in an IRS-assisted single-input multiple-output system. They developed a deep unfolding network based on a gradient descent algorithm. The model structure is inspired by DetNet \cite{Samuel_learningtodetect}. In massive MIMO, increasing the number of IRS elements and BS antennas enhances the system's capacity. However, this leads to increased training overhead to obtain accurate channel estimates within the channel coherence time. The proposed network addresses this challenge by compensating for the computational complexity. 


Awais \textit{et al.} \cite{IRScascade_2023} developed two networks based on the MAP  estimation criterion to address the challenges in IRS cascaded channel estimation. The first network is a deep denoising network designed for noise-dominated SNR regions, providing support at various noise levels. The second network is a deep unfolded half-quadratic splitting network that alternates between optimizing extrapolation and denoising for interference-dominated regions. As the SNR increases, the proposed network shows significant improvement in NMSE. Under severe channel variations, the network delivers a performance gain of 20\% in the high SNR range.

Tsai \textit{et al.} \cite{LAMP_IRS_2024} proposed a two-stage learned AMP network with row compression for joint estimation of direct and cascaded channels in IRS-aided mmWave systems. Using deep unfolding, the method integrates compressive sensing with learning to achieve reliable channel estimation with reduced training overhead, facilitating efficient IRS reconfiguration.

\subsection{IoT devices}
Joint activity detection and channel estimation are critical in grant-free random access for massive machine type communication (mMTC), where numerous IoT devices access the network sporadically without scheduling. To enhance detection under imperfect CSI, Qiang \textit{et al.} \cite{JADC_Qiang} proposed AMP-Net, an unfolded AMP based network. Each AMP iteration alternates between MMSE-based denoising and residual updates to minimize the MSE. 
The network uses four trainable scalar parameters to estimate activity probability, channel variance, regulate the denoiser, and control the state evolution.
AMP-Net achieved accurate detection of the active device by extracting the CSI from the estimated device state matrix with a $4$ dB gain over classical AMP at SNR$=15$ dB. The authors in \cite{JADC_Cellfree_2024} proposed a grant-free access protocol for mmWave cell-free massive MIMO using a hybrid structure based on learned vector AMP, reducing reliance on prior channel knowledge and improving convergence over classical AMP.
In \cite{mMTC_2024}, Ma \textit{et al.} introduced a one-phase non-coherent scheme for mMTC. The data bits are embedded in pilot sequences to enable joint activity and data detection without channel estimation. They used unfolded AMP to leverage pilot correlation.
In \cite{Massive_access_2024}, an unfolded AMP with backpropagation was applied to an asynchronous random access scheme with data length diversity, addressing activity detection, channel estimation, and data recovery in mMTC.

The authors in \cite{Mao_est_and_det} proposed an ADMM-based deep unfolding network for joint channel estimation and active user detection in massive IoT systems. 
The trainable parameters include an auxiliary variable for matrix inversion, a regularization term balancing sparsity and accuracy, and a shrinkage threshold controlling output sparsity per iteration. Similarly, in \cite{GrantfreeCS_2023}, the ADMM algorithm is unfolded to improve convergence rate and recovery accuracy in grant-free vehicular networks, particularly for BS with a limited number of antennas. 
In \cite{DU_joint_2022},  two unfolded networks were introduced for sporadic user detection and channel estimation in mMTC: ADMM-Net and VAMP-ISTA-Net. ADMM-Net unfolds a linearized ADMM \cite{linearizedADMM} to address constrained convex problems, using a single iteration of the proximal gradient method to avoid solving subproblems explicitly. VAMP-ISTA-Net unfolds a vector AMP enhanced with an ISTA-based denoiser.  Both architectures required only $5$ layers, significantly reducing the iteration count from $300$ in classical algorithms and achieving faster convergence.  Overall, the approaches in \cite{Mao_est_and_det, GrantfreeCS_2023, DU_joint_2022} tackle the multiple measurement vector sparse recovery problem using unfolded ADMM frameworks for improved reconstruction performance.

Shi \textit{et al.} \cite{Algounroll_JADC} considered grant-free massive access in an IoT network. A large number of single-antenna IoT devices sporadically transmit non-orthogonal sequences to a multi-antenna BS. An unfolded ISTA with trainable weight matrices is designed. The network trains faster due to less number of trainable parameters. Zou \textit{et al.} in \cite{random_access_2024} formulated the joint activity detection and channel estimation problem as a group-row-sparse matrix recovery problem. A proximal gradient-based method is unfolded and it achieves better convergence than the network in \cite{Algounroll_JADC}.
\begin{table*}[t!]
\centering
\renewcommand{\arraystretch}{1.5}
\caption{
\centering
Comparison of Classical vs Unfolded methods for Channel Estimation. 
\newline
$T$: Layer, $N_f$: No. of subcarriers, $N$: No. of devices, $M$: No. of BS antenna, $m$: Size of transmitted signal vector, $n$: Size of measurement vector, $L$: Pilot length, $G$: No. of discrete angular domain grid of channel clusters.}
\label{tab_channel_estimation}

\begin{tabular}{|p{1.5cm}|p{2.5cm}|p{2.5cm}|p{2cm}|p{3.5cm}|p{1.5cm}|p{1.7cm}|}
\hline
\multirow{2}{*}{\textbf{Unfolded}} & 
\multicolumn{2}{c|}{\textbf{Computational Complexity}} & 
\textbf{Number of Parameters (Unfolded)} & 
\multicolumn{3}{c|}{\textbf{Convergence speed (Number of layers)}} \\
\cline{2-3} \cline{5-7}
\textbf{Model} & \textbf{Classical} & \textbf{Unfolded} &  & \textbf{Simulation Setup} & \textbf{Classical} & \textbf{Unfolded} \\
\hline
LISTA-Net \cite{Inverse_chnnlest_2021} & ISTA: $\mathcal{O}(nm)$ & $\mathcal{O}(Tnm)$ & $n \times n + 2T$
& MIMO, $n=250$, $m=500$
& $20$ iterations & $14$ Layers \\
\hline
ADMM-Net \cite{GrantfreeCS_2023} & ADMM: $\mathcal{O}(N^{2}M)$
& $\mathcal{O}(LNM)$
& $3T$
& Massive MIMO, $L=90$, $M=4$, $N=300$
& $40$ iterations & $20$ Layers \\
\hline
ComNet \cite{Gao_ComNet} & LMMSE: $1.6$ mega FLOPs
& $0.37$ mega FLOPs
& $1.2$ MBytes
& OFDM, $N_f=64$, $64$-QAM, WINNER II channel model.
& $200$ iterations & $5$ Layers \\
\hline
SBL-CE-Net \cite{DL_Chnnel_est_2023} & Sparse Bayesian learning: $0.6115$ tera FLOPs
& $0.0189$ tera FLOPs
& $TG^2N_f$
& MIMO, $N_f=8$, $N_t=N_r=12$, $G=48$
& $100$ iterations & $3$ Layers \\
\hline
\end{tabular}
\vspace{-1em}
\end{table*}

\vspace{-3mm}

\subsection{Discussions and Takeaways}
Channel estimation in modern wireless systems, especially in massive MIMO and mmWave settings, poses distinct challenges due to high dimensionality, feedback overhead, and sparse propagation characteristics. Deep unfolding addresses these issues by embedding structural priors into learning pipelines, as demonstrated by models like LDAMP \cite{LDAMP} and LampResNet \cite{AMP_beamspace}. 
Table~\ref{tab_channel_estimation} summarizes the computational complexity, number of trainable parameters, and convergence speed (in terms of layers required for convergence) of key unfolded algorithms for channel estimation. Additionally, studies reporting complexity in FLOPs are included to highlight the implementation cost.



Nonetheless, deep unfolding for channel estimation remains sensitive to model assumptions (e.g., sparsity or statistical independence). Performance degradation in dense or correlated environments, retraining overhead, and reliance on fixed unfolding depth are key limitations. Moreover, model-agnostic unfolding and joint learning of estimation and feedback (especially in IRS and wideband settings) will be crucial for making deep unfolding a mainstream tool in physical layer design. Unlike deep unfolding models for signal detection, the practical deployment of such models for channel estimation remains limited. Moreover, their performance has yet to be thoroughly validated through real-world over the air experiments. 

\section{Precoder Design }  \label{sec::precoder_design}
This section reviews the applications of deep unfolding for precoder design used in key 6G technologies.
\vspace{-0.1in}
\subsection{Massive MIMO}
Beamforming in MIMO systems enhances signal transmission by steering signals toward the intended receivers using multiple antennas, thereby improving received SNR \cite{HBF_survey, Molisch_HBFMassiveMIMO}. Beamforming techniques can be broadly divided into three types: analog, digital, and hybrid \cite{HBF_survey}. Digital beamforming uses a digital precoding matrix 
  with one RF chain per antenna, leading to high power consumption and complexity in large arrays. Analog beamforming 
  employs phase shifters but struggles with inter-user interference. HBF \cite{HBF_2001, Sohrabi_HBF_largescale} addresses these issues by combining digital and analog methods, reducing RF chain count while maintaining performance and efficiency which is ideal for mmWave and massive MIMO in 5G and beyond.

Shlezinger \textit{et al.} \cite{Survey12_AI_MIMO} discuss HBF optimization using classical, data-driven, and deep unfolding methods, highlighting deep unfolding’s efficiency with limited RF chains. For precoder design, the weighted MMSE (WMMSE) algorithm \cite{WMMSE} maximizes the weighted sum rate via iterative updates involving matrix inversions and eigendecompositions. Given the nonconvexity of the power-constrained problem \cite{NP_hard}, a distributed inexact cyclic coordinate descent (CD) method is proposed for efficient locally optimal solutions.
	
A deep unfolded architecture, iterative algorithm induced deep unfolding network (IAIDNN) is proposed for precoder design in MU-MIMO systems in \cite{Hu_iterativeAlgo}. The iterative WMMSE algorithm is unfolded.
To improve efficiency, WMMSE is reformulated to approximate matrix inversions. A generalized chain rule in matrix form helps visualize the gradient recurrence between adjacent layers during backpropagation. Using this rule, the gradients of trainable parameters are computed effectively.
To improve convergence while avoiding matrix inversion, a downlink beamforming method using an unfolded PGD-WMMSE algorithm for multi-user multiple-input single-output (MISO) system is proposed in \cite{Matrixfree_BF_2022}. The authors replace matrix inversions with PGD and compute gradients in parallel.
Unlike IAIDNN \cite{Hu_iterativeAlgo}, this approach eliminates matrix inversion, making hardware implementation more practical. Additionally, it achieves a high weighted sum rate in fully loaded scenarios, where the number of users matches the number of transmit antennas at the BS.


In \cite{Hu_iterativeAlgo}, learnable parameters approximate matrix operations in WMMSE, but the method does not exploit the graph structure of wireless networks. To address this, in \cite{Multicell_2023} the authors unfolded WMMSE using graph convolutional networks for multicell MU-MIMO, enabling coordination with only local CSI, though with limited flexibility in learning exact transformations. In \cite{Graph_BF_2024}, the authors improved this by proposing a hybrid unfolded WMMSE that uses GNNs to learn functional mappings, reducing the number of layers and accelerating convergence. Distributed precoding across cooperative BSs was achieved in \cite{Multicell_WMMSE_precoding}, lowering complexity and backhaul load compared to centralized approaches \cite{Hu_iterativeAlgo, Matrixfree_BF_2022}. Finally, in \cite{WMMSE_2024} unfolded WMMSE is applied to massive MU-MIMO with imperfect CSI by leveraging CSI statistics, achieving faster convergence and near-optimal precoding relative to the classical WMMSE algorithm \cite{WMMSE}.

Shi \textit{et al.} \cite{Shi_DU_NN_HBF} designed a hybrid analog-digital transceiver using iterative gradient descent to obtain minimum SER (MSER) in massive MIMO systems. They unfolded the iterative gradient descent algorithm into a DNN, where step-sizes for the gradients are the learnable parameters.
The authors claim that their proposed network matches the performance of the MSER-based iterative gradient descent algorithm while significantly reducing complexity.

Gradient based methods are widely used for beamforming optimization due to their reduced complexity, and further gains have been achieved using advanced frameworks such as PGD and projected gradient ascent (PGA). In \cite{DownlinkPrecoding_2023}, a frequency selective precoding algorithm for wideband MIMO-OFDM was unfolded using PGD with trainable step sizes and projection parameters, achieving near-optimal sum rate at low complexity. Zhu \textit{et al.} \cite{BF_Het_2023} unfolded a parallel gradient projection method for MISO heterogeneous networks, where trainable step sizes and weighted sum-rate coefficients enable generalization across antenna and network configurations. PGA itself performs gradient ascent followed by projection onto the feasible set. Lavi \textit{et al.} \cite{Lavi_learn_HBF} applied unfolded PGA to hybrid precoding, with trainable step sizes for analog and digital updates; for noisy CSI, they used a projected mirror-prox scheme. Bilbao \textit{et al.} \cite{BF_2024} unfolded PGA with ZF projection for analog beamforming under constant-modulus constraints in full-duplex massive MIMO, exploiting convex ZF-feasible sets for faster convergence and improved performance. They have considered the CSI error to account for hardware impairment and non linearity in the channel. Beyond PGD and PGA, CGD has also been unfolded in \cite{EnvelopePrecoding_2020}. The authors incorporated CGD with Riemannian manifold optimization for constant-envelope precoding, reducing complexity while effectively mitigating multi-user interference.

In \cite{Fast_precoding_2023}, a fast and robust precoding network is proposed for MU-MIMO with imperfect CSI and per-antenna power constraints. The MSE minimizing geometric program and its uplink–downlink duality-based iterative solution are unfolded to create RMSED-Net, which offers robustness to channel estimation errors and achieves sum-rate performance close to the benchmark upper bound.


In \cite{Symbollevel_Precoding_2023}, an unfolded interior-point method is used for symbol-level precoding in multi-user MISO, where trainable proximal operators and Lagrange multipliers enable power minimization under SINR constraints with a 50–70\% reduction in runtime. Its main limitation is applicability only to PSK modulation.
In \cite{ADMM_Precoding_2024}, a constructive-interference-based precoding scheme unfolds the proximal Jacobian ADMM, yielding 98\% lower transmit power and significantly reduced execution time. Since ADMM suffers from biased updates in discrete optimization, iterative discrete estimation (IDE) for unbiased convergence in finite-alphabet precoding is introduced in \cite{IDE}. He \textit{et al.} \cite{FiniteAlphabet_2020} unfolded this IDE-based algorithm for massive MU-MIMO, with learnable step sizes and damping achieving improved BER across SNRs.
Finally, in \cite{BF_Est_2023} the authors proposed a learning-based channel semantic acquisition and beamforming framework for cell-free massive MIMO, where a successive over-relaxation-based beamforming algorithm is unfolded to reduce estimation error, lower complexity, and enhance robustness under imperfect CSI.

\subsection{mmWave MIMO}
mmWave MIMO systems rely on beamforming to mitigate multi-user interference while enhancing diversity and spectrum efficiency.	
Typically, the beam selection and digital precoding matrices are designed separately, leading to suboptimal solutions. To address this, Hu \textit{et al.} \cite{Hu_jointDRL} proposed a joint design approach combining DL and model-driven techniques. DRL optimizes beam selection, while a deep unfolding NN, based on iterative WMMSE, optimizes the digital precoding matrix. The goal is to maximize the system sum rate while satisfying beam selection constraints. The DRL and deep unfolding networks are trained sequentially, with one network’s output serving as the input to the other. Backpropagation and SGD are used for training and parameter updates. The authors verified the proposed framework outperforms benchmarks in sum-rate, complexity, and robustness while maintaining interpretability and scalability.

Lin \textit{et al.} \cite{LearnPrecoder_2022} unfolded the PGD algorithm for LMMSE transceiver design in uplink massive MU-MIMO with one-bit ADCs, training only the step-size per iteration to minimize system MSE and reduce training overhead.
In \cite{HBF_GAN_2021}, deep generative models are combined with unfolding to solve HBF as a bilevel optimization problem. The upper layer optimizes the digital precoder given a fixed analog precoder, and the lower layer optimizes the analog precoder given a fixed digital precoder. Trainable CNN weights map between the precoders. This unfolded bilevel approach achieves near-optimal HBF with low feedback and computational complexity.

\subsection{Next Generation Multiple Access}
Rate-splitting multiple access (RSMA) is a flexible multiple access technology that merges the strengths of space division multiple access (SDMA) and NOMA \cite{RSMA}. It uses linearly precoded rate-splitting at the transmitter and SIC at the receivers. RSMA balances interference management by partially decoding some interference while treating the rest as noise, bridging the gap between SDMA (which treats all interference as noise) and NOMA (which fully decodes interference).
In \cite{RS_BNN_2024}, the authors applied fractional programming with hyperplane fixed-point iteration to design the RSMA beamformer. The algorithm is unfolded into an NN, replacing its inner loop with a small-scale DNN to predict Lagrangian dual variables. Simulations show that the proposed approach achieves high performance while reducing computational complexity.
	
\subsection{IRS}
	
	
Chen \textit{et al.} \cite{Chen_HBF_RIS} proposed the deep unfolded WMMSE-MO (DU-WMMSE-MO) network for IRS-assisted mmWave MIMO-OFDM systems, jointly designing the HBF and IRS phase-shifting matrices. Trainable parameters include precoder and IRS phase weights and biases. The method improves IRS phase-shift design by jointly updating step-sizes, offering more flexibility than classical WMMSE-MO with Armijo backtracking line search method \cite{WMMSE_MO_original} but with higher runtime.
In \cite{IRS_beamform}, the authors studied IRS-assisted multiuser MIMO full-duplex systems, using deep unfolding to approximate the stochastic SSCA beamforming algorithm, reducing matrix inversion complexity while achieving near-optimal weighted sum rate performance.

Min \textit{et al.} \cite{IRS_QoS_2021} proposed a two-timescale unfolded primal-dual HBF method for IRS-aided multiuser MISO systems to reduce signal processing complexity and channel training overhead. By separating optimization into two timescales, the approach efficiently minimizes transmit power while meeting individual quality of service (QoS) constraints, improving computational efficiency and performance.


In \cite{FractionalIRS_2024}, a low-complexity IRS-assisted downlink MISO system is designed using an unfolded block CD algorithm, decoupling the joint optimization of transmit beamforming and IRS phase shifts into two subproblems via Lagrange dual transformation. The numerical analysis validated that the achieved weighted sum rate is comparable to that of classical approach in \cite{AO_BCD_2020} but with reduced computational complexity. In \cite{BF_IRS_2024}, an unfolded network combining WMMSE and power iteration was proposed for joint beamforming, reducing computational runtime to about 3\% of the classical approach in \cite{AO_BCD_2020}.
	

In \cite{SymbioticIRS_BF_2024}, a deep unfolding algorithm based on gradient descent is proposed for IRS-assisted MISO covert symbiotic radio systems, jointly optimizing active and passive beamforming under covert constraints, with trainable step-sizes and Lagrange multipliers. In contrast, Gangyong \textit{et al.} in \cite{IRS_RGDescent_2024} focused only on passive beamforming, using unfolded Riemannian gradient descent to optimize the IRS phase shift matrix by maximizing the channel path gain. The network in \cite{IRS_RGDescent_2024} is simpler compared to that of \cite{SymbioticIRS_BF_2024}, due to less number of parameters to train but is specific to optimizing the phase shift of the IRS.

\subsection{Terahertz}\label{tera}

6G wireless networks are expected to achieve terabit-per-second (Tbps) data rates to support ultra-reliable low-latency communication. To enhance spectral and energy efficiency, terahertz (THz)-enabled massive MIMO is a promising solution, leveraging the abundant bandwidth at THz frequencies to significantly boost data rates \cite{THz_HBF}.
	
Nguyen \textit{et al.} \cite{Nguyen_DU_HBF} proposed a deep unfolding framework that unfolds the iterations of Riemannian MO (specifically the MO-AltMin algorithm) \cite{Yu_alternating_HBF}. Instead of directly solving the original SE maximization or WMMSE minimization problems, which are highly nonconvex and computationally intensive, they reformulated the HBF design as an approximate matrix factorization problem. This factorization leads to LS formulation for analog and digital beamforming enabling a more efficient solution. By leveraging this structure, their proposed ManNet-based approach \cite{Nguyen_DU_HBF} requires significantly fewer layers to estimate the analog precoder, resulting in faster training and lower computational complexity as compared to the approach in \cite{Lavi_learn_HBF}.
	
Minghui \textit{et al.} \cite{Wu_RSMA_THz} proposed an unfolded network to enhance RSMA precoding in IRS-aided THz MU-MIMO systems under CSI imperfections. They unfold an approximate WMMSE algorithm to optimize the digital active precoding layers based on the estimated equivalent channels. The trainable set includes the transformer weights associated with the private and common streams and a regularization parameter. Compared to the original approximate WMMSE algorithm, the proposed unfolding method achieves superior precoding performance with substantially lower computational complexity.

In THz communication systems, ultra-large antenna arrays will be employed to form narrow beams and mitigate severe path loss. This objective naturally leads to the combination of massive MIMO and wideband THz, which introduces significant scalability challenges. In particular, deep unfolding algorithms for precoding must efficiently scale with the large number of antennas and multiple users, whereas existing designs have primarily focused on point-to-point massive MIMO \cite{Nguyen_DU_HBF} or single-antenna users \cite{Wu_RSMA_THz}. Moreover, the frequency selectivity of THz channels makes the design and optimization of beamformers (both analog and digital) more challenging, as they must operate across a large number of subcarriers in wideband systems. The computational complexity of deep unfolding generally grows linearly with the number of subcarriers \cite{Nguyen_DU_HBF,Wu_RSMA_THz}. This increase in complexity raises concerns about scalability in practical deployments. Moreover, THz systems face hardware constraints such as phase noise, low-resolution ADCs, and synchronization errors, which remain largely unexplored in current deep unfolding designs and present opportunities for further optimization.

\begin{table*}[t!]
\centering

\renewcommand{\arraystretch}{1.5}
\caption{
\centering
Comparison of Classical vs Unfolded methods for Precoder design. 
\newline
$T$: Layers, $I$: Outer iterations, $J$: Inner iterations, $I_{\text{net}}$: Iterations of network, $N_{RF}$: No. of RF chains, $N_t$: No. of transmit antenna, $N_r$: No. of receive antenna, $K$: No. of users, $M$: Symbol size, $N_f$: No. of subcarriers.}
\label{tab_precoder_design}
\begin{tabular}{|p{1cm}|p{3.2cm}|p{3.5cm}|p{2.5cm}|p{3.5cm}|p{1.2cm}|p{1cm}|}
\hline
\multirow{2}{*}{\textbf{Unfolded}} & 
\multicolumn{2}{c|}{\textbf{Computational Complexity} }& 
\textbf{Number of Parameters (Unfolded)} &
\multicolumn{3}{c|}{\textbf{Convergence speed (Number of layers)}} \\
\cline{2-3} \cline{5-7}
\textbf{Model} & \textbf{Classical} & \textbf{Unfolded} &  & \textbf{Simulation Setup}& \textbf{Classical} & \textbf{Unfolded} \\
\hline
IAIDNN \cite{Hu_iterativeAlgo}& WMMSE: \quad $\mathcal{O}(N_{RF}^{3})$ & $\mathcal{O}(T(K^2N_{t}N_{r}^{2}+K^{2}N_{t}^{2}N_{r}+KN_{t}^{2.37}+KN_{r}^3)$ &  $TK(3N_{r}^{2}+3M^{2}+MN_{r})+(T-1)K(3N_{t}^{2}+MN_{t})$ & Massive MIMO, $N_{t}=128$, $N_{r}=2$ and $K=30$&$100$ iterations & $7$ Layers\\
\hline
PGA-Net \cite{Lavi_learn_HBF} & PGA: \quad $\mathcal{O}(IN_{f}N_{t}(KN_{RF}+N_{t}(K+N_{RF})))$ & $\mathcal{O}(TN_{f}N_{t}(KN_{RF}+N_{t}(K+N_{RF})))$ &  $T(N_{f}+1)$ & Single cell downlink MIMO with Rayleigh fading ($N_{t}=12$, $N_{r}=1$, $K=6$ $N_{f}=8$, $N_{RF}=10$)
&$100$ iterations & $5$ Layers \\
\hline
ManNet  \cite{Nguyen_DU_HBF} & MO-AltMin algorithm: \quad $I\mathcal{O}(N_{t}N_{f}N_{RF}^{2}+J(3N_{t}N_{RF}+2N_{f}(N_{RF}^{2}+N_{RF})M))$& $(I_{\text{net}}-1)\mathcal{O}(N_{t}N_{f}N_{RF}^{2})+\mathcal{O}(N_{t}N_{f}N_{RF})+I_{\text{net}}\mathcal{O}(2N_{f}N_{RF}^{2}M+T(3N_{t}N_{RF}+2N_{f}N_{RF}M)$ & $4TI_{\text{net}}N_{t}$ & THz massive MIMO, $N_{t}=\{16,32,64,128\}$, $N_{r}=N_{RF}=M=2$ and $N_{f}=128$&$500$ iterations & $30$ Layers \\
\hline

\end{tabular}
\vspace{-1em}
\end{table*}

    \subsection{Discussions and Takeaways}
Precoding design in massive MIMO and HBF systems involves solving non-convex optimization problems under hardware constraints and channel uncertainty. 
Table~\ref{tab_precoder_design} lists some popular unfolded networks for precoding design and their performance metrics. 
Deep unfolding enables the incorporation of domain-specific constraints such as the constant modulus constraint in analog beamformers through tailored activation functions. This approach significantly reduces computational complexity by avoiding matrix inversion and accelerates convergence, making it well-suited for real-time hybrid precoding under practical hardware limitations.


Despite their benefits, current unfolded precoding models often assume fixed system dimensions and ideal CSI, which limits generalization and applications in practical systems. 
Future work should focus on scalable, adaptive unfolding frameworks that can operate reliably in time-varying, multi-user, and wideband settings. Integration with channel estimation modules and robust designs under CSI uncertainty will be key for practical deployment in next-generation wireless networks. Moreover, applying such unfolded frameworks in real systems and analyzing the performance in real experiments will be very beneficial. 

On the other hand, more emphasis should be placed on ensuring the scalability of deep unfolded precoding algorithms, especially for massive MIMO wideband systems. This requires not only unrolling conventional iterative algorithms, but also rethinking the unfolding architecture and optimization so that the models scale effectively with the array size, bandwidth, and number of users. Possible solutions include designing unfolding networks with sparse inter-layer connections \cite{Nguyen_DU_HBF} and employing adaptive pruning or approximated computations to reduce redundant operations across layers \cite{milstein2025learned,avrahami2025deep}. For distributed systems, scalability can also be achieved through hierarchical unfolding architectures that offload the main computational tasks to a central unit, as further discussed in Section~\ref{sec_distributed_unfolding}.

\section{Sensing and Communication} \label{sec::sensing_communication}
The independent operation of sensing and communication faces challenges in the 6G era due to spectrum congestion in high-frequency bands like mmWave and THz \cite{Seventy_ISAC}. This has driven the development of ISAC systems, which unify both functions using shared architectures and processing techniques \cite{ISAC_waveform}, \cite{JRC_survey}, \cite{JRCDesign}. ISAC includes active (mono-static) and passive (bi-static) sensing methods \cite{Enabling_ISAC_survey}. While ISAC enhances SE and system performance, understanding its fundamental limits is crucial. Key sensing parameter estimations have been thoroughly analyzed in \cite{Fundamental_limit_ISAC}, revealing theoretical and practical boundaries.

\subsection{MIMO ISAC}
The integration of deep unfolding in ISAC systems has shown promising advances in optimizing waveform design, beamforming, and communication-sensing trade-offs. 
In \cite{DU_constantmod_2023}, the challenge of designing constant-modulus waveforms for MU-MIMO ISAC systems was explored. Since constant-modulus waveforms help avoid signal distortion in non-linear power amplifiers, their design is crucial. However, the non-convex nature of the constant-modulus constraint makes optimization challenging. To address this, the authors employed a deep unfolded network based on PGD algorithm, ensuring efficient waveform design under these constraints.  


HBF in MIMO ISAC systems introduces complexity by balancing communication rates and sensing accuracy. The authors in \cite{JSAC_HBF_DU} addressed this by unfolding the iterative PGA algorithm, accelerating convergence and improving system performance. Similarly, the authors in \cite{ISAC_BF_2025} tackled joint radar-communication beamforming using symbol-level precoding with constructive interference constraints to enhance energy efficiency. To handle the non-convex and high-dimensional optimization, a custom iterative algorithm based on block CD and primal-dual optimization was proposed. The deep unfolded network trains step-sizes and a correction matrix for matrix inversion, reducing complexity, improving scalability, and enabling real-time implementation.

In mmWave MIMO systems, large antenna arrays allow sub-meter localization but with high computational cost. Fan \textit{et al.} \cite{DeepADMM_2023} proposed a fast direct localization method using an unfolded ADMM solver with trainable penalty, proximal parameters, and channel gain weights. Also in~\cite{Zhang_ISAC_DU_2025}, the authors have unfolded ADMM to design a transceiver for ISAC in a cluttered environment. The learnable parameters consist of the per layer step sizes in the network and the ADMM penalty coefficients. Both these approaches~\cite{DeepADMM_2023, Zhang_ISAC_DU_2025} achieve faster convergence and lower computational time than classical ADMM.

\subsection{Modern Multicarrier Modulation for ISAC}
	
A joint passive sensing and communication demodulation for OFDM is addressed in \cite{ISAC_Net_2024}. The ISAC-Net is designed by unfolding two algorithms: a denoising-based AMP algorithm for the communication module and the 2D discrete Fourier transform (DFT) for the sensing module. The step-size to update the channel and DFT
matrix form the trainable set to balance the communication demodulation and sensing accuracy. 
Furthermore, Jingcai \textit{et al.} proposed an unfolded network (FISTA-Net) for joint target parameter estimation and communication in MIMO-OFDM systems \cite{TargetEst_2023}. The fast ISTA is unfolded, with learnable parameters such as step-size for weight matrix, enabling high-precision solutions with low computational complexity. The proposed network converges with $8$ layers as compared to $2000$ iteration in fast ISTA. Also, the network achieves an NMSE of $-20.12$ dB while the classical complex ISTA achieved $-16.8$ dB.  
The problem of estimating parameters of moving targets in high-speed scenarios using OTFS modulation for ISAC systems is explored in \cite{DU_ISACOTFS_2024}. 
The ADMM iterations are unfolded (ADMM-Net) to accelerate convergence and enhance sparse recovery of target position. Regarding parameter estimation, ADMM-Net \cite{DU_ISACOTFS_2024} and FISTA-Net \cite{TargetEst_2023} reduce NMSE significantly. Also, ISAC-Net \cite{ISAC_Net_2024} and FISTA-Net \cite{TargetEst_2023} show faster convergence than iterative baselines.
	
\subsection{IRS-Assisted ISAC}
Liu \textit{et al.} addressed a robust beamforming design for an IRS-assisted ISAC system under imperfect CSI and target direction estimates in \cite{IRS_ISAC_2024}.  The objective is to minimize the sum of worst-case Cramér-Rao bounds (CRB) for all sensing targets. They formulated a bi-level optimization structure that consists of two levels: the lower-level problem finds the worst-case CSI and worst-case target sensing angles. The upper-level problem optimizes the active and passive beamforming matrices. A double-loop deep unfolding NN is designed to tackle the bi-level optimization problem. The CSI and angle uncertainties are updated via PGD in the lower-level loop. The beamforming variables are updated in the upper-level loop using PGD and Riemannian MO optimization. The proposed network effectively achieves near-optimal performance by handling statistical CSI errors and bounded target direction errors.
\begin{table*}[t!]
\centering
\renewcommand{\arraystretch}{1.5}
\caption{
\centering
Comparison of Classical vs Unfolded methods for ISAC. 
\newline
$I$: Inner Layer, $J$: Outer Layer, $N_{RF}$: No. of RF chains, $N_t$: No. of transmit antenna, $N_r$: No. of receive antenna, $N_{b}$: No. of BS antenna, $G$: No. of uniform grid location, $B$: No. of BS, $L_{b}$: No. of uniform grid of angle, $M$: No. of OFDM symbols, $N$: No. of subcarriers, $P$: No. of multipaths.}
\label{tab_ISAC}
\begin{tabular}{|p{1.83cm}|p{3.2cm}|p{3.5cm}|p{1.45cm}|p{3cm}|p{1.22cm}|p{1.2cm}|}
\hline
\multirow{2}{*}{\textbf{Unfolded}} & 
\multicolumn{2}{c|}{\textbf{Computational Complexity}} & 
\textbf{Number of Parameters (Unfolded)} & 
\multicolumn{3}{c|}{\textbf{Convergence speed (Number of layers)}} \\
\cline{2-3} \cline{5-7}
\textbf{Model} & \textbf{Classical} & \textbf{Unfolded} &  & \textbf{Simulation Setup} & \textbf{Classical} & \textbf{Unfolded} \\
\hline
Unfolded PGA \cite{JSAC_HBF_DU}
& PGA: $IJ\mathcal{O}(4N_{\text{RF}}N_{t}^{2}+3N_{t})$
& $\mathcal{O}(IJN_{t}^{2}N_{r})$
& $2IJ$
& Massive MIMO, $N_{t}=64$, $N_{r}=$, $N_{\text{RF}}=4$.
& $\sim 500$ iterations & $40$ Layers \\
\hline
Unfolded ADMM for localization \cite{DeepADMM_2023}
& Direct Source Localization \cite{DiSouL}: $\mathcal{O}((GB+\sum_{b=1}^{B}L_{b})^{3.5}\sum_{b=1}^{B}N_{b})$
& $\mathcal{O}(((BG)^{2}+(\sum_{b=1}^{B}L_{b})^{2})\sum_{b=1}^{B}N_{b})$
& $3J+B$
& 5G 3GPP Urban Macro MIMO mmWave channel, $B=4$, $N_{b}=50$, $G=900$, $L_{b}=100$.
& $30$ iterations & $5$ Layers \\
\hline
ISAC-Net \cite{ISAC_Net_2024}
& 2D FFT and cyclic cross correlation \cite{DFT_CCS_ISAC}: $\mathcal{O}(J(M^{2}+N^{2}))$
& $\mathcal{O}(TP(N_{t}N_{r}+MN+N\log(N)+M\log(M)+2N_{t}N_{r}(MN)^{2}))$
& $5J$
& MIMO OFDM ($N_{t}=N_{r}=8$), 16-QAM modulation, $M=256$, $N=1024$.
& $500$ iterations & $5$ Layers \\
\hline
\end{tabular}
\vspace{-1em}
\end{table*}

	\subsection{Discussions and Takeaways}
	Deep unfolding offers a structured and interpretable framework for tackling challenging optimization problems in ISAC systems. By unrolling classical algorithms like PGD and WMMSE into NN layers, unfolding enables the joint design of beamformers and waveforms that balance sensing accuracy and communication performance. These approaches reduce computational complexity and provide faster convergence compared to traditional iterative solutions. We summarize typical deep unfolding models for ISAC in Table~\ref {tab_ISAC} and present a comparison in terms of computational metrics. It is observed that deep unfolding exhibits advantages in constant modulus waveform design,  optimization of dual-functional beamformers, and power allocation. Indeed, the deep unfolding technique can effectively handle non-convex constraints or objective functions like SINR, sum rate, and CRB, which are infeasible with black-box models. 

In \cite{ISAC_practical}, the authors have provided techniques such as gradient calculation for estimating position that can be employed to facilitate practical deployment. However, the application of deep unfolding in ISAC is still underexplored. Indeed, most existing unfolding solutions are developed to optimize the transmit waveform in monostatic ISAC scenarios. In sensing functions, the processing of the received echo signal plays an important role in translating the observation into a decision on target detection and localization. Well-designed deep unfolding models can perform these tasks with low complexity and high accuracy, which is important for real-time sensing applications. There is limited literature addressing the practical deployment of unfolded networks in ISAC systems. 


\section{Decoding in Error Correcting Codes} \label{sec::decoding_ECC}
	
Mobile communication has evolved across generations. 1G used analog transmission (without any coding), and 2G introduced digital voice coding with convolutional and block codes. Turbo codes, adopted in 3G and 4G, offer near-Shannon limit performance with moderate complexity \cite{Turbo}. Low-density parity check (LDPC) codes, rediscovered in the 1960s, are now widely used in 4G and 5G \cite{Gallager_LDPC}, \cite{gallager_thesis}, \cite{MacKay}. Polar codes, introduced in 2009, achieve the capacity of symmetric binary-input discrete memoryless channels \cite{PolarCodes}, approaching the Shannon limit as the block length increases. They have been adopted in 5G new radio for both uplink and downlink control channels \cite{Polar_5G_control}.
BP operates on a bipartite graph known as Tanner graph, introduced by Michael Tanner \cite{Tanner}  
The min-sum and max-sum algorithms are derived from the generic sum-product decoding \cite{Factor_graph_sum_product}.  
Polar codes use successive cancellation decoding, where the received bits are decoded sequentially by leveraging frozen bits and previously decoded bits. More advanced techniques, such as successive list decoding assisted by cyclic redundancy check (CRC), further improve the performance of polar codes \cite{List_decoding}.

Wei \textit{et al.} in \cite{Wei_DUADMM_Linear} proposed an unfolded ADMM-based penalized algorithm for binary linear codes to enhance the decoding performance. The learnable ADMM decoding network (LADN) trains the ADMM penalty parameters. 
Additionally, decoding irregular binary LDPC codes via unfolded ADMM is explored in \cite{IrregularLDPC_2023}. 
The LADN \cite{Wei_DUADMM_Linear} outperforms the original ADMM-penalized decoder in block error rate for LDPC codes, while the network in \cite{IrregularLDPC_2023} outperforms the original ADMM-penalized decoder in block error rate for irregular LDPC codes.
	
In \cite{He_DU_Turbo}, the authors proposed the TurboNet architecture, to improve turbo decoding by unfolding the max-log MAP algorithm. Trainable parameters were inserted into the max-log-MAP decoding flow like weights for prior LLRs. Numerical analysis demonstarted that TurboNet outperforms classical turbo decoding based on the max-log-MAP algorithm, particularly at moderate-to-high SNRs. Furthermore, TurboNet has been validated through over the air experiments confirming its robustness to SNR mismatches. 
Moreover, the authors have extended the work by including joint detection and decoding in \cite{DU_det_2022}. A CGD-based OAMP detector (CG-OAMPNet) for the MIMO-OFDM system is proposed by unfolding EP. The robustness of CG-OAMPNet was further validated through over-the-air experiments, demonstrating practical resilience to environmental variations. In correlated channel scenario, CG-OAMPNet outperforms OAMP-Net2 \cite{He_model_MIMOdet2} by performance gain of $\sim 2$ dB at BER$=10^{-3}$. 
The authors in \cite{DL_Detection_MarkerCodes_2024} tackle the decoding challenges of marker codes in insertion and deletion channels without perfect CSI. They proposed a network called FBNet to improve detection performance; the forward-backward algorithm is unfolded for marker codes. The trainable weights are the CSI probabilities for insertion/deletion channels. FBNet performs quite close to the classical forward-backward algorithm but is robust to CSI mismatch. FBNet \cite{DL_Detection_MarkerCodes_2024} is based on RNN structure and target synchronization errors (insertions/deletions) without accurate CSI, while TurboNet \cite{He_DU_Turbo} is a deep feedforward DNN structure and targets substitution errors in turbo codes under AWGN channel.
	
Gao \textit{et al.} \cite{Polar_2020} have unfolded the polar BP algorithm to improve polar code decoding performance under one-bit quantization. The network with trainable weights for the factor graph edges outperforms standard BP with two bit quantizer by $0.2$ dB. In \cite{PDD_LDPC_2022}, a double-loop iterative decoding algorithm for LDPC codes is unfolded within a penalty dual decomposition framework. The network alternately updates primal and dual variables in a double loop structure. The penalty coefficients, penalty parameters and over relaxation factor form the trainable parameter set. The network has lower computational complexity than ADMM based decoders. Additionally, by unfolding the iterative decoding process between check nodes and variable nodes for LDPC codes, a deep unfolded normalized min-sum LDPC decoder is implemented in \cite{LDPCNN_2023}. The unfolded network outperforms the classical normalized min-sum LDPC decoder in BER and is robust across code lengths, rates and different channel conditions. The networks in \cite{PDD_LDPC_2022, LDPCNN_2023} focus on LDPC codes while the network in \cite{Polar_2020} consider polar codes. Moreover, the networks in \cite{PDD_LDPC_2022, LDPCNN_2023} use multi-SNR training for robustness.    

In \cite{Bound_BP_2024}, theoretical results for the unfolded BP decoder for Tanner codes are presented. The authors define the generalization gap of a decoder as the difference between the empirical and expected BER. They provide a set of theoretical results that bound this gap and examine its dependence on decoder complexity, which is influenced by code parameters (such as block length, message length, and variable/check node degrees), the number of decoding iterations, and the size of the training dataset.

Traditionally, channel estimation, signal detection, and channel decoding in a MIMO receiver are performed sequentially and independently. This disjoint approach leads to suboptimal performance. To address this, Sun \textit{et al.} \cite{JCDD_Sun} propose an unfolded ADMM-based receiver for joint detection and decoding. The detection and LDPC decoding components are jointly unfolded from ADMM iterations. 
Simulations show that the proposed receiver outperforms classical turbo receivers.
In \cite{Meta_2021}, Zhang \textit{et al.} proposed a joint signal detection and channel decoding for MIMO systems. Signal detection is achieved by unfolding the EP algorithm. The channel decoding is based on an unfolded turbo code decoding network, TurboNet, as proposed in \cite{He_DU_Turbo}. The damping factors in EP and decoder parameters in turbo decoding are learned in the training stage. The networks showed fast adaptation and strong BER gains in over-the-air experiments.

    \begin{table*}[t!]
\centering
\renewcommand{\arraystretch}{1.2}
\caption{
\centering
Comparison of Classical vs Unfolded methods for Decoding in Error Correcting Codes. 
\newline
$T$: Layers, $I$: Iterations, $n$: No. of variable nodes, $\Gamma_{a}$: No. of auxiliary variables,  $m$: No. of check nodes, $d_{\text{max}}$: Maximum degree of check nodes, $N_t$: No. of transmit antenna, $N_r$: No. of receive antenna, $t$: Time slot.}
\label{tab_ECC}
\begin{tabular}{|p{1.5cm}|p{3cm}|p{3.5cm}|p{1.8cm}|p{3cm}|p{1.8cm}|p{1.2cm}|}
\hline
\multirow{2}{*}{\textbf{Unfolded} }& 
\multicolumn{2}{c|}{\textbf{Computational Complexity}} & 
\textbf{Number of Parameters (Unfolded)} & 
\multicolumn{3}{c|}{\textbf{Convergence speed (Number of layers)}} \\
\cline{2-3} \cline{5-7}
\textbf{Model}& \textbf{Classical} & \textbf{Unfolded} & & \textbf{Simulation Setup} & \textbf{Classical} & \textbf{Unfolded} \\
\hline
ADMM-DL Network \cite{IrregularLDPC_2023} & ML decoder: $\mathcal{O}(2^{n})$ & $\mathcal{O}(n+\Gamma_{a})$ &  $T(n+\Gamma_{a}+1)$& Irregular LDPC code $[n=2048,m=512]$ 3GPP protocol and $[n=576,m=288]$ WiMAX protocol, BPSK modulation.& $24$ iterations &$3$ Layers \\
\hline
PDD-Net \cite{PDD_LDPC_2022} & Standard BP decoder: $\mathcal{O}(md_{\text{max}}^{2})$ & $\mathcal{O}(n+m(d_{\text{max}}-3))$&  $3T$& LDPC code $[n=256,m=128]$, BPSK modulation. & $2 \times 10^{4}$ iterations & $25$ Layers \\
\hline
JCED Net \cite{JCDD_Sun} & BP decoder: $\mathcal{O}(I(4md_{\text{max}}+md_{\text{max}}^{2}))$& $\mathcal{O}(T(8N{r}N_{t}^{2}+10N_{r}N_{t}+16N_{t}^{2}t+8N_{r}N_{t}t+(8+2d_{\text{max}})m2^{d_{\text{max}}-1}))$  &  $4T$& mmWave MIMO ($N_r=8$, $N_t=4$), regular LDPC code [$n=288$, $m=144$].&$100$ iterations & $20-25$ Layers \\
\hline

\end{tabular}
\vspace{-1em}
\end{table*}

\vspace{-1em}
    \subsection{Discussions and Takeaways}
    
Table~\ref{tab_ECC} compares various unfolded networks for decoding in error correcting codes based on computational complexity, trainable parameter count, and convergence speed (in number of layers). Deep unfolding spans their applications in modern error-correcting codes, such as LDPC, turbo, and polar codes. 
The soft probabilistic estimates are progressively refined through learnable decoding layers in deep unfolding while enforcing parity constraints, enabling near-optimal hard decisions without the need for manual hyperparameter tuning. Hence, it merges model based optimization, for structured soft-output refinement and codeword validity with data-driven learning, for adaptive penalty tuning, outperforming traditional decoders in accuracy and efficiency.

Despite these advantages, there are several limitations in existing deep unfolding architectures for decoding and channel coding. For example, the unfolded decoders in  \cite{Wei_DUADMM_Linear} 
is tied to specific code structures, requiring retraining when parameters such as code length or interleaving patterns change. Some methods demand significant training effort or hyperparameter tuning to maintain generalization. 

There are several works \cite{Meta_2021, He_DU_Turbo, DU_det_2022} that developed a prototyping system to experimentally validate the effectiveness and practical feasibility of unfolded networks in real-world channel conditions. The experimental results indicated that the unfolded network demonstrated strong robustness to environmental variations and operate effectively without the need for frequent retraining.

Future research should aim to create more adaptable unfolding architectures that can handle a variety of code structures, channel models, and SNR conditions. Joint optimization with modulation and detection, as well as hardware-aware training, may further improve the viability of deep unfolding for end-to-end error correction in next-generation wireless communication systems. 
    	\vspace{-0.1in}	
\section{Power Allocation } \label{sec::power_allocation}
	
Power allocation in wireless communication refers to the process of distributing available transmit power across different communication channels or users to optimize system performance. It aims to maximize throughput, minimize interference, and ensure fairness, while considering factors like channel conditions, user requirements, and energy efficiency for improved overall network efficiency.
Several classical approaches, such as the Dinkelbach algorithm, Lagrange dual decomposition, interference pricing, successive convex approximation (SCA), WMMSE, and matching theory, have been employed for power allocation \cite{WMMSE, Dinklebach, Lagrange_dual, Interference_pricing, SCA, Matching_theory}. However, a major drawback of these methods is their reliance on iterative algorithms, which are computationally intensive and result in long runtimes, ultimately degrading system energy efficiency. The deep unfolding approach mitigates this issue by reducing the number of iterations required to achieve a near-optimal solution, thereby enhancing computational efficiency and energy savings.    
	
Chowdhury \textit{et al.} in \cite{Chowdhury_WMMSE} unfolded WMMSE into GNN based architecture to learn a power allocation policy from the channel matrix. The key feature of their architecture is that graph topology is leveraged with GNN for efficient rate utility maximization. 
The unfolded WMMSE method integrates domain-specific elements with trainable components, which are parameterized by generated adaptive weigh of the GNN.  Hu \textit{et al.} \cite{Unsupervised_WMMSE} propose an unsupervised power allocation optimization approach for ad-hoc networks by integrating attentive graph representation with deep unfolded WMMSE. The key advantage of attentive graph representation is its ability to allow network nodes to selectively focus on the most relevant connections during learning, rather than treating all links equally. The power allocation framework is built upon this attentive graph representation, as described in \cite{GAT, EGAT}. 
The model \cite{Unsupervised_WMMSE} achieves improved interpretability and accelerated training over \cite{Chowdhury_WMMSE} while eliminating the need for supervised data. 
In \cite{Li_Graph-Based}, Li \textit{et al.} unroll the SCA algorithm into NN layers for energy-efficient power allocation in wireless interference networks. The proposed unfolded algorithm is highly adaptable, as it generalizes across varying network sizes and channel distributions. 
The power allocation policy is formulated based on the CSI matrix and the total power constraint, within a layered architecture incorporating trainable weights. Each of the networks discussed in \cite{Chowdhury_WMMSE, Unsupervised_WMMSE} and \cite{Li_Graph-Based} leverages the GNN structure in the deep unfolding approach and achieve near optimal performance of their benchmark iterative algorithm i.e, WMMSE and SCA respectively.
	\vspace{-0.1in}		
\section{Physical Layer Security} \label{sec::security}
	
Physical layer security aims to ensure secure and reliable communication while preventing eavesdropping and impersonation attacks. Wireless communication’s broadcast nature exposes it to adversaries like eavesdroppers, who intercept transmissions, and jammers, who degrade signal quality. The primary objective is to maximize the secrecy rate, ensuring high transmission rates for the intended user while minimizing information leakage to the eavesdropper. The secrecy capacity  represents the maximum rate at which secure communication can be achieved, balancing efficiency and confidentiality. It is defined as the difference between mutual information between transmit and receive signals and information leakage to the eavesdropper \cite{Jamming_1}.

Jamming attacks pose a serious threat to the reliable operation of wireless communication systems \cite{Jamming_1, Jamming_2}. A jamming attacker deliberately transmits interference or noise on the same frequencies as legitimate users. 
Marti \textit{et al.} in \cite{Marti_smart_jammer} explored solutions to combat smart jamming in massive multiuser MIMO systems. 
The authors proposed an optimization-based approach for joint jammer estimation, channel estimation, and data detection, using the BS multiantenna configuration. This method does not require prior knowledge of the jammer's behavior, making communication more resilient even during sporadic or bursty jamming.
The authors have unfolded an iterative algorithm, called as mitigation estimation and detection (MAED), that uses forward-backward splitting optimization to jointly estimate the jammer subspace, user equipment's channels, and user equipment's data within a coherence block. The soft-output MAED (SO-MAED), is the deep unfolded network. The trainable set include error precision (to represent noise in symbol estimation), gradient descent step-sizes, scaling factor and momentum weights. SO-MAED improves detection accuracy, especially in high-interference environments, and does not require knowledge of the attack type. 

There is limited literature on the application of deep unfolding techniques in physical layer security. One possible reason is that physical layer security strategies are often dynamic and must adapt to adversarial attacks. In contrast, deep unfolding methods typically rely on a structured iterative algorithm which are seldom available in scenarios such as jamming attacks, where the system behavior must adjust rapidly to unpredictable threats.

		\vspace{-0.1in}	
\section{Challenges in Developing Deep Unfolding Models and Future Research Scopes} \label{sec::challenge}
	
The extensive review of recently emerging techniques and applications of deep unfolding have demonstrated its potential in addressing complex signal processing problems with reduced complexity and runtime. Despite these advantages, the adoption of deep unfolding in practical scenarios remains limited. In this section, we outline the main challenges in developing deep unfolding models, explore the reasons behind these challenges, and propose potential improvements to make deep unfolding a more powerful and widely applicable technique in wireless communications.
		\vspace{-0.1in}

\subsection{Limitations of Deep Unfolding Techniques}

Although deep unfolding has found applications in various design problems, it also has important limitations. Most of these stem from the fact that deep unfolding is not an independent design methodology; it must be built upon existing conventional optimization frameworks. Consequently, it faces the following challenges in both development and real-world applications:

\textbf{\textit{(C1)}} \textbf{Methodological challenges:} Deep unfolding methods are inspired by the iterative procedures of conventional optimization algorithms, where each iteration is typically interpreted as a layer in a neural network. In traditional transmitter-receiver design and signal processing, optimization problems are often non-convex and highly coupled due to factors such as fractional SINR, constant-modulus constraints, or mixed-integer variables. These are typically solved by a two-layer approach: in the first layer, alternating optimization or block coordinate ascent/descent is used to decouple variables; in the second layer, convex optimization techniques such as SCA or ADMM are applied to the resulting sub-problems. While this makes the problems solvable using convex solvers (e.g., CVX), it introduces auxiliary variables and nested loops that must converge at each stage, increasing algorithmic complexity. Deep unfolding, by contrast, attempts to approximate such iterative processes in a unified, trainable architecture, making it challenging to translate multi-stage solvers into compact unfolding models.

\textbf{\textit{(C2)}} \textbf{Unfolding principle algorithms:} Deep unfolding typically enhances conventional iterative algorithms, but not all can be effectively unfolded. Complex operations such as matrix decompositions or convex solver calls (e.g., CVX, YALMIP with MOSEK or SeDuMi) are difficult to represent in differentiable NN layers. Furthermore, large-scale systems such as massive MIMO incur high complexity even for basic operations like matrix multiplications. Another issue is the fixed number of layers: iterative algorithms run until convergence, whereas unfolding models require the number of layers to be fixed a priori, often through empirical tuning. This can lead to suboptimal performance or lack of guaranteed convergence across different scenarios.

\textbf{\textit{(C3)}} \textbf{Handling design constraints and objectives:} Another major challenge is enforcing problem-specific constraints and objectives in the unfolding framework. Simple constraints (e.g., power or constant modulus) can often be imposed by normalization or suitable activation functions, but more complex QoS constraints are difficult to embed. Similarly, intricate objective functions result in complicated loss functions, especially when gradients are hard to compute for high-order or fractional functions. This complicates training and may limit the range of problems amenable to unfolding.

\textbf{\textit{(C4)}} \textbf{Sophisticated mappings:} While deep unfolding is powerful for problems with well-defined iterative solvers, many inference tasks lack such formulations. For example, in multi-modal sensing-assisted beam tracking, there is no explicit relationship between environmental sensory data (e.g., images) and the optimal beam. Since these problems cannot be expressed as tractable optimization tasks, unfolding methods are not suitable, and black-box learning models remain necessary.

\textbf{\textit{(C5)}} \textbf{Real-world applications:} Deep unfolding architectures, while promising in theory, often require considerable network depth and parameter complexity. This creates challenges for real-time inference under stringent latency and throughput constraints, especially in high-dimensional systems such as massive MIMO-OFDM. Hardware-wise, unfolding designs are not naturally aligned with area-efficient or parallelizable architectures, leading to inefficient silicon utilization and high power consumption. Moreover, practical deployments face dynamic channel conditions, mobility, and hardware impairments that often violate the assumptions made in unfolding-based designs. As a result, models trained under idealized conditions may suffer significant performance degradation in practice.

The above five challenges (\textbf{\textit{C1}}--\textbf{\textit{C5}}) illustrate scenarios where deep unfolding is either not applicable or offers limited performance-complexity benefits. These include the absence of suitable iterative algorithms (\textbf{\textit{C1}}, \textbf{\textit{C2}}, \textbf{\textit{C4}}), the presence of highly complicated constraints or objectives (\textbf{\textit{C3}}), and practical imperfections in real-world propagation environments and hardware (\textbf{\textit{C5}}). It is clear that deep unfolding is not a universal replacement for conventional algorithms or black-box machine learning models, but rather a complementary tool most effective when problem structures are well matched to unfolding principles. 

At the same time, several strategies can mitigate the negative impacts of these limitations in practice. Robustness can be improved by training on realistic datasets that incorporate mobility, propagation variability, and hardware impairments. Hybrid designs, in which unfolding is applied only to specific modules (e.g., replacing matrix inversion or accelerating convergence) while other parts rely on conventional or data-driven methods, can achieve better tradeoffs between interpretability and performance. Finally, hardware-aware unfolding architectures tailored for parallelism and energy efficiency can help overcome deployment challenges in large-scale systems. Moreover, the applicability of deep unfolding can be enhanced by overcoming those challenges, as will be discussed next.

\subsection{Development of Deep Unfolding Techniques}
	
\subsubsection{Coexistence of Deep Unfolding Models, Conventional Solvers, and Black-Box Models} Deep unfolding doesn't need to fully replace conventional algorithms or black-box machine learning models. Instead, these approaches can coexist, leveraging their combined strengths. In hybrid designs, deep unfolding can (i) reduce the number of sub-problems, (ii) lower the complexity of solving each sub-problem, and (iii) enhance the learning process in black-box models. Benefits (i) and (ii) correspond to challenges \textbf{\textit{C1}} and \textbf{\textit{C2}}, while (iii) relates to \textbf{\textit{C4}}.
	
Benefit (i) arises because deep unfolding can rely on techniques like PGA/PGD, which avoids transforming the original problem into convex sub-problems. Benefit (ii) comes from the fact that unfolded models typically exhibit lower complexity compared to methods like SCA, and their runtime is usually shorter than that of solvers like CVX. For example, in the AO framework (discussed in \textbf{\textit{C1}}), the most challenging sub-problem often dominates the overall complexity. By applying an efficient deep unfolding model to this sub-problem, the overall efficiency improves, while simpler sub-problems can be handled by conventional solvers like CVX. Benefit (iii) is frequently seen when deep unfolding is used to preprocess input data for black-box models, as in multimodal sensing applications. The combination of deep unfolding structures with conventional solvers and black-box models leads to simpler yet more efficient designs.
	
\subsubsection{Deep Unfolding for Multi-Objective Learning} Most existing deep unfolding models are developed for single-objective designs, such as maximizing throughput, SINR, or minimizing BER. However, in many scenarios, system designs involve multiple objectives. Traditional approaches often formulate these as constrained problems, such as maximizing energy efficiency under a QoS constraint (e.g., throughput or SINR). In  ISAC designs, weighted sums of communication and sensing performance metrics are  used when the goal is to balance both functions without prioritizing one over the other. The complexity grows in multi-static ISAC systems, where distributed sensing receivers have different performance metrics depending on their functions, such as signal clutter noise ratio (SCNR), mutual information, CRB, or probability of detection. 
	
These designs are often difficult to solve using traditional constrained formulations, as the problem can easily become infeasible if constraints are not properly tuned. However, by incorporating all design metrics into the objective function with appropriate weights, to form a multi-objective design, the feasibility of the problem can be significantly improved \cite{liu2018mu, JSAC_HBF_DU}. This multiobjective formulation also makes it easier to develop unfolding models, optimizing multiple objectives simultaneously, and improving performance trade-offs. However, developing unfolding models for multi-objective designs presents several challenges. First, as the number of terms in the objective function increases, training the unfolding DNN becomes more difficult. This is because the loss function is typically set to mirror the objective function (or its negative in maximization problems). Furthermore, balancing the changes of these objectives with respect to the design variables across layers is crucial \cite{JSAC_HBF_DU, ma2024model}, necessitating careful design of the unfolding models.

\subsubsection{Scalable Distributed Deep Unfolding Architectures}
\label{sec_distributed_unfolding}

While deep unfolding models provide advantages in terms of complexity due to faster convergence compared to conventional iterative algorithms, they can still face scalability issues due to challenge \textbf{\textit{C2}}, especially when deployed on platforms with constrained computational resources such as edge devices or distributed access points (APs). Future wireless networks will evolve into large-scale distributed systems, such as cell-free massive MIMO, that integrate advanced sensing, communications, and computing capabilities. However, most existing AI models, both data-driven and model-driven, are designed for centralized scenarios where they are executed at a powerful central node, making them poorly suited for distributed deployments. This highlights the need for scalable distributed unfolding architectures. 

One potential solution is to develop distributed deep unfolding frameworks. Instead of executing all the unfolded layers at every AP or edge device, the layers can be partitioned across multiple platforms for parallel execution, thereby reducing latency and computational burden. For example, to design $L$ precoders for $L$ APs, a CPU with strong computational resources can process the majority of the layers and generate high-quality beamformers. These beamformers are then sent to the APs as initial solutions, which are refined locally by lightweight sub-unfolding models requiring only a few layers. This ensures reliable beamformers at the APs while maintaining scalability by reducing their computational load. Conversely, in the uplink, a distributed unfolding model can be deployed at the APs to generate initial solutions, which are then forwarded to the CPU for final refinement by a larger unfolding model. 

Such a hierarchical CPU-AP cooperation framework directly addresses scalability challenges in distributed deployments by lowering the per-node computational cost, reducing fronthaul/backhaul signaling overhead, and enabling heterogeneous APs to operate with resource-aware lightweight sub-models. In practice, this can be further enhanced by tailoring the unfolding depth at each AP to its available hardware resources and by adopting split-learning style execution, where unfolding layers are adaptively divided between CPU and APs. In this way, reliable symbol detection and beamforming can be achieved with scalable complexity across distributed networks.

\subsubsection{Deep Unfolding to Enhance Black-box Solution} As its name suggests, a black-box machine learning model operates with a black-box nature, making it unexplainable and resistant to re-structuring beyond tuning parameters like the number of layers or nodes. The performance and efficiency of such models are heavily dependent on the quality of input data. Deep unfolding can contribute by performing pre-processing steps to improve the input data quality. One example is in multi-modal sensing-assisted communication tasks, where data from distributed cameras, LiDAR, and radar sensors are used to enhance communication performance, such as in beam tracking and blockage prediction. However, significant challenges arise in collecting, transferring, and processing large sensory data. Specifically, high-resolution camera images generate considerable traffic overhead and latency, impacting real-time communication processing.	
Deep unfolding models can be particularly useful in pre-processing such high resolution images. In image processing, deep unfolding has been successfully applied to tasks such as denoising, super-resolution, and image reconstruction, allowing for high-quality outcomes with fewer iterations than conventional algorithms. For example, unfolded algorithms based on ISTA for sparse coding have been shown to significantly accelerate image reconstruction while maintaining high-quality results \cite{ista_superresolution, ista_compressedsensing, ista_denoising}. This makes deep unfolding a valuable tool for handling large-scale sensory data in real-time communication systems. 
	

\subsubsection{Evaluation Protocols for Deep Unfolding}
\label{sec_evaluation_protocols}
To facilitate the development of deep unfolding techniques, it is necessary to establish standardized evaluation protocols. A natural and broadly applicable criterion is the \textit{convergence profile}, i.e., how quickly an unfolding model achieves satisfactory performance within a limited number of layers \cite{Nguyen_DU_HBF,JSAC_HBF_DU,ma2024model,Lavi_learn_HBF,Matrixfree_BF_2022,Matrixfree_MU_MIMO_2022,nguyen2023deep}. This metric reflects both computational complexity (since fewer layers correspond to fewer iterations and lower runtime) and model performance. Convergence profiles also allow direct comparison with conventional iterative algorithms, highlighting the performance complexity trade-offs of unfolding designs.

Another important aspect is the methodology for employing different learnable \textit{hyperparameters}. Most unfolding techniques learn scalar parameters such as step sizes \cite{Nguyen_DU_HBF,JSAC_HBF_DU,ma2024model,Lavi_learn_HBF,Matrixfree_BF_2022,Matrixfree_MU_MIMO_2022,nguyen2023deep}, while others consider richer structures such as trainable matrices \cite{Hu_iterativeAlgo}. Although the specific parameters vary, they share the common effect of shaping convergence speed and stability, which again makes the convergence profile a valid and unifying evaluation criterion. By contrast, training convergence criteria (e.g., number of epochs or early stopping) are less critical, as unfolding models typically converge quickly \cite{Nguyen_DU_HBF}, and this detail is often omitted in the literature. Furthermore, reproducible open-source implementations of deep unfolding solutions with clear and consistent settings would not only enable fair comparisons but also facilitate the development of new unfolding models. In Table~\ref{tab:open_unfolding}, we present several available implementations for representative tasks.

\hypersetup{
    colorlinks=true,
    linkcolor=blue,
    citecolor=blue,
    urlcolor=blue
}

\begin{table}[t!]
\centering
\caption{{Open-source implementations of deep unfolding techniques.
The GitHub links are embedded directly in model names as hyperlinks.}}
\label{tab:open_unfolding}

\begin{threeparttable}
\renewcommand{\arraystretch}{1.25}   
\setlength{\tabcolsep}{8pt}          

\begin{tabularx}{0.97\linewidth}{@{}p{3.1cm}X@{}}
\toprule
\textbf{Design Problem} & \textbf{Representative Works} \\ 
\midrule

\textbf{Theoretical convergence of LISTA} &
\href{https://github.com/VITA-Group/LISTA-CPSS}{\textcolor{blue}{\underline{\textit{LISTA}}}}~\cite{LISTA_theoretical} \\
\textbf{Detection} &
\href{https://github.com/neevsamuel/DeepMIMODetection}{\textcolor{blue}{\underline{\textit{DetNet}}}}~\cite{Samuel_learningtodetect},
\href{https://github.com/nhanng9115/Deep-Learning-Aided-Tabu-Search-Detection-for-Large-MIMO-Systems}{\textcolor{blue}{\underline{\textit{FSNet}}}}~\cite{FSNet},
\href{https://github.com/mehrdadkhani/MMNet}{\textcolor{blue}{\underline{\textit{MMNet}}}}~\cite{MMNet},
\href{https://github.com/skhobahi/LoRD-Net}{\textcolor{blue}{\underline{\textit{LoRD-Net}}}}~\cite{LoRD_2021},
\href{https://github.com/STARainZ/CG-OAMP-NET}{\textcolor{blue}{\underline{\textit{CG-OAMP-Net}}}}~\cite{DU_det_2022} \\

\textbf{Channel Estimation} &
\href{https://github.com/EricGJB/SBL_unfolding_CE}{\textcolor{blue}{\underline{\textit{SBL-Net}}}}~\cite{DL_Chnnel_est_2023} \\

\textbf{Precoding} &
\href{https://github.com/lpkg/WMMSE-deep-unfolding?tab=readme-ov-file\#ourpaper}{\textcolor{blue}{\underline{\textit{Unfolded WMMSE}}}}~\cite{Matrixfree_BF_2022} \\

\textbf{Hybrid Beamforming} &
\href{https://github.com/nhanng9115/Deep_unfolding_Hybrid_beamforming}{\textcolor{blue}{\underline{\textit{ManNet, Sub-ManNet}}}}~\cite{Nguyen_DU_HBF},
\href{https://github.com/nhanng9115/AI-Empowered-Hybrid-MIMO-Beamforming}{\textcolor{blue}{\underline{\textit{Unfolded PGA}}}}~\cite{Lavi_learn_HBF},
\href{https://github.com/levyohad/Power-Aware-Deep-Unfolding-for-Beamforming}{\textcolor{blue}{\underline{\textit{Modular PGA}}}}~\cite{Power_beamforming_2025} \\

\textbf{ISAC} &
\href{https://github.com/nhanng9115/Joint-Communications-and-Sensing-Hybrid-Beamforming-Design-via-Deep-Unfolding}{\textcolor{blue}{\underline{\textit{Unfolded PGA-ISAC}}}}~\cite{JSAC_HBF_DU} \\

\textbf{Decoding in Error Correcting Codes} &
\href{https://github.com/sunyi0101/JCDDNet}{\textcolor{blue}{\underline{\textit{JCED-Net}}}}~\cite{JCDD_Sun},
\href{https://github.com/tjuxiaofeng/A-Model-Driven-Deep-Learning-Method-for-Normalized-Min-Sum-LDPC-Decoding}{\textcolor{blue}{\underline{\textit{NMS-Net}}}}~\cite{LDPC_unfold_conference} \\

\bottomrule
\end{tabularx}
\vspace{-1em}
\end{threeparttable}
\end{table}

	\vspace{-0.11in}	
	
\section{Conclusions} \label{sec::conclusion}
The integration of classical algorithmic principles with DL techniques has led to the development of deep unfolding, a method that overcomes the limitations associated with both classical algorithms and standalone DL models in wireless communication. 
This article offers a comprehensive survey of deep unfolding techniques, examining their application across various facets of the physical layer in wireless communication systems. Key applications, such as signal detection, channel estimation, precoder design, ISAC, decoding for error-correcting codes, power allocation, and physical layer security, are explored, with an emphasis on technologies anticipated for beyond-5G networks. The comparisons of the unfolded networks have also been given, which reflect the performance gains over the benchmark iterative algorithms. Also, we have analyzed the advantages and limitations of the design of various unfolding models. Moreover, we have explored the practical feasibility of deep unfolding networks across key application areas of wireless communications. The survey highlights the strengths of deep unfolding in addressing complex challenges, particularly in achieving low-latency, high-accuracy performance in dynamic environments. We conclude by discussing the key challenges in developing deep unfolding models, including issues related to scalability, computational efficiency, and robustness across diverse scenarios. Finally, this paper presents potential advancements to enhance the adaptability and resilience of deep unfolding, with the goal of solidifying its role as a robust and versatile technique for next-generation wireless communications.

    \vspace{-0.16in}	
	\bibliographystyle{IEEEtran}
	\footnotesize
	\bibliography{mybib}

@INPROCEEDINGS{Baptiste,
  author={Chatelier, Baptiste and Magoaroul, Luc Le and Redieteab, Getachew},
  booktitle={IEEE ICC 2023}, 
  title={{Efficient Deep Unfolding for SISO-OFDM Channel Estimation}}, 
  year={2023},
  volume={},
  number={},
  pages={3450-3455},
  doi={10.1109/ICC45041.2023.10278825}}

@ARTICLE{ISTA_ADMM_optim,
  author={Shah, Shaik Basheeruddin and Pradhan, Pradyumna and Pu, Wei and Randhi, Ramunaidu and Rodrigues, Miguel R. D. and Eldar, Yonina C.},
  journal={IEEE Trans. Signal Process.}, 
  title={{Optimization Guarantees of Unfolded ISTA and ADMM Networks With Smooth Soft-Thresholding}}, 
  year={2024},
  volume={72},
  number={},
  pages={3272-3286},
  doi={10.1109/TSP.2024.3412981}}

@ARTICLE{Zhang_ISAC_DU_2025,
  author={Zhang, Jifa and Zhu, Yongxu and Zhao, Nan and Jin, Shi and Wang, Xianbin and Ng, Derrick Wing Kwan and Al-Dhahir, Naofal},
  journal={IEEE Trans. Wirel. Commun.}, 
  title={{Deep Unfolding Learning Aided ISAC Transceiver Design}}, 
  year={2025},
  volume={24},
  number={10},
  pages={8681-8695},
  doi={10.1109/TWC.2025.3568367}}

@article{Arikawa,
author = {Manabu Arikawa and Kazunori Hayashi},
journal = {Opt. Express},
number = {16},
pages = {23478--23494},
publisher = {Optica Publishing Group},
title = {{Adaptive equalization of transmitter and receiver IQ skew by multi-layer linear and widely linear filters with deep unfolding}},
volume = {28},
month = {Aug},
year = {2020},
doi = {10.1364/OE.395361}}

@article{StabilityGen_oliver,
author = {Bousquet, Olivier and Elisseeff, Andr\'{e}},
title = {{Stability and generalization}},
year = {2002},
issue_date = {3/1/2002},
publisher = {JMLR.org},
volume = {2},
issn = {1532-4435},
url = {https://doi.org/10.1162/153244302760200704},
doi = {10.1162/153244302760200704},
journal = {J. Mach. Learn. Res.},
month = mar,
pages = {499–526},
numpages = {28}
}

@ARTICLE{Converge_ISAC,
  author={Yang, Jiapan and Ai, Bo and Chen, Wei and Yang, Songjie and Wang, Ning and Yuen, Chau},
  journal={IEEE J. Sel. Areas Commun.}, 
  title={{Deep Unfolding-Based Sensing-Assisted Channel Estimation With Imperfect Radar Arrays}}, 
  year={2025},
  volume={},
  number={},
  pages={1-1},
  doi={10.1109/JSAC.2025.3610425}}

@article{avrahami2025deep,
  title={{Deep Unfolding with Approximated Computations for Rapid Optimization}},
  author={Avrahami, Dvir and Milstein, Amit and Chaux, Caroline and Routtenberg, Tirza and Shlezinger, Nir},
  journal={arXiv preprint arXiv:2509.00782},
  year={2025}
}

@article{FISTA_amir,
author = {Beck, Amir and Teboulle, Marc},
title = {{A Fast Iterative Shrinkage-Thresholding Algorithm for Linear Inverse Problems}},
journal = {SIAM Journal on Imaging Sciences},
volume = {2},
number = {1},
pages = {183-202},
year = {2009},
doi = {10.1137/080716542}
}

@ARTICLE{Survey_DU_24,
  author={Liu, Ya-Feng and Chang, Tsung-Hui and Hong, Mingyi and Wu, Zheyu and Man-Cho So, Anthony and Jorswieck, Eduard A. and Yu, Wei},
  journal={IEEE J. Sel. Areas Commun.}, 
  title={{A Survey of Recent Advances in Optimization Methods for Wireless Communications}}, 
  year={2024},
  volume={42},
  number={11},
  pages={2992-3031},
  doi={10.1109/JSAC.2024.3443759}}

@inproceedings{milstein2025learned,
  title={{Learned Approximated Optimization for Rapid Low-Complexity Hybrid Beamforming Design}},
  author={Milstein, Amit and Yablonka, Tomer and Shlezinger, Nir},
  booktitle={IEEE ICASSP},
  pages={1--5},
  year={2025},
  organization={}
}

@misc{gen_error,
      title={{Compressive Sensing and Neural Networks from a Statistical Learning Perspective}}, 
      author={Arash Behboodi and Holger Rauhut and Ekkehard Schnoor},
      year={2021},
      eprint={2010.15658},
      archivePrefix={arXiv},
      primaryClass={math.ST},
      url={https://arxiv.org/abs/2010.15658}, 
}

@techreport{3gpp36.104,
  title        = {{Evolved Universal Terrestrial Radio Access (E-UTRA); Base Station (BS) Radio Transmission and Reception}},
  institution  = {3rd Generation Partnership Project (3GPP)},
  number       = {TS 36.104},
  type         = {Technical Specification},
  version      = {Release 8},
  year         = {2021},
  note         = {Accessed: Apr. 24, 2021},
  url          = {https://www.3gpp.org/ftp/Specs/archive/36_series/36.104/}
}

@inproceedings{gen_rademecher,
    author = {Avner Shultzman and Eyar Azar and Rodrigues, Miguel R.D. and Eldar, Yonina C.},
    title = {{Generalization and Estimation Error Bounds for Model-based Neural Networks}},
    booktitle = {Proc. Int. Conf. Learn. Represent},
    year = {2023}
}

@ARTICLE{DFT_CCS_ISAC,
  author={Wei, Zhiqing and Qu, Hanyang and Jiang, Wangjun and Han, Kaifeng and Wu, Huici and Feng, Zhiyong},
  journal={IEEE Trans. Green Commun. Netw.}, 
  title={{Iterative Signal Processing for Integrated Sensing and Communication Systems}}, 
  year={2023},
  volume={7},
  number={1},
  pages={401-412},
  }

@ARTICLE{DiSouL,
  author={Garcia, Nil and Wymeersch, Henk and Larsson, Erik G. and Haimovich, Alexander M. and Coulon, Martial},
  journal={IEEE Trans. Signal Process.}, 
  title={{Direct Localization for Massive MIMO}}, 
  year={2017},
  volume={65},
  number={10},
  pages={2475-2487},
  }

@ARTICLE{Power_beamforming_2025,
  author={Levy, Ohad and Shlezinger, Nir},
  journal={IEEE Trans. Commun.}, 
  title={{Rapid and Power-Aware Learned Optimization for Modular Receive Beamforming}}, 
  year={2025},
  volume={73},
  number={9},
  pages={7136-7150},
  }

@INPROCEEDINGS{LDPC_unfold_conference,
  author={Wang, Qing and Wang, Shunfu and Fang, Haoyu and Chen, Leian and Chen, Luyong and Guo, Yuzhang},
  booktitle={2020 IEEE ICC Workshops}, 
  title={{A Model-Driven Deep Learning Method for Normalized Min-Sum LDPC Decoding}}, 
  year={2020},
  volume={},
  number={},
  pages={1-6},
  }

@ARTICLE{ISTA_theoretical,
  author={Scarlett, Jonathan and Heckel, Reinhard and Rodrigues, Miguel R. D. and Hand, Paul and Eldar, Yonina C.},
  journal={IEEE J. Sel. Areas Inf. Theory}, 
  title={{Theoretical Perspectives on Deep Learning Methods in Inverse Problems}}, 
  year={2022},
  volume={3},
  number={3},
      pages={433-453},
  }

@inproceedings{LISTA_theoretical,
author = {Chen, Xiaohan and Liu, Jialin and Wang, Zhangyang and Yin, Wotao},
title = {{Theoretical linear convergence of unfolded ISTA and its practical weights and thresholds}},
year = {2018},
volume={31},
booktitle = {Proc. Conf. Neural Inf. Process. Syst.},
pages = {9079–9089}
}

@inproceedings{ALISTA_theoretical,
author = {Chen, Xiaohan and Liu, Jialin and Wang, Zhangyang and Yin, Wotao},
title = {{Hyperparameter tuning is all you need for LISTA}},
year = {2021},
booktitle = {Proc. Conf. Neural Inf. Process. Syst.},
pages = {11678–11689},
}

@ARTICLE{DECONET_learning,
  author={Kouni, Vicky and Panagakis, Yannis},
  journal={IEEE Trans. Signal Process.}, 
  title={{DECONET: An Unfolding Network for Analysis-Based Compressed Sensing With Generalization Error Bounds}}, 
  year={2023},
  volume={71},
  number={},
  pages={1938-1951},
  doi={10.1109/TSP.2023.3272286}}

@article{L2O_survey,
author = {Chen, Xiaohan and Liu, Jialin and Yin, Wotao},
    year = {2024},
    month = {05},
    pages = {},
    title = {{Learning to optimize: A tutorial for continuous and mixed-integer optimization}},
    volume = {67},
    journal = {Science China Mathematics},
    doi = {10.1007/s11425-023-2293-3}
}

@ARTICLE{ISAC_practical,
  author={Hu, Zhixiang and Liu, An and Xu, Wenkang and Quek, Tony Q. S. and Zhao, Minjian},
  journal={IEEE J. Sel. Areas Commun.}, 
  title={{A Stochastic Particle Variational Bayesian Inference Inspired Deep-Unfolding Network for Sensing Over Wireless Networks}}, 
  year={2024},
  volume={42},
  number={10},
  pages={2832-2846},
  doi={10.1109/JSAC.2024.3414626}}

@ARTICLE{HBF_2001,
  author={Heath, R.W. and Sandhu, S. and Paulraj, A.},
  journal={IEEE Commun. Lett.}, 
  title={{Antenna Selection for Spatial Multiplexing Systems with :Linear Receivers}}, 
  year={2001},
  volume={5},
  number={4},
  pages={142-144},
  doi={10.1109/4234.917094}}

@ARTICLE{EP_1,
  author={Céspedes, Javier and Olmos, Pablo M. and Sánchez-Fernández, Matilde and Perez-Cruz, Fernando},
  journal={IEEE Trans. Commun.}, 
  title={{Expectation Propagation Detection for High-Order High-Dimensional MIMO Systems}}, 
  year={2014},
  volume={62},
  number={8},
  pages={2840-2849},
  doi={10.1109/TCOMM.2014.2332349}}

@ARTICLE{EP_2,
  author={Santos, Irene and Murillo-Fuentes, Juan José and Boloix-Tortosa, Rafael and Arias-de-Reyna, Eva and Olmos, Pablo M.},
  journal={IEEE Trans. Commun.}, 
  title={{Expectation Propagation as Turbo Equalizer in ISI Channels}}, 
  year={2017},
  volume={65},
  number={1},
  pages={360-370},
  doi={10.1109/TCOMM.2016.2616141}}

@ARTICLE{MassiveMIMO_det,
  author={Albreem, Mahmoud A. and Juntti, Markku and Shahabuddin, Shahriar},
  journal={IEEE Commun. Surveys Tuts.}, 
  title={{Massive MIMO Detection Techniques: A Survey}}, 
  year={2019},
  volume={21},
  number={4},
  pages={3109-3132},
  doi={10.1109/COMST.2019.2935810}}

@ARTICLE{OTFS_withADMM,
  author={Li, Xuefeng and Shan, Chengzhao and Zhao, Honglin and Yuan, Weijie and Zhang, Ruoyu},
  journal={IEEE Commun. Lett.}, 
  title={{Iterative Detection for Orthogonal Time Frequency Space Modulation with ADMM}}, 
  year={2025},
  volume={},
  number={},
  pages={1-1},
  doi={10.1109/LCOMM.2025.3549525}}

@ARTICLE{MMNet,
  author={Khani, Mehrdad and Alizadeh, Mohammad and Hoydis, Jakob and Fleming, Phil},
  journal={IEEE Trans. Wireless Commun.}, 
  title={{Adaptive Neural Signal Detection for Massive MIMO}}, 
  year={2020},
  volume={19},
  number={8},
  pages={5635-5648},
  doi={10.1109/TWC.2020.2996144}}

@INPROCEEDINGS{He_model_MIMOdet1,
  author={He, Hengtao and Wen, Chao-Kai and Jin, Shi and Li, Geoffrey Ye},
  booktitle={2018 IEEE GlobalSIP}, 
  title={{A Model-Driven Deep Learning Network for MIMO Detection}}, 
  year={2018},
  volume={},
  number={},
  pages={584-588},
  doi={10.1109/GlobalSIP.2018.8646357}}

@ARTICLE{He_model_MIMOdet2,
  author={He, Hengtao and Wen, Chao-Kai and Jin, Shi and Li, Geoffrey Ye},
  journal={IEEE Trans. Signal Process.}, 
  title={{Model-Driven Deep Learning for MIMO Detection}}, 
  year={2020},
  volume={68},
  number={},
  pages={1702-1715},
  doi={10.1109/TSP.2020.2976585}}

@INPROCEEDINGS{TPGD,
  author={Takabe, Satoshi and Imanishi, Masayuki and Wadayama, Tadashi and Hayashi, Kazunori},
  booktitle={ICC 2019 }, 
  title={{Deep Learning-Aided Projected Gradient Detector for Massive Overloaded MIMO Channels}}, 
  year={2019},
  volume={},
  number={},
  pages={1-6},
  doi={10.1109/ICC.2019.8761049}}

@article{Samuel_learningtodetect,
  author={Samuel, Neev and Diskin, Tzvi and Wiesel, Ami},
  journal={IEEE Trans. Signal Process.}, 
  title={{Learning to Detect}}, 
  year={2019},
  volume={67},
  number={10},
  pages={2554-2564},
  doi={10.1109/TSP.2019.2899805}}

@ARTICLE{FSNet,
  author={Nguyen, Nhan Thanh and Lee, Kyungchun},
  journal={IEEE Trans. Wireless Commun.}, 
  title={{Deep Learning-Aided Tabu Search Detection for Large MIMO Systems}}, 
  year={2020},
  volume={19},
  number={6},
  pages={4262-4275},
  doi={10.1109/TWC.2020.2981919}}

@ARTICLE{Multicell_WMMSE_precoding,
  author={Yi, Chenyang and Xu, Wei and Sun, Yan and Ng, Derrick Wing Kwan and Wang, Zhaocheng},
  journal={IEEE Trans. Cogn. Commun. Netw.}, 
  title={{Distributed Multiuser Precoding via Unfolding WMMSE for Multicell MIMO Downlinks}}, 
  year={2024},
  volume={},
  number={},
  pages={1-1},
  doi={10.1109/TCCN.2024.3511956}}

@article{Hu_iterativeAlgo,
  author={Hu, Qiyu and Cai, Yunlong and Shi, Qingjiang and Xu, Kaidi and Yu, Guanding and Ding, Zhi},
  journal={IEEE Trans. Wireless Commun.}, 
  title={{Iterative Algorithm Induced Deep-Unfolding Neural Networks: Precoding Design for Multiuser MIMO Systems}}, 
  year={2021},
  volume={20},
  number={2},
  pages={1394-1410},
  doi={10.1109/TWC.2020.3033334}}

@article{Chowdhury_WMMSE,
  author={Chowdhury, Arindam and Verma, Gunjan and Rao, Chirag and Swami, Ananthram and Segarra, Santiago},
  journal={IEEE Trans. Wireless Commun.}, 
  title={{Unfolding WMMSE Using Graph Neural Networks for Efficient Power Allocation}}, 
  year={2021},
  volume={20},
  number={9},
  pages={6004-6017},
  doi={10.1109/TWC.2021.3071480}}

@INPROCEEDINGS{LAMP,
  author={Borgerding, Mark and Schniter, Philip},
  booktitle={2016 IEEE GlobalSIP}, 
  title={{Onsager-corrected deep learning for sparse linear inverse problems}}, 
  year={2016},
  volume={},
  number={},
  pages={227-231},
  doi={10.1109/GlobalSIP.2016.7905837}}

@ARTICLE{LDAMP,
  author={He, Hengtao and Wen, Chao-Kai and Jin, Shi and Li, Geoffrey Ye},
  journal={IEEE Wireless Commun. Lett.}, 
  title={{Deep Learning-Based Channel Estimation for Beamspace mmWave Massive MIMO Systems}}, 
  year={2018},
  volume={7},
  number={5},
  pages={852-855},
  doi={10.1109/LWC.2018.2832128}}

@ARTICLE{AMP_beamspace,
  author={Wei, Yi and Zhao, Ming-Min and Zhao, Minjian and Lei, Ming and Yu, Quan},
  journal={IEEE Wireless Commun. Lett.}, 
  title={{An AMP-Based Network With Deep Residual Learning for mmWave Beamspace Channel Estimation}}, 
  year={2019},
  volume={8},
  number={4},
  pages={1289-1292},
  doi={10.1109/LWC.2019.2916786}}

@ARTICLE{Liao_MassiveMIMODEt,
  author={Liao, Jieyu and Zhao, Junhui and Gao, Feifei and Li, Geoffrey Ye},
  journal={IEEE Commun. Lett.}, 
  title={{A Model-Driven Deep Learning Method for Massive MIMO Detection}}, 
  year={2020},
  volume={24},
  number={8},
  pages={1724-1728},
  doi={10.1109/LCOMM.2020.2989672}}

@ARTICLE{Tan_massiveMIMO_det,
  author={Tan, Xiaosi and Xu, Weihong and Sun, Kai and Xu, Yunhao and Be’ery, Yair and You, Xiaohu and Zhang, Chuan},
  journal={IEEE Trans. Veh. Technol.}, 
  title={{Improving Massive MIMO Message Passing Detectors With Deep Neural Network}}, 
  year={2020},
  volume={69},
  number={2},
  pages={1267-1280},
  doi={10.1109/TVT.2019.2960763}}

@ARTICLE{Hu_jointDRL,
  author={Hu, Qiyu and Liu, Yanzhen and Cai, Yunlong and Yu, Guanding and Ding, Zhi},
  journal={IEEE J. Sel. Areas Commun.}, 
  title={{Joint Deep Reinforcement Learning and Unfolding: Beam Selection and Precoding for mmWave Multiuser MIMO With Lens Arrays}}, 
  year={2021},
  volume={39},
  number={8},
  pages={2289-2304},
  doi={10.1109/JSAC.2021.3087233}}

@ARTICLE{Sohrabi_HBF_largescale,
  author={Sohrabi, Foad and Yu, Wei},
  journal={IEEE J. Sel. Topics Signal Process.}, 
  title={{Hybrid Digital and Analog Beamforming Design for Large-Scale Antenna Arrays}}, 
  year={2016},
  volume={10},
  number={3},
  pages={501-513},
  doi={10.1109/JSTSP.2016.2520912}}

@ARTICLE{Ye_powerDL,
  author={Ye, Hao and Li, Geoffrey Ye and Juang, Biing-Hwang},
  journal={IEEE Wireless Commun. Lett.}, 
  title={{Power of Deep Learning for Channel Estimation and Signal Detection in OFDM Systems}}, 
  year={2018},
  volume={7},
  number={1},
  pages={114-117},
  doi={10.1109/LWC.2017.2757490}}

@ARTICLE{Huang_DLPhyLayer,
  author={Huang, Hongji and Guo, Song and Gui, Guan and Yang, Zhen and Zhang, Jianhua and Sari, Hikmet and Adachi, Fumiyuki},
  journal={IEEE Wirel. Commun.}, 
  title={{Deep Learning for Physical-Layer 5G Wireless Techniques: Opportunities, Challenges and Solutions}}, 
  year={2020},
  volume={27},
  number={1},
  pages={214-222},  
  doi={10.1109/MWC.2019.1900027}}

@ARTICLE{Nguyen_DU_HBF,
  author={Nguyen, Nhan Thanh and Ma, Mengyuan and Lavi, Ortal and Shlezinger, Nir and Eldar, Yonina C. and Swindlehurst, A. Lee and Juntti, Markku},
  journal={IEEE Trans. Signal Process.}, 
  title={{Deep Unfolding Hybrid Beamforming Designs for THz Massive MIMO Systems}}, 
  year={2023},
  volume={71},
  number={},
  pages={3788-3804},
  doi={10.1109/TSP.2023.3322852}}

@ARTICLE{Shi_DU_NN_HBF,
  author={Shi, Shuhan and Cai, Yunlong and Hu, Qiyu and Champagne, Benoit and Hanzo, Lajos},
  journal={IEEE Trans. Veh. Technol.}, 
  title={{Deep-Unfolding Neural-Network Aided Hybrid Beamforming Based on Symbol-Error Probability Minimization}}, 
  year={2023},
  volume={72},
  number={1},
  pages={529-545},
  doi={10.1109/TVT.2022.3201961}}

@ARTICLE{Lavi_learn_HBF,
  author={Lavi, Ortal and Shlezinger, Nir},
  journal={IEEE Trans. Commun.}, 
  title={{Learn to Rapidly and Robustly Optimize Hybrid Precoding}}, 
  year={2023},
  volume={71},
  number={10},
  pages={5814-5830},
  doi={10.1109/TCOMM.2023.3292472}}

@ARTICLE{Jamming_1,
  author={Huo, Yan and Tian, Yuqi and Ma, Liran and Cheng, Xiuzhen and Jing, Tao},
  journal={IEEE Wirel. Commun.}, 
  title={{Jamming Strategies for Physical Layer Security}}, 
  year={2018},
  volume={25},
  number={1},
  pages={148-153},
  doi={10.1109/MWC.2017.1700015}}

@INPROCEEDINGS{Jamming_2,
  author={Vaishnavi, K. N. and Khorvi, Shubham Dashrath and Kishore, Rajalekshmi and Gurugopinath, Sanjeev},
  booktitle={2021 ICT}, 
  title={{A Survey on Jamming Techniques in Physical Layer Security and Anti-Jamming Strategies for 6G}}, 
  year={2021},
  volume={},
  number={},
  pages={174-179},
  doi={10.1109/ICT52184.2021.9511465}}

@ARTICLE{Marti_smart_jammer,
  author={Marti, Gian and Kölle, Torben and Studer, Christoph},
  journal={IEEE Trans. Signal Process.}, 
  title={{Mitigating Smart Jammers in Multi-User MIMO}}, 
  year={2023},
  volume={71},
  number={},
  pages={756-771},
  doi={10.1109/TSP.2023.3246226}}

@ARTICLE{Yassine_mpNet,
  author={Yassine, Taha and Le Magoarou, Luc},
  journal={IEEE Trans. Wireless Commun.}, 
  title={{mpNet: Variable Depth Unfolded Neural Network for Massive MIMO Channel Estimation}}, 
  year={2022},
  volume={21},
  number={7},
  pages={5703-5714},
  doi={10.1109/TWC.2022.3142737}}

@ARTICLE{Guo_CSIfeedbackMassiveMIMO,
  author={Guo, Jianhua and Wang, Lei and Li, Feng and Xue, Jiang},
  journal={IEEE Commun. Lett.}, 
  title={{CSI Feedback With Model-Driven Deep Learning of Massive MIMO Systems}}, 
  year={2022},
  volume={26},
  number={3},
  pages={547-551},
  doi={10.1109/LCOMM.2021.3138927}}

@ARTICLE{Compressive_Sensing_survey,
  author={Hassan, Kais and Masarra, Mohammad and Zwingelstein, Marie and Dayoub, Iyad},
  journal={IEEE Open J. Commun. Soc.}, 
  title={{Channel Estimation Techniques for Millimeter-Wave Communication Systems: Achievements and Challenges}}, 
  year={2020},
  volume={1},
  number={},
  pages={1336-1363},
  doi={10.1109/OJCOMS.2020.3015394}}

@ARTICLE{Hu_DDPG,
  author={Hu, Qiyu and Shi, Shuhan and Cai, Yunlong and Yu, Guanding},
  journal={IEEE Trans. Signal Process.}, 
  title={{DDPG-Driven Deep-Unfolding With Adaptive Depth for Channel Estimation With Sparse Bayesian Learning}}, 
  year={2022},
  volume={70},
  number={},
  pages={4665-4680},
  doi={10.1109/TSP.2022.3207269}}

@ARTICLE{DDPG,
  author={Nguyen, Khoi Khac and Duong, Trung Q. and Vien, Ngo Anh and Le-Khac, Nhien-An and Nguyen, Long D.},
  journal={IEEE Access}, 
  title={{Distributed Deep Deterministic Policy Gradient for Power Allocation Control in D2D-Based V2V Communications}}, 
  year={2019},
  volume={7},
  number={},
  pages={164533-164543},
  doi={10.1109/ACCESS.2019.2952411}}

@ARTICLE{He_learning_estimate_RIS,
  author={He, Jiguang and Wymeersch, Henk and Di Renzo, Marco and Juntti, Markku},
  journal={IEEE Wireless Commun. Lett.}, 
  title={{Learning to Estimate RIS-Aided mmWave Channels}}, 
  year={2022},
  volume={11},
  number={4},
  pages={841-845},
  doi={10.1109/LWC.2022.3147250}}

@ARTICLE{Li_Graph-Based,
  author={Li, Boning and Verma, Gunjan and Segarra, Santiago},
  journal={IEEE Trans. Wireless Commun.}, 
  title={{Graph-Based Algorithm Unfolding for Energy-Aware Power Allocation in Wireless Networks}}, 
  year={2023},
  volume={22},
  number={2},
  pages={1359-1373},
  doi={10.1109/TWC.2022.3204486}}

@ARTICLE{Li_ExpectOTFS,
  author={Li, Shuo and Ding, Chao and Xiao, Lixia and Zhang, Xufan and Liu, Guanghua and Jiang, Tao},
  journal={IEEE Trans. Veh. Technol.}, 
  title={{Expectation Propagation Aided Model Driven Learning for OTFS Signal Detection}}, 
  year={2023},
  volume={72},
  number={9},
  pages={12407-12412},
  doi={10.1109/TVT.2023.3268231}}

@ARTICLE{Molisch_HBFMassiveMIMO,
  author={Molisch, Andreas F. and Ratnam, Vishnu V. and Han, Shengqian and Li, Zheda and Nguyen, Sinh Le Hong and Li, Linsheng and Haneda, Katsuyuki},
  journal={IEEE Commun. Mag.}, 
  title={{Hybrid Beamforming for Massive MIMO: A Survey}}, 
  year={2017},
  volume={55},
  number={9},
  pages={134-141},
  doi={10.1109/MCOM.2017.1600400}}

@ARTICLE{Zhao_underwater_OFDM,
  author={Zhao, Hao and Yang, Cui and Xu, Yalu and Ji, Fei and Wen, Miaowen and Chen, Yankun},
  journal={IEEE Trans. Veh. Technol.}, 
  title={{Model-Driven Based Deep Unfolding Equalizer for Underwater Acoustic OFDM Communications}}, 
  year={2023},
  volume={72},
  number={5},
  pages={6056-6067},
  doi={10.1109/TVT.2022.3230143}}

@ARTICLE{Survey6_Monga_AlgoUnroll,
  author={Monga, Vishal and Li, Yuelong and Eldar, Yonina C.},
  journal={IEEE Signal Process. Mag.}, 
  title={{Algorithm Unrolling: Interpretable, Efficient Deep Learning for Signal and Image Processing}}, 
  year={2021},
  volume={38},
  number={2},
  pages={18-44},
  doi={10.1109/MSP.2020.3016905}}

@ARTICLE{Mashhadi_DL_pilot_chhEst,
  author={Mashhadi, Mahdi Boloursaz and Gündüz, Deniz},
  journal={IEEE Trans. Wireless Commun.}, 
  title={{Pruning the Pilots: Deep Learning-Based Pilot Design and Channel Estimation for MIMO-OFDM Systems}}, 
  year={2021},
  volume={20},
  number={10},
  pages={6315-6328},
  doi={10.1109/TWC.2021.3073309}}

@ARTICLE{Gao_ComNet,
  author={Gao, Xuanxuan and Jin, Shi and Wen, Chao-Kai and Li, Geoffrey Ye},
  journal={IEEE Commun. Lett.}, 
  title={{ComNet: Combination of Deep Learning and Expert Knowledge in OFDM Receivers}}, 
  year={2018},
  volume={22},
  number={12},
  pages={2627-2630},
  doi={10.1109/LCOMM.2018.2877965}}

@ARTICLE{Chen_HBF_RIS,
  author={Chen, Kuan-Ming and Chang, Hsin-Yuan and Chang, Ronald Y. and Chung, Wei-Ho},
  journal={IEEE Wireless Commun. Lett.}, 
  title={{Deep Unfolded Hybrid Beamforming in Reconfigurable Intelligent Surface Aided mmWave MIMO-OFDM Systems}}, 
  year={2024},
  volume={13},
  number={4},
  pages={1118-1122},
  doi={10.1109/LWC.2024.3362399}}

@ARTICLE{WMMSE_MO_original,
  author={Zhao, Xingyu and Lin, Tian and Zhu, Yu and Zhang, Jun},
  journal={IEEE Trans. Wireless Commun.}, 
  title={{Partially-Connected Hybrid Beamforming for Spectral Efficiency Maximization via a Weighted MMSE Equivalence}}, 
  year={2021},
  volume={20},
  number={12},
  pages={8218-8232},
  doi={10.1109/TWC.2021.3091524}}

@ARTICLE{Wu_RSMA_THz,
  author={Wu, Minghui and Gao, Zhen and Huang, Yang and Xiao, Zhenyu and Ng, Derrick Wing Kwan and Zhang, Zhaoyang},
  journal={IEEE J. Sel. Areas Commun. }, 
  title={{Deep Learning-Based Rate-Splitting Multiple Access for Reconfigurable Intelligent Surface-Aided Tera-Hertz Massive MIMO}}, 
  year={2023},
  volume={41},
  number={5},
  pages={1431-1451},
  doi={10.1109/JSAC.2023.3240781}}

@ARTICLE{RSMA,
  author={Y.  Mao  and B.  Clerckx and V.  O. K. Li},
  journal={EURASIP J Wirel Commun Netw}, 
  title={{Rate-splitting Multiple Access for Downlink Communication Systems: Bridging, Generalizing, and Outperforming SDMA and NOMA}}, 
  year={2018},
  volume={1},
  pages={133},
  doi={https://doi.org/10.48550/arXiv.1710.11018}}

@ARTICLE{THz_HBF,
  author={Dai, Linglong and Tan, Jingbo and Chen, Zhi and Poor, H. Vincent},
  journal={IEEE Trans. Wireless Commun.}, 
  title={{Delay-Phase Precoding for Wideband THz Massive MIMO}}, 
  year={2022},
  volume={21},
  number={9},
  pages={7271-7286},
  doi={10.1109/TWC.2022.3157315}}

@INPROCEEDINGS{Turbo,
  author={Berrou, C. and Glavieux, A. and Thitimajshima, P.},
  booktitle={Proc. of ICC '93}, 
  title={{Near Shannon limit error-correcting coding and decoding: Turbo-codes. 1}}, 
  year={1993},
  volume={2},
  number={},
  pages={1064-1070 vol.2},
  doi={10.1109/ICC.1993.397441}}

@ARTICLE{Factor_graph_sum_product,
  author={Kschischang, F.R. and Frey, B.J. and Loeliger, H.-A.},
  journal={IEEE Trans. Inf. Theory}, 
  title={{Factor graphs and the sum-product algorithm}}, 
  year={2001},
  volume={47},
  number={2},
  pages={498-519},
  doi={10.1109/18.910572}}

@ARTICLE{Polar_5G_control,
  author={Kaykac Egilmez, Zeynep B. and Xiang, Luping and Maunder, Robert G. and Hanzo, Lajos},
  journal={IEEE Commun. Surveys Tuts.}, 
  title={{The Development, Operation and Performance of the 5G Polar Codes}}, 
  year={2020},
  volume={22},
  number={1},
  pages={96-122},
  doi={10.1109/COMST.2019.2960746}}

@ARTICLE{List_decoding,
  author={Tal, Ido and Vardy, Alexander},
  journal={IEEE Trans. Inf. Theory}, 
  title={{List Decoding of Polar Codes}}, 
  year={2015},
  volume={61},
  number={5},
  pages={2213-2226},
  doi={10.1109/TIT.2015.2410251}}

@phdthesis{gallager_thesis,
  author       = {Robert G. Gallager},
  title        = {{Low-Density Parity-Check Codes}},
  school       = {MIT},
  year         = {1962},
  address      = {Cambridge, MA},
  type         = {Ph.D. Thesis},
  url          = {https://dspace.mit.edu/handle/1721.1/13799}
  
}

@Article{sph,
  author   = {McCormick, S. Thomas and Peis, Britta and Scheidweiler, Robert and Vallentin, Frank},
  journal  = {SIAM Journal on Discrete Mathematics},
  title    = {{A Polynomial Time Algorithm for Solving the Closest Vector Problem in Zonotopal Lattices}},
  year     = {2021},
  number   = {4},
  pages    = {2345-2356},
  volume   = {35},
    doi      = {10.1137/20M1382258},  
}

@ARTICLE{Gallager_LDPC,
  author={Gallager, R.},
  journal={IRE trans. inf. theory }, 
  title={{Low-density parity-check codes}}, 
  year={1962},
  volume={8},
  number={1},
  pages={21-28},
  doi={10.1109/TIT.1962.1057683}}

@INPROCEEDINGS{MacKay,
  author={MacKay, D.J.C.},
  booktitle={IEEE Int. Symp. Inf. Theory  Proc.}, 
  title={{Good Error-Correcting Codes Based on Very Sparse Matrices}}, 
  year={1997},
  volume={42},
  number={2},
  pages={399--431},
  doi={10.1109/ISIT.1997.613028}}

@ARTICLE{Wei_DUADMM_Linear,
  author={Wei, Yi and Zhao, Ming-Min and Zhao, Min-Jian and Lei, Ming},
  journal={IEEE Commun. Lett.}, 
  title={{ADMM-Based Decoder for Binary Linear Codes Aided by Deep Learning}}, 
  year={2020},
  volume={24},
  number={5},
  pages={1028-1032},
  doi={10.1109/LCOMM.2020.2974199}}

@ARTICLE{He_DU_Turbo,
  author={He, Yunfeng and Zhang, Jing and Jin, Shi and Wen, Chao-Kai and Li, Geoffrey Ye},
  journal={IEEE Trans. Commun.}, 
  title={{Model-Driven DNN Decoder for Turbo Codes: Design, Simulation, and Experimental Results}}, 
  year={2020},
  volume={68},
  number={10},
  pages={6127-6140},
  doi={10.1109/TCOMM.2020.3010964}}

@book{6G_guide,
    author = {Wei Jiang and Fa Long Luo},
    isbn = {9781119847496 },
    title = {{6G Key Technologies: A Comprehensive Guide}} ,
    publisher = {Wiley-IEEE Press } ,
    year = {2023} 
}

@ARTICLE{PD_CD_NOMA,
  author={Islam, S. M. Riazul and Avazov, Nurilla and Dobre, Octavia A. and Kwak, Kyung-sup},
  journal={IEEE Commun. Surveys Tuts.}, 
  title={{Power-Domain Non-Orthogonal Multiple Access (NOMA) in 5G Systems: Potentials and Challenges}}, 
  year={2017},
  volume={19},
  number={2},
  pages={721-742},
  doi={10.1109/COMST.2016.2621116}}

@ARTICLE{MA_6G,
  author={Clerckx, Bruno and Mao, Yijie and Yang, Zhaohui and Chen, Mingzhe and Alkhateeb, Ahmed and Liu, Liang and Qiu, Min and Yuan, Jinhong and Wong, Vincent W. S. and Montojo, Juan},
  journal={Proc. IEEE }, 
  title={{Multiple Access Techniques for Intelligent and Multifunctional 6G: Tutorial, Survey, and Outlook}}, 
  year={2024},
  volume={112},
  number={7},
  pages={832-879},
  doi={10.1109/JPROC.2024.3409428}}

@ARTICLE{mmWaveBF_challenge,
  author={Kebede, Tewelgn and Wondie, Yihenew and Steinbrunn, Johannes and Kassa, Hailu Belay and Kornegay, Kevin T.},
  journal={IEEE Access}, 
  title={{Precoding and Beamforming Techniques in mmWave-Massive MIMO: Performance Assessment}}, 
  year={2022},
  volume={10},
  number={},
  pages={16365-16387},
  doi={10.1109/ACCESS.2022.3149301}}

@ARTICLE{Tanner,
  author={Tanner, R.},
  journal={IEEE Trans. Inf. Theory}, 
  title={{A recursive approach to low complexity codes}}, 
  year={1981},
  volume={27},
  number={5},
  pages={533-547},
  doi={10.1109/TIT.1981.1056404}}

@book{LMIMO,
    author = {Chockalingam, A. and Rajan, B. Sundar},
    isbn = {1107026652},
    title = {{Large MIMO Systems}},
    publisher = {Cambridge University Press } ,
    year = {2014} 
}

@ARTICLE{SD1,
  author={B. Hassibi and H. Vikalo},
  journal={IEEE Trans. Signal Processing}, 
  title={{On the sphere decoding algorithm I. Expected complexity}}, 
  year={2005},
  volume={53},
  number={8},
  pages={2806-2818},
  doi={10.1109/TSP.2005.850352}}

@ARTICLE{SD2,
  author={B. Hassibi and H. Vikalo},
  journal={IEEE Trans. Signal Processing}, 
  title={{On the sphere decoding algorithm II. Generalizations,
second-order statistics, and applications to communications}}, 
  year={2005},
  volume={53},
  number={8},
  pages={2819-2834},
  doi={10.1109/TSP.2005.850350}}

@ARTICLE{Mandloi_ISD,
  author={Mandloi, Manish and Bhatia, Vimal},
  journal={IEEE Commun. Lett.}, 
  title={{Low-Complexity Near-Optimal Iterative Sequential Detection for Uplink Massive MIMO Systems}}, 
  year={2017},
  volume={21},
  number={3},
  pages={568-571},
  doi={10.1109/LCOMM.2016.2637366}}

@ARTICLE{Wei_ConjugateDescent,
  author={Wei, Yi and Zhao, Ming-Min and Hong, Mingyi and Zhao, Min-Jian and Lei, Ming},
  journal={IEEE Trans. Signal Process.}, 
  title={{Learned Conjugate Gradient Descent Network for Massive MIMO Detection}}, 
  year={2020},
  volume={68},
  number={},
  pages={6336-6349},
  doi={10.1109/TSP.2020.3035832}}

@ARTICLE{PolarCodes,
  author={Arikan, Erdal},
  journal={IEEE Trans. Inf. Theory}, 
  title={{Channel Polarization: A Method for Constructing Capacity-Achieving Codes for Symmetric Binary-Input Memoryless Channels}}, 
  year={2009},
  volume={55},
  number={7},
  pages={3051-3073},
  doi={10.1109/TIT.2009.2021379}}

@ARTICLE{JCDD_Sun,
  author={Sun, Yi and Shen, Hong and Li, Bingqing and Xu, Wei and Zhu, Pengcheng and Hu, Nan and Zhao, Chunming},
  journal={IEEE Trans. Wireless Commun.}, 
  title={{Trainable Joint Channel Estimation, Detection and Decoding for MIMO URLLC Systems}}, 
  year={2024},
  volume={},
  number={},
  pages={1-1},
  doi={10.1109/TWC.2024.3388603}}

@ARTICLE{Sparse_CDMA,
  author={Takabe, Satoshi and Yamauchi, Yuki and Wadayama, Tadashi},
  journal={IEEE Access}, 
  title={{Deep-Unfolded Sparse CDMA: Multiuser Detector and Sparse Signature Design}}, 
  year={2021},
  volume={9},
  number={},
  pages={40027-40038},
  doi={10.1109/ACCESS.2021.3064558}}

@INPROCEEDINGS{LDS,
  author={Jinho Choi},
  booktitle={Eighth IEEE ISSSTA}, 
  title={{Low density spreading for multicarrier systems}}, 
  year={2004},
  volume={},
  number={},
  pages={575-578},
  doi={10.1109/ISSSTA.2004.1371765}}

@INPROCEEDINGS{SCDMA,
  author={Yoshida, Mika and Tanaka, Toshiyuki},
  booktitle={2006 IEEE ISIT}, 
  title={{Analysis of Sparsely-Spread CDMA via Statistical Mechanics}}, 
  year={2006},
  volume={},
  number={},
  pages={2378-2382},
  doi={10.1109/ISIT.2006.262014}}

@ARTICLE{JADC_Qiang,
  author={Qiang, Yiyang and Shao, Xiaodan and Chen, Xiaoming},
  journal={IEEE Commun. Lett.}, 
  title={{A Model-Driven Deep Learning Algorithm for Joint Activity Detection and Channel Estimation}}, 
  year={2020},
  volume={24},
  number={11},
  pages={2508-2512},
  doi={10.1109/LCOMM.2020.3011571}}

@INPROCEEDINGS{MP_donoho,
  author={Donoho, David L. and Maleki, Arian and Montanari, Andrea},
  booktitle={2010 IEEE ITW}, 
  title={{Message passing algorithms for compressed sensing: I. motivation and construction}}, 
  year={2010},
  volume={},
  number={},
  pages={1-5},
  doi={10.1109/ITWKSPS.2010.5503193}}

@ARTICLE{Mao_est_and_det,
  author={Mao, Zhendong and Liu, Xiqing and Peng, Mugen and Chen, Zijie and Wei, Guiming},
  journal={IEEE Internet Things J.}, 
  title={{Joint Channel Estimation and Active-User Detection for Massive Access in Internet of Things—A Deep Learning Approach}}, 
  year={2022},
  volume={9},
  number={4},
  pages={2870-2881},
  doi={10.1109/JIOT.2021.3097133}}

@ARTICLE{IRS_beamform,
  author={Liu, Yanzhen and Hu, Qiyu and Cai, Yunlong and Yu, Guanding and Li, Geoffrey Ye},
  journal={IEEE Trans. Wireless Commun.}, 
  title={{Deep-Unfolding Beamforming for Intelligent Reflecting Surface Assisted Full-Duplex Systems}}, 
  year={2022},
  volume={21},
  number={7},
  pages={4784-4800},
  doi={10.1109/TWC.2021.3133296}}

@ARTICLE{HBF_survey,
  author={Ahmed, Irfan and Khammari, Hedi and Shahid, Adnan and Musa, Ahmed and Kim, Kwang Soon and De Poorter, Eli and Moerman, Ingrid},
  journal={IEEE Commun. Surveys Tuts. }, 
  title={{A Survey on Hybrid Beamforming Techniques in 5G: Architecture and System Model Perspectives}}, 
  year={2018},
  volume={20},
  number={4},
  pages={3060-3097},
  doi={10.1109/COMST.2018.2843719}}

@book{Goodfellow,
  title={{Deep Learning}},
  author={Ian J. Goodfellow and Yoshua Bengio and Aaron Courville},
  year={2016},
  publisher={MIT Press}
}

@ARTICLE{6G,
  author={Jiang, Wei and Han, Bin and Habibi, Mohammad Asif and Schotten, Hans Dieter},
  journal={IEEE Open J. Commun. Soc.}, 
  title={{The Road Towards 6G: A Comprehensive Survey}}, 
  year={2021},
  volume={2},
  number={},
  pages={334-366},
  doi={10.1109/OJCOMS.2021.3057679}}

@ARTICLE{5G,
  author={Agiwal, Mamta and Roy, Abhishek and Saxena, Navrati},
  journal={IEEE Commun. Surveys Tuts. }, 
  title={{Next Generation 5G Wireless Networks: A Comprehensive Survey}}, 
  year={2016},
  volume={18},
  number={3},
  pages={1617-1655},
  doi={10.1109/COMST.2016.2532458}}

@ARTICLE{BF_DNN,
  author={Huang, Hao and Peng, Yang and Yang, Jie and Xia, Wenchao and Gui, Guan},
  journal={IEEE Trans. Veh. Technol.}, 
  title={{Fast Beamforming Design via Deep Learning}}, 
  year={2020},
  volume={69},
  number={1},
  pages={1065-1069},
  doi={10.1109/TVT.2019.2949122}}

@ARTICLE{Chnnl_DNN,
  author={Bai, Qinbo and Wang, Jintao and Zhang, Yue and Song, Jian},
  journal={IEEE Trans. Cogn. Commun. Netw.}, 
  title={{Deep Learning-Based Channel Estimation Algorithm Over Time Selective Fading Channels}}, 
  year={2020},
  volume={6},
  number={1},
  pages={125-134},
  doi={10.1109/TCCN.2019.2943455}}

@ARTICLE{Survey2_Phy,
  author={He, Hengtao and Jin, Shi and Wen, Chao-Kai and Gao, Feifei and Li, Geoffrey Ye and Xu, Zongben},
  journal={IEEE Wirel. Commun.}, 
  title={{Model-Driven Deep Learning for Physical Layer Communications}}, 
  year={2019},
  volume={26},
  number={5},
  pages={77-83},
  doi={10.1109/MWC.2019.1800447}}

@ARTICLE{Survey7_Redefine,
  author={Jagannath, Anu and Jagannath, Jithin and Melodia, Tommaso},
  journal={IEEE Trans. Artif. Intell.}, 
  title={{Redefining Wireless Communication for 6G: Signal Processing Meets Deep Learning With Deep Unfolding}}, 
  year={2021},
  volume={2},
  number={6},
  pages={528-536},
  doi={10.1109/TAI.2021.3108129}}

@article{Survey1_DL_Phy_layer,
author = {Wang, Tianqi and Wen, Chao-Kai and Wang, Hanqing and Gao, Feifei and Jiang, Tao and Jin, Shi},
year = {2017},
month = {10},
pages = {92-111},
title = {{Deep Learning for Wireless Physical Layer: Opportunities and Challenges}},
volume = {14},
journal = {China Communications},
doi = {10.1109/CC.2017.8233654}
}

@INPROCEEDINGS{Survey4_DU,
  author={Balatsoukas-Stimming, Alexios and Studer, Christoph},
  booktitle={2019 IEEE SiPS}, 
  title={{Deep Unfolding for Communications Systems: A Survey and Some New Directions}}, 
  year={2019},
  volume={},
  number={},
  pages={266-271},
  doi={10.1109/SiPS47522.2019.9020494}}

@ARTICLE{Survey11_ML_Optim,
  author={Shi, Yandong and Lian, Lixiang and Shi, Yuanming and Wang, Zixin and Zhou, Yong and Fu, Liqun and Bai, Lin and Zhang, Jun and Zhang, Wei},
  journal={IEEE Commun. Surveys Tuts.}, 
  title={{Machine Learning for Large-Scale Optimization in 6G Wireless Networks}}, 
  year={2023},
  volume={25},
  number={4},
  pages={2088-2132},
  doi={10.1109/COMST.2023.3300664}}

@inproceedings{Gregor,
    author = {Gregor, Karol and LeCun, Yann},
    title = {{Learning fast approximations of sparse coding}},
    year = {2010},
    isbn = {9781605589077},
    publisher = {Omnipress},
    address = {Madison, WI, USA},
    booktitle = {27th ICML},
    pages = {399–406},
    numpages = {8},
    series = {ICML'10}
}

@ARTICLE{MIMO_det_survey,
  author={Yang, Shaoshi and Hanzo, Lajos},
  journal={IEEE Commun. Surveys Tuts.}, 
  title={{Fifty Years of MIMO Detection: The Road to Large-Scale MIMOs}}, 
  year={2015},
  volume={17},
  number={4},
  pages={1941-1988},
  doi={10.1109/COMST.2015.2475242}}

@ARTICLE{SP_DNN,
  author={Nguyen, Nhan Thanh and Lee, Kyungchun and DaiIEEE, Huaiyu},
  journal={IEEE Trans. Wireless Commun.}, 
  title={{Application of Deep Learning to Sphere Decoding for Large MIMO Systems}}, 
  year={2021},
  volume={20},
  number={10},
  pages={6787-6803},
  doi={10.1109/TWC.2021.3076527}}

@article{Moment_PGD,
    title = {On the momentum term in gradient descent learning algorithms},
    journal = {Neural Networks},
    volume = {12},
    number = {1},
    pages = {145-151},
    year = {1999},
    issn = {0893-6080},
    doi = {https://doi.org/10.1016/S0893-6080(98)00116-6},
    author = {Ning Qian},
}

@ARTICLE{MomentNet_2024,
  author={Yun, Sangbu and Moon, Seungsik and Jeon, Yo-Seb and Lee, Youngjoo},
  journal={IEEE Wireless Commun. Lett.}, 
  title={{Intelligent MIMO Detection With Momentum-Induced Unfolded Layers}}, 
  year={2024},
  volume={13},
  number={3},
  pages={879-883},
  doi={10.1109/LWC.2023.3348933}}

@ARTICLE{HS_2024,
  author={Takabe, Satoshi and Abe, Takashi},
  journal={IEEE Wireless Commun. Lett.}, 
  title={{Hubbard–Stratonovich Detector for Simple Trainable MIMO Signal Detection}}, 
  year={2024},
  volume={13},
  number={3},
  pages={701-705},
  doi={10.1109/LWC.2023.3340172}}

@ARTICLE{TPG_2021,
  author={Takabe, Satoshi and Imanishi, Masayuki and Wadayama, Tadashi and Hayakawa, Ryo and Hayashi, Kazunori},
  journal={IEEE Access}, 
  title={{Trainable Projected Gradient Detector for Massive Overloaded MIMO Channels: Data-Driven Tuning Approach}}, 
  year={2019},
  volume={7},
  number={},
  pages={93326-93338},
  doi={10.1109/ACCESS.2019.2927997}}

@book{Preconditioner, 
   
    title={{Matrix Preconditioning Techniques and Applications}}, 
    publisher={Cambridge University Press}, 
    author={Chen, Ke}, 
    year={2005}, 
     
}

@ARTICLE{CG_2024,
  author={Olutayo, Toluwaleke and Champagne, Benoit},
  journal={ IEEE Open J. Veh. Technol.}, 
  title={{Dynamic Conjugate Gradient Unfolding for Symbol Detection in Time-Varying Massive MIMO}}, 
  year={2024},
  volume={5},
  number={},
  pages={792-806},
  doi={10.1109/OJVT.2024.3410834}}

@ARTICLE{EM_2024,
  author={Shao, Mingjie and Ma, Wing-Kin and Liu, Junbin and Huang, Zihao},
  journal={IEEE Trans. Signal Process.}, 
  title={{Accelerated and Deep Expectation Maximization for One-Bit MIMO-OFDM Detection}}, 
  year={2024},
  volume={72},
  number={},
  pages={1094-1113},
  doi={10.1109/TSP.2024.3359083}}

@ARTICLE{SO_2024,
  author={Ullah, Arif and Choi, Wooyeol and Berhane, Teklu Merhawit and Sambo, Yusuf and Imran, Muhammad Ali},
  journal={IEEE Trans. Veh. Technol.}, 
  title={{Soft-Output Deep LAS Detection for Coded MIMO Systems: A Learning-Aided LLR Approximation}}, 
  year={2024},
  volume={},
  number={},
  pages={1-15},
  doi={10.1109/TVT.2024.3391614}}

@ARTICLE{PGD_2024,
  author={He, Lanxin and Wang, Zheng and Yang, Shaoshi and Liu, Tao and Huang, Yongming},
  journal={IEEE Trans. Wireless Commun.}, 
  title={{Generalizing Projected Gradient Descent for Deep-Learning-Aided Massive MIMO Detection}}, 
  year={2024},
  volume={23},
  number={3},
  pages={1827-1839},
  doi={10.1109/TWC.2023.3292124}}

@ARTICLE{MP_GNN_2024,
  author={He, Hengtao and Yu, Xianghao and Zhang, Jun and Song, Shenghui and Letaief, Khaled B.},
  journal={IEEE Trans. Wireless Commun.}, 
  title={{Message Passing Meets Graph Neural Networks: A New Paradigm for Massive MIMO Systems}}, 
  year={2024},
  volume={23},
  number={5},
  pages={4709-4723},
  doi={10.1109/TWC.2023.3321667}}

@ARTICLE{AL_2023,
  author={Zilberstein, Nicolas and Dick, Chris and Doost-Mohammady, Rahman and Sabharwal, Ashutosh and Segarra, Santiago},
  journal={IEEE Trans. Wireless Commun.}, 
  title={{Annealed Langevin Dynamics for Massive MIMO Detection}}, 
  year={2023},
  volume={22},
  number={6},
  pages={3762-3776},
  doi={10.1109/TWC.2022.3221057}}

@ARTICLE{PINN_1,
  author={Han, Qiaojian and Zhang, Haixia and Li, Yueheng and Yuan, Dongfeng and Ma, Xiaoyan},
  journal={IEEE Wirel. Commun.}, 
  title={{Physics Informed Digital Twin for RIS-Assisted Wireless Communication System}}, 
  year={2025},
  volume={32},
  number={3},
  pages={106-112},
  doi={10.1109/MWC.003.2400418}}

@ARTICLE{PINN_2,
  author={Shi, Jianyang and Liu, Yu and Luo, Zhiteng and Li, Ziwei and Shen, Chao and Zhang, Junwen and Wang, Guangxu and Chi, Nan},
  journal={J. Light. Technol.}, 
  title={{Simplified Neural Network With Physics-Informed Module in MIMO Visible Light Communication Systems}}, 
  year={2024},
  volume={42},
  number={1},
  pages={57-68},
  doi={10.1109/JLT.2023.3305196}}

@ARTICLE{DTL_2023,
  author={Wang, Xiaoming and Zhang, Dongcai and Chen, Bingcen and Liu, Ting and Xin, Yuanxue and Xu, Youyun},
  journal={IEEE Trans. Veh. Technol.}, 
  title={{Deep Transfer Learning for Model-Driven Signal Detection in MIMO-NOMA Systems}}, 
  year={2023},
  volume={72},
  number={10},
  pages={13039-13054},
  doi={10.1109/TVT.2023.3274561}}

@ARTICLE{Robust_2023,
  author={Sun, Yi and Shen, Hong and Xu, Wei and Hu, Nan and Zhao, Chunming},
  journal={IEEE Trans. Commun.}, 
  title={{Robust MIMO Detection With Imperfect CSI: A Neural Network Solution}}, 
  year={2023},
  volume={71},
  number={10},
  pages={5877-5892},
  doi={10.1109/TCOMM.2023.3299974}}

@ARTICLE{DU_2023,
  author={Berra, Salah and Chakraborty, Sourav and Dinis, Rui and Shahabuddin, Shahriar},
  journal={IEEE Access}, 
  title={{Deep Unfolding of Chebyshev Accelerated Iterative Method for Massive MIMO Detection}}, 
  year={2023},
  volume={11},
  number={},
  pages={52555-52569},
  doi={10.1109/ACCESS.2023.3279350}}

@ARTICLE{Adaptive_2022,
  author={Zhang, Jing and Wen, Chao-Kai and Jin, Shi},
  journal={IEEE J. Sel. Areas Commun. }, 
  title={{Adaptive MIMO Detector Based on Hypernetwork: Design, Simulation, and Experimental Test}}, 
  year={2022},
  volume={40},
  number={1},
  pages={65-81},
  doi={10.1109/JSAC.2021.3126064}}

@ARTICLE{DU_det_2022,
  author={Zhou, Xingyu and Zhang, Jing and Syu, Chen-Wei and Wen, Chao-Kai and Zhang, Jun and Jin, Shi},
  journal={IEEE Trans. Commun.}, 
  title={{Model-Driven Deep Learning-Based MIMO-OFDM Detector: Design, Simulation, and Experimental Results}}, 
  year={2022},
  volume={70},
  number={8},
  pages={5193-5207},
  doi={10.1109/TCOMM.2022.3186404}}

@BOOK{linearizedADMM,
  author={Parikh, Neal and Boyd, Stephen},
  title={{Proximal Algorithms}},
  year={2014},
  publisher={Now Foundations and Trends},
  volume={},
  number={},
  pages={},
  doi={10.1561/2400000003}}

@ARTICLE{DU_joint_2022,
  author={Johnston, Jeremy and Wang, Xiaodong},
  journal={IEEE Trans. Wireless Commun.}, 
  title={{Model-Based Deep Learning for Joint Activity Detection and Channel Estimation in Massive and Sporadic Connectivity}}, 
  year={2022},
  volume={21},
  number={11},
  pages={9806-9817},
  doi={10.1109/TWC.2022.3179600}}

@ARTICLE{DU_sparse_2022,
  author={Datta, Arijit and Nema, Aneesh and Bhatia, Vimal},
  journal={IEEE Trans. Veh. Technol.}, 
  title={{Deep Unfolded Sparse Refinement Network Based Detection in Uplink Massive MIMO}}, 
  year={2022},
  volume={71},
  number={6},
  pages={6825-6830},
  doi={10.1109/TVT.2022.3166399}}

@ARTICLE{Meta_2021,
  author={Zhang, Jing and He, Yunfeng and Li, Yu-Wen and Wen, Chao-Kai and Jin, Shi},
  journal={IEEE Trans. Wireless Commun.}, 
  title={{Meta Learning-Based MIMO Detectors: Design, Simulation, and Experimental Test}}, 
  year={2021},
  volume={20},
  number={2},
  pages={1122-1137},
  doi={10.1109/TWC.2020.3030882}}

@ARTICLE{LoRD_2021,
  author={Khobahi, Shahin and Shlezinger, Nir and Soltanalian, Mojtaba and Eldar, Yonina C.},
  journal={IEEE Trans. Signal Process.}, 
  title={{LoRD-Net: Unfolded Deep Detection Network With Low-Resolution Receivers}}, 
  year={2021},
  volume={69},
  number={},
  pages={5651-5664},
  doi={10.1109/TSP.2021.3117503}}

@ARTICLE{DeepSIC,
  author={Shlezinger, Nir and Fu, Rong and Eldar, Yonina C.},
  journal={IEEE Trans. Wireless Commun.}, 
  title={{DeepSIC: Deep Soft Interference Cancellation for Multiuser MIMO Detection}}, 
  year={2021},
  volume={20},
  number={2},
  pages={1349-1362},
  doi={10.1109/TWC.2020.3032663}}

@ARTICLE{nML,
  author={Choi, Junil and Mo, Jianhua and Heath, Robert W.},
  journal={IEEE Trans. Commun.}, 
  title={{Near Maximum-Likelihood Detector and Channel Estimator for Uplink Multiuser Massive MIMO Systems With One-Bit ADCs}}, 
  year={2016},
  volume={64},
  number={5},
  pages={2005-2018},
  doi={10.1109/TCOMM.2016.2545666}}

@ARTICLE{ADC_imp,
  author={Andrews, Jeffrey G. and Buzzi, Stefano and Choi, Wan and Hanly, Stephen V. and Lozano, Angel and Soong, Anthony C. K. and Zhang, Jianzhong Charlie},
  journal={IEEE J. Sel. Areas Commun. }, 
  title={{What Will 5G Be?}}, 
  year={2014},
  volume={32},
  number={6},
  pages={1065-1082},
  doi={10.1109/JSAC.2014.2328098}}

@ARTICLE{AlgoParam_2021,
  author={Chen, Hang and Yao, Guoqiang and Hu, Jianhao},
  journal={IEEE Trans. Veh. Technol.}, 
  title={{Algorithm Parameters Selection Method With Deep Learning for EP MIMO Detector}}, 
  year={2021},
  volume={70},
  number={10},
  pages={10146-10156},
  doi={10.1109/TVT.2021.3103568}}

@ARTICLE{Approx_EP,
  author={Tan, Xiaosi and Ueng, Yeong-Luh and Zhang, Zaichen and You, Xiaohu and Zhang, Chuan},
  journal={IEEE Trans. Veh. Technol.}, 
  title={{A Low-Complexity Massive MIMO Detection Based on Approximate Expectation Propagation}}, 
  year={2019},
  volume={68},
  number={8},
  pages={7260-7272},
  doi={10.1109/TVT.2019.2924952}}

@ARTICLE{EP_2021,
  author={Ge, Yingmeng and Tan, Xiaosi and Ji, Zhenhao and Zhang, Zaichen and You, Xiaohu and Zhang, Chuan},
  journal={IEEE Wireless Commun. Lett.}, 
  title={{Improving Approximate Expectation Propagation Massive MIMO Detector With Deep Learning}}, 
  year={2021},
  volume={10},
  number={10},
  pages={2145-2149},
  doi={10.1109/LWC.2021.3095117}}

@ARTICLE{CMD_2021,
  author={Beck, Edgar and Bockelmann, Carsten and Dekorsy, Armin},
  journal={IEEE Trans. Commun.}, 
  title={{CMDNet: Learning a Probabilistic Relaxation of Discrete Variables for Soft Detection With Low Complexity}}, 
  year={2021},
  volume={69},
  number={12},
  pages={8214-8227},
  doi={10.1109/TCOMM.2021.3114682}}

@article{ADMM,
    url = {http://dx.doi.org/10.1561/2200000016},
    year = {2011},
    volume = {3},
    journal = {Foundations and Trends® in Machine Learning},
    title = {{Distributed Optimization and Statistical Learning via the Alternating Direction Method of Multipliers}},
    doi = {10.1561/2200000016},
    issn = {1935-8237},
    number = {1},
    pages = {1-122},
    author = {Stephen Boyd and Neal Parikh and Eric Chu and Borja Peleato and Jonathan Eckstein}
}

@ARTICLE{DLwithCSI_2019,
  author={Qing, Chaojin and Cai, Bin and Yang, Qingyao and Wang, Jiafan and Huang, Chuan},
  journal={IEEE Access}, 
  title={{Deep Learning for CSI Feedback Based on Superimposed Coding}}, 
  year={2019},
  volume={7},
  number={},
  pages={93723-93733},
  doi={10.1109/ACCESS.2019.2928049}}

@ARTICLE{CSIrecoveryMIMO_ISTA_2020,
  author={Wang, Yiyun and Chen, Xiaohui and Yin, Huarui and Wang, Weidong},
  journal={IEEE Commun. Lett.}, 
  title={{Learnable Sparse Transformation-Based Massive MIMO CSI Recovery Network}}, 
  year={2020},
  volume={24},
  number={7},
  pages={1468-1471},
  doi={10.1109/LCOMM.2020.2981448}}

@inproceedings{
liu2018alista,
title={{{ALISTA}: Analytic Weights Are As Good As Learned Weights in {LISTA}}},
author={Jialin Liu and Xiaohan Chen and Zhangyang Wang and Wotao Yin},
booktitle={International Conference on Learning Representations},
year={2019},
url={https://openreview.net/forum?id=B1lnzn0ctQ},
}

@ARTICLE{LORA_2024,
  author={Hu, Zhengyang and Liu, Guanzhang and Xie, Qi and Xue, Jiang and Meng, Deyu and Gündüz, Deniz},
  journal={IEEE Trans. Wireless Commun.}, 
  title={{A Learnable Optimization and Regularization Approach to Massive MIMO CSI Feedback}}, 
  year={2024},
  volume={23},
  number={1},
  pages={104-116},
  doi={10.1109/TWC.2023.3275990}}

@ARTICLE{OTFS_Est_2024,
  author={Yue, Yang and Shi, Jia and Li, Zan and Hu, Junfan and Tie, Zhuangzhuang},
  journal={IEEE Commun. Lett.}, 
  title={{Model-Driven Deep Learning Assisted Detector for OTFS With Channel Estimation Error}}, 
  year={2024},
  volume={28},
  number={4},
  pages={842-846},
  doi={10.1109/LCOMM.2024.3362970}}

@ARTICLE{mMTC_2024,
  author={Ma, Zhe and Wu, Wen and Gao, Feifei and Shen, Xuemin},
  journal={IEEE Trans. Wireless Commun.}, 
  title={{Model-Driven Deep Learning for Non-Coherent Massive Machine-Type Communications}}, 
  year={2024},
  volume={23},
  number={3},
  pages={2197-2211},
  doi={10.1109/TWC.2023.3296218}}

@ARTICLE{OTFS_chnest,
  author={Wang, Qian and Chen, Xingke and Tao, Qin and Qian, Li Ping and Kam, Pooi-Yuen and Wu, Yuan},
  journal={IEEE Trans. Veh. Technol.}, 
  title={{Model-Driven Channel Estimation Network for Orthogonal Time-Frequency Space Systems}}, 
  year={2025},
  volume={},
  number={},
  pages={1-5},
  doi={10.1109/TVT.2025.3548639}}

@ARTICLE{Massive_access_2024,
  author={Bai, Yanna and Chen, Wei and Ai, Bo and Popovski, Petar},
  journal={IEEE Trans. Wireless Commun.}, 
  title={{Deep Learning for Asynchronous Massive Access With Data Frame Length Diversity}}, 
  year={2024},
  volume={23},
  number={6},
  pages={5529-5540},
  doi={10.1109/TWC.2023.3326952}}

@ARTICLE{LAMP_IRS_2024,
  author={Tsai, Wen-Chiao and Chen, Chi-Wei and Wu, An-Yeu},
  journal={IEEE Trans. Veh. Technol.}, 
  title={{Deep Unfolding-Based Channel Estimation for IRS-Aided mmWave Systems via Two-Stage LAMP Network with Row Compression}}, 
  year={2024},
  volume={},
  number={},
  pages={1-14},
  doi={10.1109/TVT.2024.3419845}}

@ARTICLE{BF_2024,
  author={Bilbao, Iñigo and Iradier, Eneko and Montalbán, Jon and Angueira, Pablo and Thanh Nguyen, Nhan and Juntti, Markku},
  journal={IEEE Open J. Commun. Soc.}, 
  title={{Deep Unfolding-Powered Analog Beamforming for In-Band Full-Duplex}}, 
  year={2024},
  volume={5},
  number={},
  pages={3753-3761},
  doi={10.1109/OJCOMS.2024.3417349}}

@ARTICLE{chnnlest_2024,
  author={Zhao, Yangliu and Teng, Yinglei and Liu, An and Wang, Xiaojuan and Lau, Vincent K. N.},
  journal={IEEE Trans. Wireless Commun.}, 
  title={{Joint UL/DL Dictionary Learning and Channel Estimation via Two-Timescale Optimization in Massive MIMO Systems}}, 
  year={2024},
  volume={23},
  number={3},
  pages={2369-2382},
  doi={10.1109/TWC.2023.3297995}}

@ARTICLE{MM_Chnnel_est,
  author={Liu, Fangqing and Shang, Xiaolei and Zhu, Heng},
  journal={IEEE Trans. Wireless Commun.}, 
  title={{Efficient Majorization-Minimization-Based Channel Estimation for One-Bit Massive MIMO Systems}}, 
  year={2021},
  volume={20},
  number={6},
  pages={3444-3457},
  doi={10.1109/TWC.2021.3050498}}

@article{BayesianCS,
    author = {Ji, Shihao and Xue, Ya and Carin, L.},
    title = {{Bayesian Compressive Sensing}},
    year = {2008},
    issue_date = {June 2008},
    publisher = {IEEE Press},
    volume = {56},
    number = {6},
    issn = {1053-587X},
    url = {https://doi.org/10.1109/TSP.2007.914345},
    doi = {10.1109/TSP.2007.914345},
    journal = {Trans. Sig. Proc.},
    month = {jun},
    pages = {2346–2356},
    numpages = {11}
    }

@ARTICLE{random_access_2024,
  author={Zou, Yinan and Zhou, Yong and Chen, Xu and Eldar, Yonina C.},
  journal={IEEE Trans. Wireless Commun.}, 
  title={{Proximal Gradient-Based Unfolding for Massive Random Access in IoT Networks}}, 
  year={2024},
  volume={},
  number={},
  pages={1-1},
  doi={10.1109/TWC.2024.3416375}}

@ARTICLE{WMMSE_2024,
  author={Wang, Kexuan and Liu, An},
  journal={IEEE Wireless Commun. Lett.}, 
  title={{Robust WMMSE-Based Precoder With Practice-Oriented Design for Massive MU-MIMO}}, 
  year={2024},
  volume={13},
  number={7},
  pages={1858-1862},
  doi={10.1109/LWC.2024.3393484}}

@ARTICLE{ChnnelEst_2023,
  author={He, Hengtao and Wang, Rui and Jin, Weijie and Jin, Shi and Wen, Chao-Kai and Li, Geoffrey Ye},
  journal={IEEE Trans. Wireless Commun.}, 
  title={{Beamspace Channel Estimation for Wideband Millimeter-Wave MIMO: A Model-Driven Unsupervised Learning Approach}}, 
  year={2023},
  volume={22},
  number={3},
  pages={1808-1822},
  doi={10.1109/TWC.2022.3206773}}

@ARTICLE{DL_Chnnel_est_2023,
  author={Gao, Jiabao and Zhong, Caijun and Li, Geoffrey Ye and Soriaga, Joseph B. and Behboodi, Arash},
  journal={IEEE Trans. Commun.}, 
  title={{Deep Learning-Based Channel Estimation for Wideband Hybrid MmWave Massive MIMO}}, 
  year={2023},
  volume={71},
  number={6},
  pages={3679-3693},
  doi={10.1109/TCOMM.2023.3258484}}

@ARTICLE{BF_Est_2023,
  author={Gao, Zhen and Liu, Shicong and Su, Yu and Li, Zhongxiang and Zheng, Dezhi},
  journal={IEEE J. Sel. Topics Signal Process. }, 
  title={{Hybrid Knowledge-Data Driven Channel Semantic Acquisition and Beamforming for Cell-Free Massive MIMO}}, 
  year={2023},
  volume={17},
  number={5},
  pages={964-979},
  doi={10.1109/JSTSP.2023.3299175}}

@ARTICLE{DeepADMM_2023,
  author={Fan, Wenzhe and Liu, Shengheng and Li, Chunguo and Huang, Yongming},
  journal={IEEE Wireless Commun. Lett.}, 
  title={{Fast Direct Localization for Millimeter Wave MIMO Systems via Deep ADMM Unfolding}}, 
  year={2023},
  volume={12},
  number={4},
  pages={748-752},
  doi={10.1109/LWC.2023.3243157}}

@INPROCEEDINGS{JADC_Cellfree_2024,
  author={Li, Tao and Jiang, Yanxiang and Huang, Yige and Zhu, Pengcheng and Zheng, Fu-Chun and Wang, Dongming},
  booktitle={2024 IEEE ICC Workshops}, 
  title={{Model-Based Deep Learning for Massive Access in mmWave Cell-Free Massive MIMO System}}, 
  year={2024},
  volume={},
  number={},
  pages={828-833},
  doi={10.1109/ICCWorkshops59551.2024.10615768}}

@ARTICLE{Recovery_2023,
  author={Chen, Jianqiao and Meng, Fanyang and Ma, Nan and Xu, Xiaodong and Zhang, Ping},
  journal={IEEE Trans. Veh. Technol.}, 
  title={{Model-Driven Deep Learning-Based Sparse Channel Representation and Recovery for Wideband mmWave Massive MIMO Systems}}, 
  year={2023},
  volume={72},
  number={12},
  pages={16788-16793},
  doi={10.1109/TVT.2023.3291693}}

@ARTICLE{ChannelAMP_2021,
  author={Ma, Xisuo and Gao, Zhen and Gao, Feifei and Di Renzo, Marco},
  journal={IEEE J. Sel. Areas Commun. }, 
  title={{Model-Driven Deep Learning Based Channel Estimation and Feedback for Millimeter-Wave Massive Hybrid MIMO Systems}}, 
  year={2021},
  volume={39},
  number={8},
  pages={2388-2406},
  doi={10.1109/JSAC.2021.3087269}}

@ARTICLE{IRScascade_2023,
  author={Awais, Muhammad and Khan, Mubasher Ahmed and Kim, Yun Hee},
  journal={IEEE Wireless Commun. Lett.}, 
  title={{Deep Denoising and Unfolding for IRS Cascaded Channel Estimation With Element-Grouping}}, 
  year={2023},
  volume={12},
  number={10},
  pages={1726-1730},
  doi={10.1109/LWC.2023.3290204}}

@ARTICLE{DU_CSI_FB_2023,
  author={Cao, Zheng and Guo, Jiajia and Wen, Chao-Kai and Jin, Shi},
  journal={IEEE Wireless Commun. Lett.}, 
  title={{Deep-Unfolding-Based Bit-Level CSI Feedback in Massive MIMO Systems}}, 
  year={2023},
  volume={12},
  number={2},
  pages={371-375},
  doi={10.1109/LWC.2022.3227315}}

@ARTICLE{GrantfreeCS_2023,
  author={Dang, Xiaobing and Xiang, Wei and Yuan, Lei and Yang, Yuan and Wang, Eric and Huang, Tao},
  journal={IEEE Trans. Intell. Transp. Syst. }, 
  title={{Deep Unfolding Scheme for Grant-Free Massive-Access Vehicular Networks}}, 
  year={2023},
  volume={24},
  number={12},
  pages={14443-14452},
  doi={10.1109/TITS.2023.3296452}}

@ARTICLE{DownlinkPrecoding_2023,
  author={Xu, Qian and Sun, Jianyong and Xu, Zongben},
  journal={IEEE Trans. Veh. Technol.}, 
  title={{Robust Frequency Selective Precoding for Downlink Massive MIMO in 5G Broadband System}}, 
  year={2023},
  volume={72},
  number={12},
  pages={15941-15952},
  doi={10.1109/TVT.2023.3293848}}

@ARTICLE{Tanner_decoding_2023,
  author={Liu, Jiaai and Wang, Xiaodong},
  journal={IEEE Trans. Wireless Commun.}, 
  title={{Tanner-Graph-Based Massive Multiple Access—Transmission and Decoding Schemes}}, 
  year={2023},
  volume={22},
  number={11},
  pages={8468-8482},
  doi={10.1109/TWC.2023.3263488}}

@ARTICLE{TrainableChnnlest_2023,
  author={Zheng, Peicong and Lyu, Xuantao and Gong, Yi},
  journal={IEEE Wireless Commun. Lett.}, 
  title={{Trainable Proximal Gradient Descent-Based Channel Estimation for mmWave Massive MIMO Systems}}, 
  year={2023},
  volume={12},
  number={10},
  pages={1781-1785},
  doi={10.1109/LWC.2023.3294184}}

@ARTICLE{LearnPrecoder_2022,
  author={Lin, Qi and Shen, Hong and Zhao, Chunming},
  journal={IEEE Wireless Commun. Lett.}, 
  title={{Learning Linear MMSE Precoder for Uplink Massive MIMO Systems With One-Bit ADCs}}, 
  year={2022},
  volume={11},
  number={10},
  pages={2235-2239},
  doi={10.1109/LWC.2022.3198286}}

@ARTICLE{BasispursuitChnnlEst_2022,
  author={Wu, Pengxia and Cheng, Julian},
  journal={IEEE Trans. Wireless Commun.}, 
  title={{Deep Unfolding Basis Pursuit: Improving Sparse Channel Reconstruction via Data-Driven Measurement Matrices}}, 
  year={2022},
  volume={21},
  number={10},
  pages={8090-8105},
  doi={10.1109/TWC.2022.3164091}}

@ARTICLE{RIS_chnnlEst_MU_2021,
  author={Wei, Li and Huang, Chongwen and Alexandropoulos, George C. and Yuen, Chau and Zhang, Zhaoyang and Debbah, Mérouane},
  journal={IEEE Trans. Commun.}, 
  title={{Channel Estimation for RIS-Empowered Multi-User MISO Wireless Communications}}, 
  year={2021},
  volume={69},
  number={6},
  pages={4144-4157},
  doi={10.1109/TCOMM.2021.3063236}}

@ARTICLE{IRS_QoS_2021,
  author={Zhao, Ming-Min and Liu, An and Wan, Yubo and Zhang, Rui},
  journal={IEEE Trans. Wireless Commun.}, 
  title={{Two-Timescale Beamforming Optimization for Intelligent Reflecting Surface Aided Multiuser Communication With QoS Constraints}}, 
  year={2021},
  volume={20},
  number={9},
  pages={6179-6194},
  doi={10.1109/TWC.2021.3072382}}

@ARTICLE{Inverse_chnnlest_2021,
  author={Chen, Wei and Zhang, Bowen and Jin, Shi and Ai, Bo and Zhong, Zhangdui},
  journal={IEEE J. Sel. Areas Commun. }, 
  title={{Solving Sparse Linear Inverse Problems in Communication Systems: A Deep Learning Approach With Adaptive Depth}}, 
  year={2021},
  volume={39},
  number={1},
  pages={4-17},
  doi={10.1109/JSAC.2020.3036959}}

@ARTICLE{HBF_GAN_2021,
  author={Balevi, Eren and Andrews, Jeffrey G.},
  journal={IEEE Trans. Wireless Commun.}, 
  title={{Unfolded Hybrid Beamforming With GAN Compressed Ultra-Low Feedback Overhead}}, 
  year={2021},
  volume={20},
  number={12},
  pages={8381-8392},
  doi={10.1109/TWC.2021.3092350}}

@ARTICLE{FiniteAlphabet_2020,
  author={He, Hengtao and Zhang, Mengjiao and Jin, Shi and Wen, Chao-Kai and Li, Geoffrey Ye},
  journal={IEEE Commun. Lett.}, 
  title={{Model-Driven Deep Learning for Massive MU-MIMO With Finite-Alphabet Precoding}}, 
  year={2020},
  volume={24},
  number={10},
  pages={2216-2220},
  doi={10.1109/LCOMM.2020.3002082}}

@ARTICLE{EnvelopePrecoding_2020,
  author={He, Yunfeng and He, Hengtao and Wen, Chao-Kai and Jin, Shi},
  journal={IEEE Wireless Commun. Lett.}, 
  title={{Model-Driven Deep Learning for Massive Multiuser MIMO Constant Envelope Precoding}}, 
  year={2020},
  volume={9},
  number={11},
  pages={1835-1839},
  doi={10.1109/LWC.2020.3005027}}

@ARTICLE{Chnnel_det_OFDM_2019,
  author={Zhang, Jing and Wen, Chao-Kai and Jin, Shi and Li, Geoffrey Ye},
  journal={IEEE Access}, 
  title={{Artificial Intelligence-Aided Receiver for a CP-Free OFDM System: Design, Simulation, and Experimental Test}}, 
  year={2019},
  volume={7},
  number={},
  pages={58901-58914},
  doi={10.1109/ACCESS.2019.2914928}}

@ARTICLE{Graph_BF_2024,
  author={Chowdhury, Arindam and Verma, Gunjan and Swami, Ananthram and Segarra, Santiago},
  journal={IEEE Trans. Wireless Commun.}, 
  title={{Deep Graph Unfolding for Beamforming in MU-MIMO Interference Networks}}, 
  year={2024},
  volume={23},
  number={5},
  pages={4889-4903},
  doi={10.1109/TWC.2023.3323207}}

@ARTICLE{SymbioticIRS_BF_2024,
  author={He, Xiuli and Xu, Hongbo and Wang, Ji and Xie, Wenwu and Li, Xingwang and Nallanathan, Arumugam},
  journal={IEEE Trans. Veh. Technol.}, 
  title={{Joint Active and Passive Beamforming in RIS-Assisted Covert Symbiotic Radio Based on Deep Unfolding}}, 
  year={2024},
  volume={},
  number={},
  pages={1-6},
  doi={10.1109/TVT.2024.3393724}}

@ARTICLE{FractionalIRS_2024,
  author={Xia, Wenchao and Jiang, Yajing and Zhao, Ben and Zhao, Haitao and Zhu, Hongbo},
  journal={IEEE Wireless Commun. Lett.}, 
  title={{Deep Unfolded Fractional Programming-Based Beamforming in RIS-Aided MISO Systems}}, 
  year={2024},
  volume={13},
  number={2},
  pages={515-519},
  doi={10.1109/LWC.2023.3334031}}

@ARTICLE{AO_BCD_2020,
  author={Guo, Huayan and Liang, Ying-Chang and Chen, Jie and Larsson, Erik G.},
  journal={IEEE Trans. Wirel. Commun.}, 
  title={{Weighted Sum-Rate Maximization for Reconfigurable Intelligent Surface Aided Wireless Networks}}, 
  year={2020},
  volume={19},
  number={5},
  pages={3064-3076}
  }

@ARTICLE{IRS_ISAC_2024,
  author={Liu, Wanxian and Xu, Hongbo and He, Xiuli and Ye, Yuchen and Zhou, Aizhi},
  journal={IEEE Access}, 
  title={{Bi-Level Deep Unfolding Based Robust Beamforming Design for IRS-Assisted ISAC System}}, 
  year={2024},
  volume={12},
  number={},
  pages={76663-76672},
  doi={10.1109/ACCESS.2024.3406527}}

@ARTICLE{RS_BNN_2024,
  author={Wang, Yiwen and Mao, Yijie and Ji, Sijie},
  journal={IEEE Trans. Veh. Technol.}, 
  title={{RS-BNN: A Deep Learning Framework for the Optimal Beamforming Design of Rate-Splitting Multiple Access}}, 
  year={2024},
  volume={},
  number={},
  pages={1-6},
  doi={10.1109/TVT.2024.3423002}}

@ARTICLE{BF_Het_2023,
  author={Zhu, Minghe and Chang, Tsung-Hui and Hong, Mingyi},
  journal={IEEE Trans. Wireless Commun.}, 
  title={{Learning to Beamform in Heterogeneous Massive MIMO Networks}}, 
  year={2023},
  volume={22},
  number={7},
  pages={4901-4915},
  doi={10.1109/TWC.2022.3230662}}

@ARTICLE{Fast_precoding_2023,
  author={Xu, Jing and Kang, Chaohui and Xue, Jiang and Zhang, Yizhai},
  journal={IEEE Trans. Cogn. Commun. Netw.}, 
  title={{A Fast Deep Unfolding Learning Framework for Robust MU-MIMO Downlink Precoding}}, 
  year={2023},
  volume={9},
  number={2},
  pages={359-372},
  doi={10.1109/TCCN.2023.3235763}}

@ARTICLE{Symbollevel_Precoding_2023,
  author={Mohammad, Abdullahi and Masouros, Christos and Andreopoulos, Yiannis},
  journal={IEEE Open J. Commun. Soc.}, 
  title={{An Unsupervised Deep Unfolding Framework for Robust Symbol-Level Precoding}}, 
  year={2023},
  volume={4},
  number={},
  pages={1075-1090},
  doi={10.1109/OJCOMS.2023.3270455}}

@ARTICLE{NP_hard,
  author={Liu, Ya-Feng and Dai, Yu-Hong and Luo, Zhi-Quan},
  journal={IEEE Trans. Signal Process.}, 
  title={{Coordinated Beamforming for MISO Interference Channel: Complexity Analysis and Efficient Algorithms}}, 
  year={2011},
  volume={59},
  number={3},
  pages={1142-1157},
  doi={10.1109/TSP.2010.2092772}}

@ARTICLE{Matrixfree_BF_2022,
  author={Pellaco, Lissy and Bengtsson, Mats and Jaldén, Joakim},
  journal={IEEE Open J. Commun. Soc.}, 
  title={{Matrix-Inverse-Free Deep Unfolding of the Weighted MMSE Beamforming Algorithm}}, 
  year={2022},
  volume={3},
  number={},
  pages={65-81},
  doi={10.1109/OJCOMS.2021.3139858}}

@ARTICLE{Matrixfree_MU_MIMO_2022,
  author={Pellaco, Lissy and Jaldén, Joakim},
  journal={IEEE Trans. Signal Process.}, 
  title={{A Matrix-Inverse-Free Implementation of the MU-MIMO WMMSE Beamforming Algorithm}}, 
  year={2022},
  volume={70},
  number={},
  pages={6360-6375},
  doi={10.1109/TSP.2023.3238275}}

@ARTICLE{IrregularLDPC_2023,
  author={Guo, Xiaomeng and Chang, Tsung-Hui and Wang, Yongchao},
  journal={IEEE Commun. Lett.}, 
  title={{Model-Driven Deep Learning ADMM Decoder for Irregular Binary LDPC Codes}}, 
  year={2023},
  volume={27},
  number={2},
  pages={571-575},
  doi={10.1109/LCOMM.2022.3223114}}

@ARTICLE{LDPCNN_2023,
  author={Wang, Qing and Liu, Qing and Wang, Shunfu and Chen, Leian and Fang, Haoyu and Chen, Luyong and Guo, Yuzhang and Wu, Zhiqiang},
  journal={IEEE Trans. Cogn. Commun. Netw.}, 
  title={{Normalized Min-Sum Neural Network for LDPC Decoding}}, 
  year={2023},
  volume={9},
  number={1},
  pages={70-81},
  doi={10.1109/TCCN.2022.3212438}}

@ARTICLE{Polar_2020,
  author={Gao, Jian and Niu, Kai and Dong, Chao},
  journal={IEEE Access}, 
  title={{Learning to Decode Polar Codes With One-Bit Quantizer}}, 
  year={2020},
  volume={8},
  number={},
  pages={27210-27217},
  doi={10.1109/ACCESS.2020.2971526}}

@ARTICLE{PDD_LDPC_2022,
  author={Liu, Yihao and Zhao, Ming-Min and Wang, Chan and Lei, Ming and Zhao, Min-Jian},
  journal={IEEE Commun. Lett.}, 
  title={{PDD-Based Decoder for LDPC Codes With Model-Driven Neural Networks}}, 
  year={2022},
  volume={26},
  number={11},
  pages={2532-2536},
  doi={10.1109/LCOMM.2022.3199747}}

@ARTICLE{Bound_BP_2024,
  author={Adiga, Sudarshan and Xiao, Xin and Tandon, Ravi and Vasić, Bane and Bose, Tamal},
  journal={IEEE Trans. Inf. Theory}, 
  title={{Generalization Bounds for Neural Belief Propagation Decoders}}, 
  year={2024},
  volume={70},
  number={6},
  pages={4280-4296},
  doi={10.1109/TIT.2024.3361388}}

@ARTICLE{Algounroll_JADC,
  author={Shi, Yandong and Choi, Hayoung and Shi, Yuanming and Zhou, Yong},
  journal={IEEE Trans. Wireless Commun.}, 
  title={{Algorithm Unrolling for Massive Access via Deep Neural Networks With Theoretical Guarantee}}, 
  year={2022},
  volume={21},
  number={2},
  pages={945-959},
  doi={10.1109/TWC.2021.3100500}}

@ARTICLE{Dinklebach,
  author={Cheung, Kent Tsz Kan and Yang, Shaoshi and Hanzo, Lajos},
  journal={IEEE Trans. Commun.}, 
  title={{Achieving Maximum Energy-Efficiency in Multi-Relay OFDMA Cellular Networks: A Fractional Programming Approach}}, 
  year={2013},
  volume={61},
  number={7},
  pages={2746-2757},
  doi={10.1109/TCOMM.2013.13.120727}}

@ARTICLE{Lagrange_dual,
  author={Ng, Derrick Wing Kwan and Lo, Ernest S. and Schober, Robert},
  journal={IEEE Trans. Wireless Commun.}, 
  title={{Energy-Efficient Resource Allocation in OFDMA Systems with Hybrid Energy Harvesting Base Station}}, 
  year={2013},
  volume={12},
  number={7},
  pages={3412-3427},
  doi={10.1109/TWC.2013.052813.121589}}

@ARTICLE{Efficient_QAM_ADMM,
  author={Zhang, Quan and Wang, Jiangtao and Wang, Yongchao},
  journal={IEEE Trans. Wireless Commun.}, 
  title={{Efficient QAM Signal Detector for Massive MIMO Systems via PS/DPS-ADMM Approaches}}, 
  year={2022},
  volume={21},
  number={10},
  pages={8859-8871},
  doi={10.1109/TWC.2022.3170510}}

@ARTICLE{Interference_pricing,
  author={Ding, Zhiguo and Perlaza, Samir M. and Esnaola, Inaki and Poor, H. Vincent},
  journal={IEEE Trans. Wireless Commun.}, 
  title={{Power Allocation Strategies in Energy Harvesting Wireless Cooperative Networks}}, 
  year={2014},
  volume={13},
  number={2},
  pages={846-860},
  doi={10.1109/TWC.2013.010213.130484}}

@ARTICLE{SCA,
  author={Yang, Zhaohui and Chen, Mingzhe and Saad, Walid and Xu, Wei and Shikh-Bahaei, Mohammad and Poor, H. Vincent and Cui, Shuguang},
  journal={IEEE Trans. Wireless Commun.}, 
  title={{Energy-Efficient Wireless Communications With Distributed Reconfigurable Intelligent Surfaces}}, 
  year={2022},
  volume={21},
  number={1},
  pages={665-679},
  doi={10.1109/TWC.2021.3098632}}

@ARTICLE{Matching_theory,
  author={Liang, Wei and Ding, Zhiguo and Li, Yonghui and Song, Lingyang},
  journal={IEEE Trans. Commun.}, 
  title={{User Pairing for Downlink Non-Orthogonal Multiple Access Networks Using Matching Algorithm}}, 
  year={2017},
  volume={65},
  number={12},
  pages={5319-5332},
  doi={10.1109/TCOMM.2017.2744640}}

@ARTICLE{Unsupervised_WMMSE,
  author={Hu, Haifeng and Xie, Zhefei and Shi, Hongkui and Liu, Bin and Zhao, Haitao and Gui, Guan},
  journal={IEEE Trans. Veh. Technol.}, 
  title={{Unsupervised Power Allocation Based on Combination of Edge Aggregated Graph Attention Network with Deep Unfolded WMMSE}}, 
  year={2024},
  volume={},
  number={},
  pages={1-15},
  doi={10.1109/TVT.2024.3430353}}

@inproceedings{GAT,
  title={{Graph attention networks}},
  author={Velickovic, Petar and Cucurull, Guillem and Casanova, Arantxa and Romero, Adriana and Li{\`o}, Pietro and Bengio, Yoshua},
  booktitle={6th ICLR},
  year={2018},
  url={https://arxiv.org/abs/1710.10903}
}

@article{EGAT,
  title={{EGRET}: {Edge Aggregated Graph Attention Networks and Transfer Learning Improve Protein–Protein Interaction Site Prediction}},
  author={Mahbub, Sazan and Shamsuzzoha Bayzid, Md},
  journal={Briefings in Bioinformatics},
  volume={23},
  number={2},
  pages={bbab578},
  year={2022},
  publisher={Oxford University Press},
  doi={10.1093/bib/bbab578}
}

@ARTICLE{Survey9_DU_Optim,
  author={Shlezinger, Nir and Eldar, Yonina C. and Boyd, Stephen P.},
  journal={IEEE Access}, 
  title={{Model-Based Deep Learning: On the Intersection of Deep Learning and Optimization}}, 
  year={2022},
  volume={10},
  number={},
  pages={115384-115398},
  doi={10.1109/ACCESS.2022.3218802}}

@ARTICLE{learn_convex,
  author={Agrawal, Akshay and Barratt, Shane and Boyd, Stephen},
  journal={IEEE/CAA Journal of Automatica Sinica}, 
  title={Learning Convex Optimization Models}, 
  year={2021},
  volume={8},
  number={8},
  pages={1355-1364},
  doi={10.1109/JAS.2021.1004075}}

@inproceedings{learn_convex_control,
  title = {{Learning Convex Optimization Control Policies}},
  author = {Agrawal, Akshay and Barratt, Shane and Boyd, Stephen and Stellato, Bartolomeo},
  booktitle ={Proceedings of the 2nd Conf. on Learning for Dynamics and Control},
  pages = {361-373},
  year = 	 {2020},
  volume = {120},
  series = {Proceedings of Machine Learning Research},
  year = {2020},
  publisher =  {PMLR},
  url = {https://proceedings.mlr.press/v120/agrawal20a.html} 
}

@INPROCEEDINGS{Survey8_Model_guidelines,
  author={Shlezinger, Nir and Whang, Jay and Eldar, Yonina C. and Dimakis, Alexandros G.},
  booktitle={2021 IEEE DSLW}, 
  title={{Model-Based Deep Learning: Key Approaches and Design Guidelines}}, 
  year={2021},
  volume={},
  number={},
  pages={1-6},
  doi={10.1109/DSLW51110.2021.9523403}}

@ARTICLE{Survey5_Wireless_Both,
  author={Zappone, Alessio and Di Renzo, Marco and Debbah, Mérouane},
  journal={IEEE Trans. Commun.}, 
  title={{Wireless Networks Design in the Era of Deep Learning: Model-Based, AI-Based, or Both?}}, 
  year={2019},
  volume={67},
  number={10},
  pages={7331-7376},
  doi={10.1109/TCOMM.2019.2924010}}

@ARTICLE{Survey12_AI_MIMO,
  author={Shlezinger, Nir and Ma, Mengyuan and Lavi, Ortal and Nguyen, Nhan Thanh and Eldar, Yonina C. and Juntti, Markku},
  journal={IEEE Veh. Technol. Mag.}, 
  title={{Artificial Intelligence-Empowered Hybrid Multiple-Input/Multiple-Output Beamforming: Learning to Optimize for High-Throughput Scalable MIMO}}, 
  year={2024},
  volume={},
  number={},
  pages={2-11},
  doi={10.1109/MVT.2024.3396927}}

@book{Survey_10Eldar_MLandWC,
    author={Yonina C. Eldar, Andrea Goldsmith. Deniz Gündüz and H. Vincent Poor},
    title={{Machine Learning and Wireless Communications}}, 
    publisher={Cambridge University Press}, 
    year={2022}}

@INPROCEEDINGS{DU_ISACOTFS_2024,
  author={Lin, Weizhi and Zheng, Haifeng and Feng, Xinxin and Chen, Youjia},
  booktitle={2024 IEEE WCNC}, 
  title={{Deep Unfolding Network for Target Parameter Estimation in OTFS-based ISAC Systems}}, 
  year={2024},
  volume={},
  number={},
  pages={1-6},
  doi={10.1109/WCNC57260.2024.10571155}}

@ARTICLE{JRC_survey,
  author={Feng, Zhiyong and Fang, Zixi and Wei, Zhiqing and Chen, Xu and Quan, Zhi and Ji, Danna},
  journal={China Communications}, 
  title={{Joint radar and communication: A survey}}, 
  year={2020},
  volume={17},
  number={1},
  pages={1-27},
  doi={10.23919/JCC.2020.01.001}}

@ARTICLE{ISAC_waveform,
  author={Zhou, Wenxing and Zhang, Ruoyu and Chen, Guangyi and Wu, Wen},
  journal={IEEE Open J. Commun. Soc.}, 
  title={{Integrated Sensing and Communication Waveform Design: A Survey}}, 
  year={2022},
  volume={3},
  number={},
  pages={1930-1949},
  doi={10.1109/OJCOMS.2022.3215683}}

@ARTICLE{ISAC_Net_2024,
  author={Jiang, Wangjun and Ma, Dingyou and Wei, Zhiqing and Feng, Zhiyong and Zhang, Ping and Peng, Jinlin},
  journal={IEEE Trans. Commun.}, 
  title={{ISAC-NET: Model-driven Deep Learning for Integrated Passive Sensing and Communication}}, 
  year={2024},
  volume={},
  number={},
  pages={1-1},
  doi={10.1109/TCOMM.2024.3375818}}

@ARTICLE{Enabling_ISAC_survey,
  author={Zhang, J. Andrew and Rahman, Md. Lushanur and Wu, Kai and Huang, Xiaojing and Guo, Y. Jay and Chen, Shanzhi and Yuan, Jinhong},
  journal={IEEE Commun. Surveys Tuts.}, 
  title={{Enabling Joint Communication and Radar Sensing in Mobile Networks—A Survey}}, 
  year={2022},
  volume={24},
  number={1},
  pages={306-345},
  doi={10.1109/COMST.2021.3122519}}

@INPROCEEDINGS{TargetEst_2023,
  author={Lin, Jingcai and Zheng, Haifeng and Feng, Xinxin},
  booktitle={2023 WCSP}, 
  title={{Target Parameter Estimation with Deep Unfolding Networks for MIMO-OFDM Based Integrated Sensing and Communication Systems}}, 
  year={2023},
  volume={},
  number={},
  pages={371-376},
  doi={10.1109/WCSP58612.2023.10404645}}

@ARTICLE{Seventy_ISAC,
  author={Liu, Fan and Zheng, Le and Cui, Yuanhao and Masouros, Christos and Petropulu, Athina P. and Griffiths, Hugh and Eldar, Yonina C.},
  journal={IEEE Signal Process. Mag.}, 
  title={{Seventy Years of Radar and Communications: The road from separation to integration}}, 
  year={2023},
  volume={40},
  number={5},
  pages={106-121},
  doi={10.1109/MSP.2023.3272881}}

@ARTICLE{Survey3_DL_in_PhyLayer,
  author={Qin, Zhijin and Ye, Hao and Li, Geoffrey Ye and Juang, Biing-Hwang Fred},
  journal={IEEE Wirel. Commun.}, 
  title={{Deep Learning in Physical Layer Communications}}, 
  year={2019},
  volume={26},
  number={2},
  pages={93-99},
  doi={10.1109/MWC.2019.1800601}}

@ARTICLE{Fundamental_limit_ISAC,
  author={Liu, An and Huang, Zhe and Li, Min and Wan, Yubo and Li, Wenrui and Han, Tony Xiao and Liu, Chenchen and Du, Rui and Tan, Danny Kai Pin and Lu, Jianmin and Shen, Yuan and Colone, Fabiola and Chetty, Kevin},
  journal={IEEE Commun. Surveys Tuts.}, 
  title={{A Survey on Fundamental Limits of Integrated Sensing and Communication}}, 
  year={2022},
  volume={24},
  number={2},
  pages={994-1034},
  doi={10.1109/COMST.2022.3149272}}

@ARTICLE{JRCDesign,
  author={Liu, Fan and Masouros, Christos and Petropulu, Athina P. and Griffiths, Hugh and Hanzo, Lajos},
  journal={IEEE Trans. Commun.}, 
  title={{Joint Radar and Communication Design: Applications, State-of-the-Art, and the Road Ahead}}, 
  year={2020},
  volume={68},
  number={6},
  pages={3834-3862},
  doi={10.1109/TCOMM.2020.2973976}}

@unpublished{DU_constantmod_2023,
  author = {Krishnananthalingam, Prashanth and Nguyen, Nhan Thanh and Juntti, Markku},
  year = {2023},
  month = {06},
  pages = {},
  title = {{Deep Unfolding Enabled Constant Modulus Waveform Design for Joint Communications and Sensing}},
  url={https://arxiv.org/abs/2306.14702},
}

@article{JSAC_HBF_DU,
    author = {Nguyen, Nhan Thanh and Nguyen, Ly and Shlezinger, Nir and Eldar, Yonina and Swindlehurst, A. and Juntti, Markku},
    journal ={IEEE J. Sel. Top. Signal Proces.},
    year = {2024},
    month = {07},
    pages = {},
    title = {{Joint Communications and Sensing Hybrid Beamforming Design via Deep Unfolding}} 
}

@ARTICLE{CSI_ISTA_2024,
  author={Zhang, Yangyang and Zhang, Xichang and Liu, Yi},
  journal={IEEE Trans. Veh. Technol.}, 
  title={{Lightweight Differential Frameworks for CSI Feedback in Time-Varying Massive MIMO Systems}}, 
  year={2024},
  volume={73},
  number={5},
  pages={6878-6893},
  doi={10.1109/TVT.2023.3345938}}

@ARTICLE{Generalized_AMP,
  author={Zhao, Yan and Xiao, Yue and Yang, Ping and Dong, Binhong and Shi, Rong and Deng, Ke},
  journal={IEEE Trans. Veh. Technol.}, 
  title={{Generalized Approximate Message Passing Aided Frequency Domain Turbo Equalizer for Single-Carrier Spatial Modulation}}, 
  year={2018},
  volume={67},
  number={4},
  pages={3630-3634},
  doi={10.1109/TVT.2017.2774042}}

@ARTICLE{MP_detector_2024,
  author={Zeng, Yuwei and Ge, Yingmeng and Tan, Xiaosi and Ji, Zhenhao and Zhang, Zaichen and You, Xiaohu and Zhang, Chuan},
  journal={IEEE Trans. Veh. Technol.}, 
  title={{A Deep-Learning-Aided Message Passing Detector for MIMO SC-FDMA}}, 
  year={2024},
  volume={73},
  number={7},
  pages={10767-10771},
  doi={10.1109/TVT.2024.3366244}
}

@ARTICLE{IRS_RGDescent_2024,
  author={Zhu, Gangyong and Hu, Jinfeng and Zhong, Kai and Cheng, Xin and Song, Ziyun},
  journal={IEEE Signal Process. Lett. }, 
  title={{Sum-Path-Gain Maximization for IRS-Aided MIMO Communication System via Riemannian Gradient Descent Network}}, 
  year={2024},
  volume={31},
  number={},
  pages={51-55},
  doi={10.1109/LSP.2023.3340611}}

@ARTICLE{ADMM_Precoding_2024,
  author={Yang, Junwen and Li, Ang and Liao, Xuewen and Masouros, Christos},
  journal={IEEE Trans. Veh. Technol.}, 
  title={{ADMM-SLPNet: A Model-Driven Deep Learning Framework for Symbol-Level Precoding}}, 
  year={2024},
  volume={73},
  number={1},
  pages={1376-1381},
  doi={10.1109/TVT.2023.3301241}}

@ARTICLE{CSIFeedback_DU_2024,
  author={Liu, Zhenyu and Wang, Li and Xu, Lianming and Ding, Zhi},
  journal={IEEE Trans. Wireless Commun.}, 
  title={{Deep Learning for Efficient CSI Feedback in Massive MIMO: Adapting to New Environments and Small Datasets}}, 
  year={2024},
  volume={},
  number={},
  pages={1-1},
  doi={10.1109/TWC.2024.3390583}}

@ARTICLE{DL_Detection_MarkerCodes_2024,
  author={Ma, Guochen and Jiao, Xiaopeng and Mu, Jianjun and Han, Hui and Yang, Yaming},
  journal={IEEE Trans. Commun.}, 
  title={{Deep Learning-Based Detection for Marker Codes over Insertion and Deletion Channels}}, 
  year={2024},
  volume={},
  number={},
  pages={1-1},
  doi={10.1109/TCOMM.2024.3394039}}

@ARTICLE{ChannelEst_mmWave_MassiveMIMO_2024,
  author={Wu, Pengxia and Cheng, Julian and Eldar, Yonina C. and Cioffi, John M.},
  journal={IEEE Trans. Commun.}, 
  title={{Learned Trimmed-Ridge Regression for Channel Estimation in Millimeter-Wave Massive MIMO}}, 
  year={2024},
  volume={},
  number={},
  pages={1-1},
  doi={10.1109/TCOMM.2024.3440875}}

@ARTICLE{Multicell_2023,
  author={Schynol, Lukas and Pesavento, Marius},
  journal={IEEE J. Sel. Areas Commun. }, 
  title={{Coordinated Sum-Rate Maximization in Multicell MU-MIMO With Deep Unrolling}}, 
  year={2023},
  volume={41},
  number={4},
  pages={1120-1134},
  doi={10.1109/JSAC.2023.3242716}}

@ARTICLE{WMMSE,
  author={Shi, Qingjiang and Razaviyayn, Meisam and Luo, Zhi-Quan and He, Chen},
  journal={IEEE Trans. Signal Process.}, 
  title={{An Iteratively Weighted MMSE Approach to Distributed Sum-Utility Maximization for a MIMO Interfering Broadcast Channel}}, 
  year={2011},
  volume={59},
  number={9},
  pages={4331-4340},
  doi={10.1109/TSP.2011.2147784}}

@ARTICLE{IDE,
  author={Wang, Chang-Jen and Wen, Chao-Kai and Jin, Shi and Tsai, Shang-Ho},
  journal={IEEE Trans. Wireless Commun.}, 
  title={{Finite-Alphabet Precoding for Massive MU-MIMO With Low-Resolution DACs}}, 
  year={2018},
  volume={17},
  number={7},
  pages={4706-4720},
  doi={10.1109/TWC.2018.2830343}}

@ARTICLE{IntelligentRadio,
  author={Pham, Quoc-Viet and Nguyen, Nhan Thanh and Huynh-The, Thien and Bao Le, Long and Lee, Kyungchun and Hwang, Won-Joo},
  journal={IEEE Access}, 
  title={{Intelligent Radio Signal Processing: A Survey}}, 
  year={2021},
  volume={9},
  number={},
  pages={83818-83850},
  doi={10.1109/ACCESS.2021.3087136}}

@ARTICLE{OFDM_basics,
  author={Weinstein, S. and Ebert, P.},
  journal={IEEE Trans. Commun.}, 
  title={{Data Transmission by Frequency-Division Multiplexing Using the Discrete Fourier Transform}}, 
  year={1971},
  volume={19},
  number={5},
  pages={628-634},
  doi={10.1109/TCOM.1971.1090705}
}

@ARTICLE{ISAC_BF_2025,
  author={Zhang, Jianjun and Masouros, Christos and Liu, Fan and Huang, Yongming and Swindlehurst, A. Lee},
  journal={IEEE J. Sel. Topics Signal Process. }, 
  title={{Low-Complexity Joint Radar-Communication Beamforming: From Optimization to Deep Unfolding}}, 
  year={2025},
  volume={},
  number={},
  pages={1-16},
  doi={10.1109/JSTSP.2024.3522787}}

@ARTICLE{BF_IRS_2024,
  author={Jin, Weijie and Zhang, Jing and Wen, Chao-Kai and Jin, Shi and Li, Xiao and Han, Shuangfeng},
  journal={IEEE Trans. Wireless Commun.}, 
  title={{Low-Complexity Joint Beamforming for RIS-Assisted MU-MISO Systems Based on Model-Driven Deep Learning}}, 
  year={2024},
  volume={23},
  number={7},
  pages={6968-6982},
  doi={10.1109/TWC.2023.3336742}}

@ARTICLE{Inexact_ADMM_det_2021,
  author={Kim, Minsik and Park, Daeyoung},
  journal={IEEE Trans. Wireless Commun.}, 
  title={{Learnable MIMO Detection Networks Based on Inexact ADMM}}, 
  year={2021},
  volume={20},
  number={1},
  pages={565-576},
  doi={10.1109/TWC.2020.3026471}}

@ARTICLE{Yu_alternating_HBF,
  author={Yu, Xianghao and Shen, Juei-Chin and Zhang, Jun and Letaief, Khaled B.},
  journal={IEEE J. Sel. Topics Signal Process. }, 
  title={{Alternating Minimization Algorithms for Hybrid Precoding in Millimeter Wave MIMO Systems}}, 
  year={2016},
  volume={10},
  number={3},
  pages={485-500},
  doi={10.1109/JSTSP.2016.2523903}}

@ARTICLE{liu2018mu,
  author={Liu, Fan and Masouros, Christos and Li, Ang and Sun, Huafei and Hanzo, Lajos},
  journal={IEEE Trans. Wireless Commun.}, 
  title={{MU-MIMO Communications With MIMO Radar: From Co-Existence to Joint Transmission}}, 
  year={2018},
  volume={17},
  number={4},
  pages={2755-2770},
  }

@inproceedings{ma2024model,
  title={Model-Based Machine Learning for Max-Min Fairness Beamforming Design in {JCAS} Systems},
  author={Ma, Mengyuan and Fang, Tianyu and Shlezinger, Nir and Swindlehurst, AL and Juntti, Markku and Nguyen, Nhan},
  booktitle={IEEE Int. Conf. Acoustics, Speech Signal Process. (ICASSP)},
  year={2025}
}

@inproceedings{nguyen2023deep,
  title={Deep unfolding-enabled hybrid beamforming design for {mmWave} massive {MIMO} systems},
  author={Nguyen, Nhan and Ma, Mengyuan and Shlezinger, Nir and Eldar, Yonina C and Swindlehurst, A Lee and Juntti, Markku},
  booktitle={IEEE Int. Conf. Acoustics, Speech Signal Process. (ICASSP)},
  year={2023}
}

@ARTICLE{ista_superresolution,
  author={Deng, Xin and Dragotti, Pier Luigi},
  journal={IEEE Trans. Image Process.}, 
  title={{Deep Coupled ISTA Network for Multi-Modal Image Super-Resolution}}, 
  year={2020},
  volume={29},
  number={},
  pages={1683-1698},
  doi={10.1109/TIP.2019.2944270}}

@ARTICLE{ista_compressedsensing,
  author={Zheng, Ziyang and Dai, Wenrui and Xue, Duoduo and Li, Chenglin and Zou, Junni and Xiong, Hongkai},
  journal={IEEE Trans. Pattern Anal. Mach. Intell.}, 
  title={{Hybrid ISTA: Unfolding ISTA With Convergence Guarantees Using Free-Form Deep Neural Networks}}, 
  year={2023},
  volume={45},
  number={3},
  pages={3226-3244},
  doi={10.1109/TPAMI.2022.3172214}}

@ARTICLE{ista_denoising,
  author={Zhang, Jian and Zhang, Zhenyu and Xie, Jingfen and Zhang, Yongbing},
  journal={IEEE J. Sel. Topics Signal Process.}, 
  title={{High-Throughput Deep Unfolding Network for Compressive Sensing MRI}}, 
  year={2022},
  volume={16},
  number={4},
  pages={750-761},
  doi={10.1109/JSTSP.2022.3170227}}
\end{document}